\documentclass[pdflatex,sn-mathphys-num]{sn-jnl}

\usepackage{graphicx}%
\usepackage{multirow}%
\usepackage{amsmath,amssymb,amsfonts}%
\usepackage{amsthm}%
\usepackage{mathrsfs}%
\usepackage[title]{appendix}%
\usepackage{xcolor}%
\usepackage{textcomp}%
\usepackage{manyfoot}%
\usepackage{booktabs}%
\usepackage{float}
\usepackage{algorithm}%
\usepackage{algorithmicx}%
\usepackage{algpseudocode}%
\usepackage{listings}%
\usepackage{subcaption}
\usepackage{cleveref}
\usepackage{siunitx}%
\usepackage{microtype}

\theoremstyle{thmstyleone}%
\theoremstyle{thmstyletwo}%

\theoremstyle{thmstylethree}%

\begin{document}

\title[Article Title]{Uncovering EEG Patterns Consistently Associated with Cybersickness Discomfort Using Deep Learning Interpretability Maps}


\author*[1]{\fnm{Jacqueline} \sur{Yau}}\email{jhyau2@illinois.edu}


\author[2]{\fnm{Evan G.} \sur{Center}}\email{
evan.center@oulu.fi}

\author[3]{\fnm{Paweł} \sur{Augustynowicz}}\email{pawel.augustynowicz@kul.pl}

\author[1,2]{\fnm{Steven M.} \sur{LaValle}}\email{steven.lavalle@oulu.fi}

\author[2]{\fnm{Timo} \sur{Ojala}}\email{timo.ojala@oulu.fi}

\author[1]{\fnm{Wenzhen} \sur{Yuan}}\email{yuanwz@illinois.edu}

\author[1,5]{\fnm{Nancy M.} \sur{Amato}}\email{namato@illinois.edu}

\author[3]{\fnm{Paweł} \sur{Stróżak}}\email{pawel.strozak@kul.pl}

\author[1]{\fnm{Minje} \sur{Kim}}\email{minje@illinois.edu}

\author[4,5]{\fnm{Kara D.} \sur{Federmeier}}\email{kfederme@illinois.edu}

\author*[1]{\fnm{Katherine J.} \sur{Mimnaugh}}\email{kmimnau2@illinois.edu}

\affil*[1]{\orgdiv{Siebel School of Computing and Data Science}, \orgname{University of Illinois Urbana-Champaign}, \orgaddress{ \city{Urbana}, \state{IL}, \country{USA}}}

\affil[2]{\orgdiv{Faculty of Information Technology and Electrical Engineering}, \orgname{University of Oulu}, \orgaddress{\city{Oulu}, \country{Finland}}}

\affil[3]{\orgdiv{Department of Experimental Psychology, Institute of Psychology}, \orgname{The John Paul II Catholic University of Lublin}, \orgaddress{\city{Lublin}, \country{Poland}}}

\affil[4]{\orgdiv{Department of Psychology}, \orgname{University of Illinois Urbana-Champaign}, \orgaddress{\city{Champaign}, \state{IL}, \country{USA}}}

\affil[5]{\orgdiv{Beckman Institute for Advanced Science and Technology}, \orgname{University of Illinois Urbana-Champaign}, \orgaddress{\city{Urbana}, \state{IL}, \country{USA}}}




\abstract{
Uncomfortable sensations similar to motion sickness, called cybersickness, can develop when using Virtual Reality (VR) head-mounted displays. Cybersickness poses a hindrance to greater use of VR technology. Brain activity recorded using electroencephalogram (EEG) can be used to unintrusively detect cybersickness and the discomfort it causes.  To intervene for mitigation,
machine learning algorithms that can extract meaningful signals related to cybersickness discomfort 
from the rest of the brain data will be required. 
In this work, we determined which features in EEG data were most helpful for classifying cybersickness-related discomfort by building a framework with neural networks and interpretability maps. 
Using brain data from two separate auditory event-related potential (ERP) cybersickness user studies, 
we extracted which spatio-temporal EEG features (from sensor locations and time steps) were most important for discomfort classification. For the first dataset (n=29), across 120 runs of our framework with three different neural networks over multiple random seeds, the models consistently pointed to scalp locations in left frontal, midline central, and right parietal areas as 
most helpful in determining if 
EEG data belonged to someone who was experiencing cybersickness discomfort. Similar locations were also tagged in a second dataset (n=36), showing cross-dataset generalizability of our findings across time, VR stimulus, and participant sample. Early periods in the extracted ERP time windows, approximately between 80 and 260 milliseconds post-stimulus, were marked as important for model classification in both datasets.
These results help clarify 
which tagged features can be used for cybersickness-related (and potentially other types of) discomfort classification with EEG in the future. 
We will provide our code [code released on acceptance]. Project webpage at: \href{https://eeg-discomfort-nnilc.github.io/}{https://eeg-discomfort-nnilc.github.io/}. 
}




\maketitle

\begin{figure}[htbp]
    \centering

    \begin{subfigure}[b]{0.49\textwidth}
        \centering
        \includegraphics[width=\textwidth]{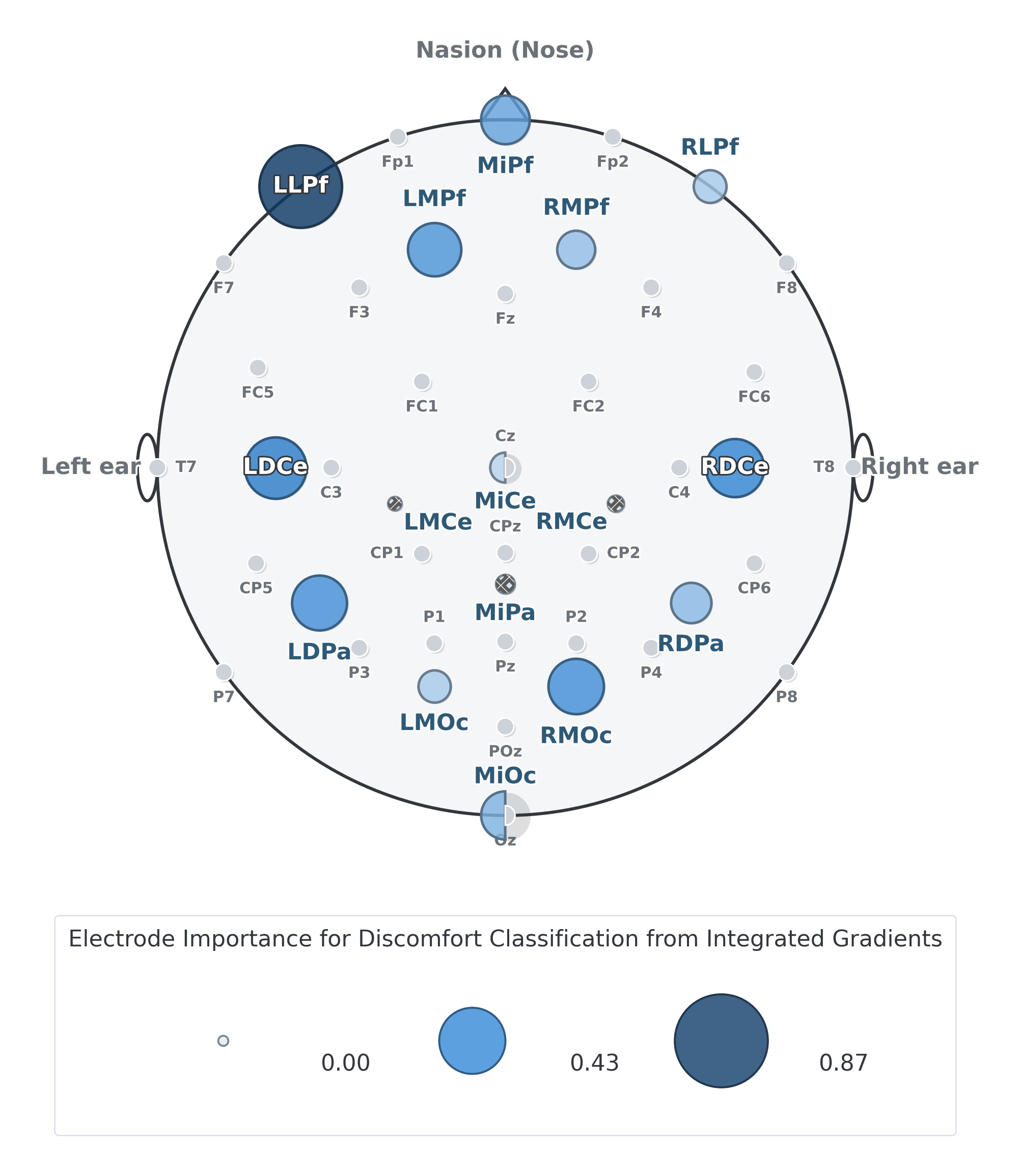}
        \caption{Discomfort Group (ERPVR)}
        \label{fig:erpvr-discomfort}
    \end{subfigure}
    \hfill
    \begin{subfigure}[b]{0.49\textwidth}
        \centering
        \includegraphics[width=\textwidth]{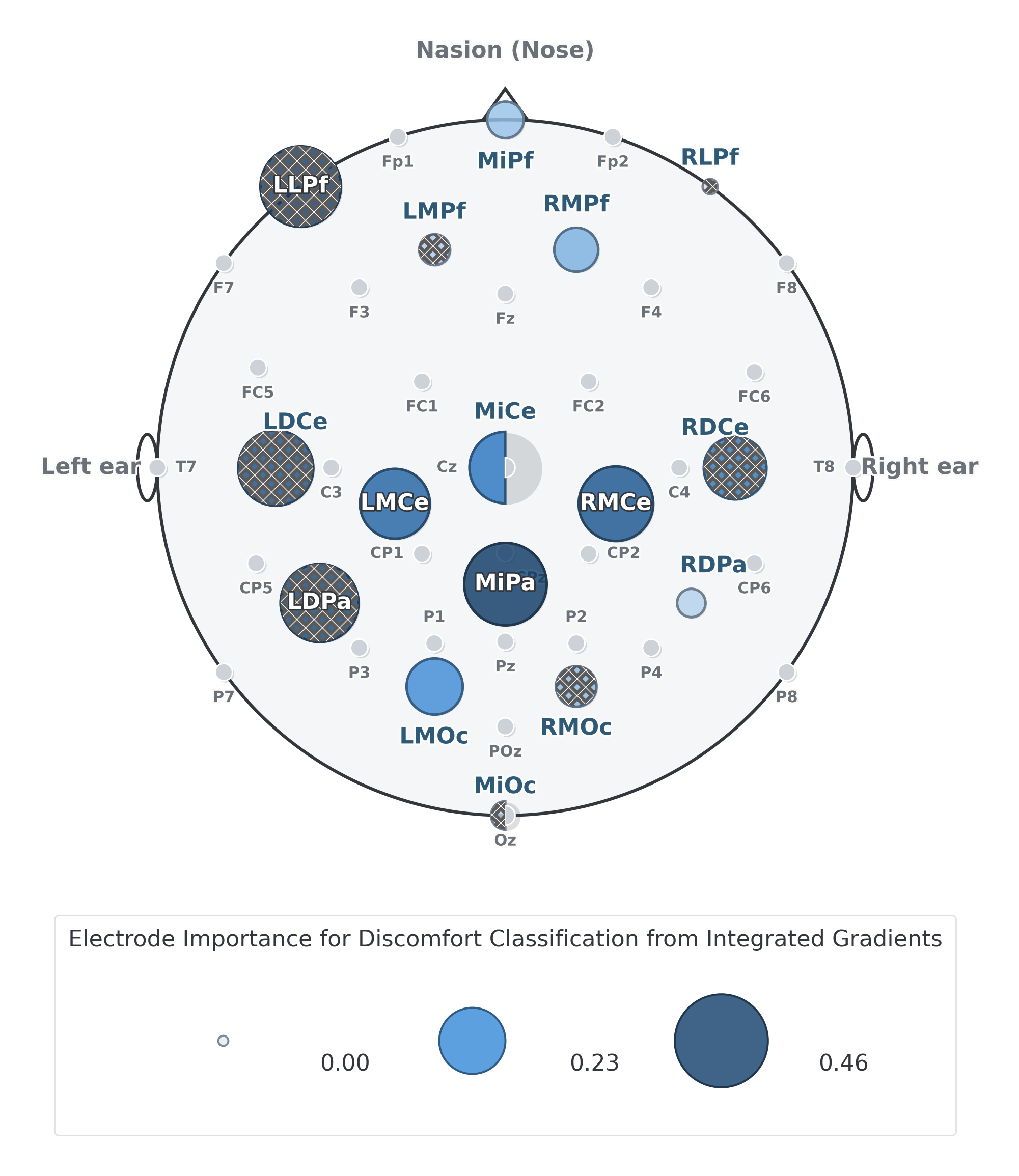}
        \caption{No Discomfort Group (ERPVR)}
        \label{fig:erpvr-no-discomfort}
    \end{subfigure}

    \vskip\baselineskip

    \begin{subfigure}[b]{0.49\textwidth}
        \centering
        \includegraphics[width=\textwidth]{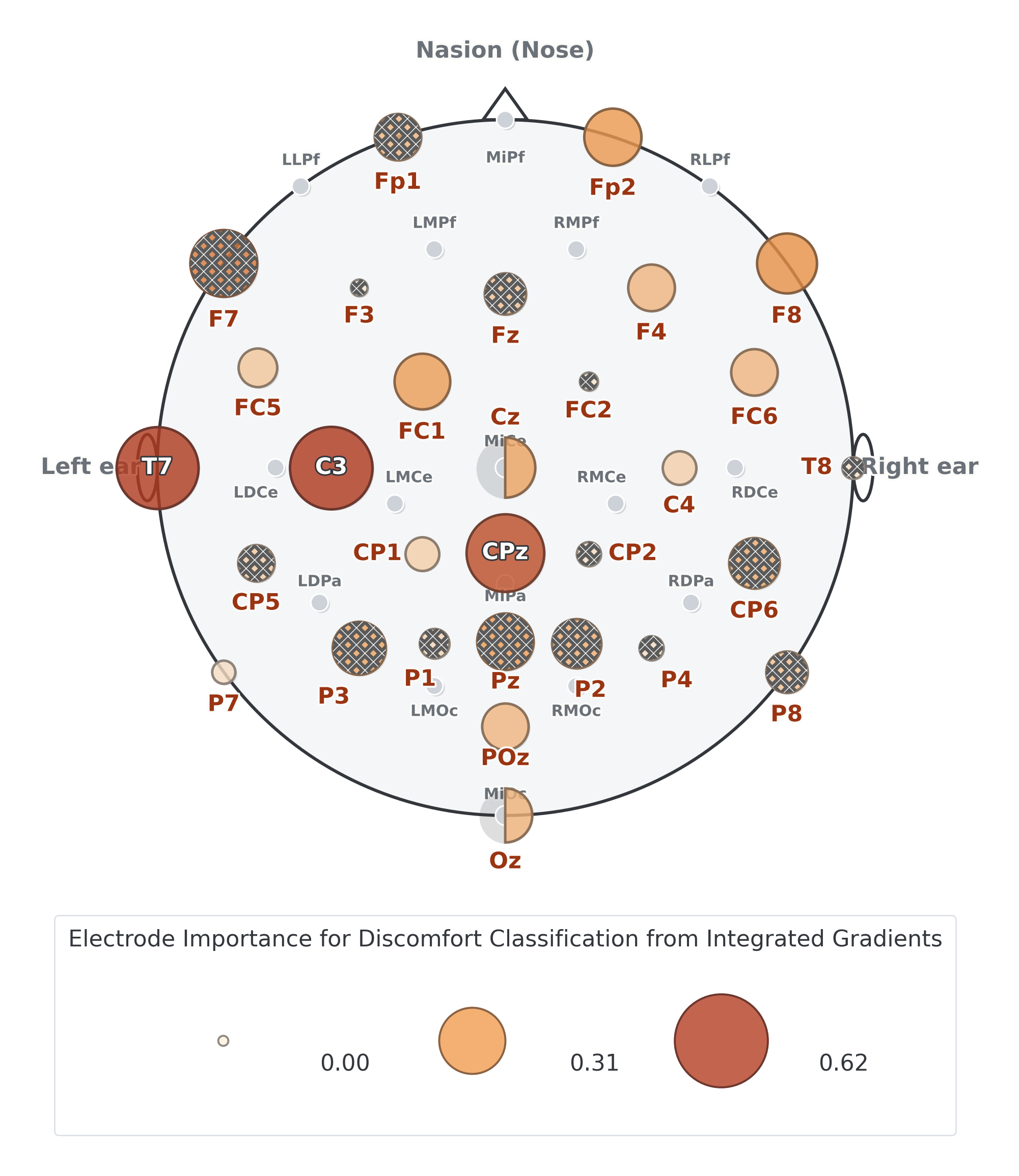}
        \caption{High Discomfort Group (VectionVR)}
        \label{fig:vectionvr-high}
    \end{subfigure}
    \hfill
    \begin{subfigure}[b]{0.49\textwidth}
        \centering
        \includegraphics[width=\textwidth]{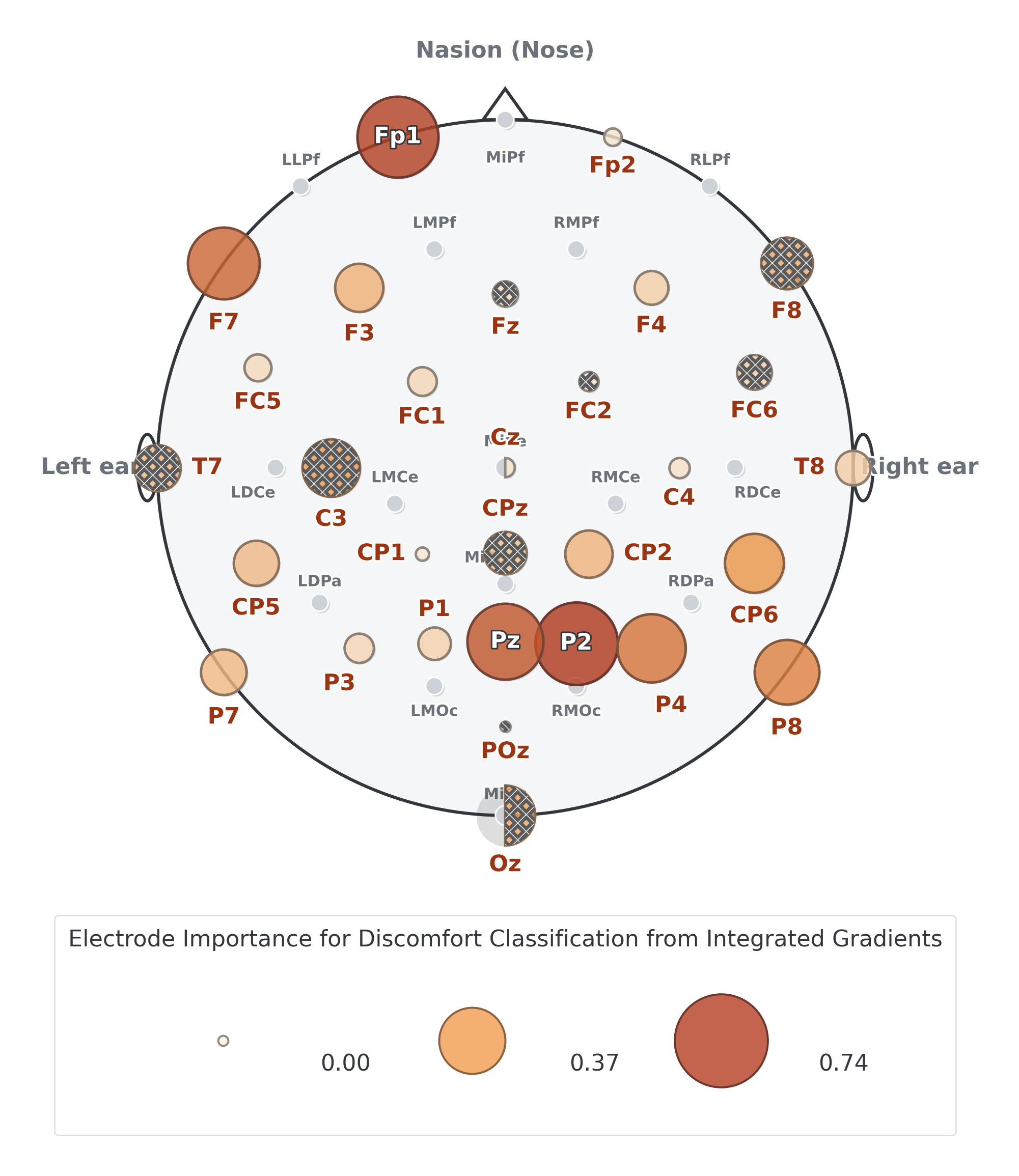}
        \caption{Low Discomfort Group (VectionVR)}
        \label{fig:vectionvr-low}
    \end{subfigure}

    \caption{Importance of electrode features for discomfort classification across all runs. ERPVR (top) used a 16-channel approximately equidistant montage, and VectionVR (bottom) used a modified 30-channel 10-20 layout. Circle sizes and colors are based on calculated integrated gradients. Circles with crosshatch indicate that feature was negatively contributing to model classification at that magnitude. The top three most important electrodes have labels inside their circles. 
    These integrated gradients are the same as in Figures~\ref{fig:erpvr_continuous_ig_sum} and~\ref{fig:vectionvr-ig-normmax}.}
    \label{fig:electrode-montages-all}
\end{figure}

\section{Introduction}\label{sec1}

Virtual reality (VR) is an effective medium for training surgeons, educating students, treating pain, and improving mental health~\cite{lavalle2023virtual}. However, the unpleasant symptoms that develop from using VR devices, called cybersickness, preclude long duration applications and makes it difficult for some individuals to use VR at all~\cite{Chang_Kim_Yoo_2020}.  
At the moment, there are few options for precise passive cybersickness detection. Thus, there has been substantial effort in the scientific community to find methods to unintrusively detect cybersickness 
using brain activity recorded with  electroencephalogram (EEG)~\cite{Chang_Billinghurst_Yoo_2023}. 

Given the complexity of brain data, it is desirable to find features indicative of cybersickness that can be used by machine learning models (ML) to improve classification. 
If we can determine which features to focus the models on, it is possible that cybersickness and/or the discomfort that it causes could be detected as it occurs. 
It might even be possible to detect discomfort from brain activity even before VR users report it, because pain-related sensory processing can precede conscious awareness~\cite{Bastuji2016,Garcia-Larrea2018}. However, diagnostic patterns in brain activity could occur at different time points, electrode locations, band power changes, or some combination of the three. Determining which of these features is predictive of discomfort requires overcoming considerable hurdles, including limited datasets for training models, data with high inter-subject variability, and confounds in the signal from unrelated brain activity.

In this article, we describe an extensive investigation of EEG data using a number of supervised deep learning methods, and present the resulting EEG features consistently found to be associated with cybersickness-related discomfort. In overcoming the aforementioned obstacles, we are able to offer several important contributions. First, we provide a feature extraction framework (and its code) that highlights which input features from brain data provide the highest attribution for model classification. 
While much previous ML work has investigated changes in power in frequency bands like alpha and beta (see~\cite{Chang_Billinghurst_Yoo_2023,Yang_Kasabov_Cakmak_2022,Yildirim_2020} for review), here we have built an explainable classification method for EEG data from event-related potential (ERP) studies, a much less studied area with better comparison across samples. 
Second, although the attribution maps displayed complex interactions across electrodes and time, multiple models across multiple random seeds showed consistent results pointing to the same subset of electrodes as being the most helpful in classifying cybersickness-related discomfort in one dataset (called ERPVR): LLPf, LDCe, RDCe, RMOc, and LDPa for samples with discomfort and MiPa, RMCe, LMCe, MiCe and LMOc for samples without discomfort. We also found related patterns in the second dataset (VectionVR): C3, T7, CPz, F8, and Fp2 for high levels of discomfort and P2, Fp1, Pz, F7, and P4 for low levels of discomfort (see Figure~\ref{fig:electrode-montages-all}).  
The datasets were collected at different times, in different countries, with different VR headsets, different EEG systems, different electrode locations, and different stimuli in virtual reality. 
Thus, the results presented here provide a contribution to the scientific literature on important features in auditory ERP data that could be used as tagged features for detection of cybersickness discomfort.

\section{Background}

\subsection{Electroencephalography and Event-Related Potentials} \label{EEG-background-sec}


EEG data recorded at the scalp reflect voltage fluctuations generated primarily by summed postsynaptic activity of populations of synchronously active neurons~\cite{Luck_2014}. These signals reflect sensory, motor, cognitive, and other ongoing neural processes. 
Because EEG reflects many simultaneous processes, activity unrelated to the process of interest contributes ``noise" relative to the signal being studied, even when biological, equipment, and environmental artifacts have been well controlled. One method to address this issue of noise involves repeatedly invoking patterns of brain activity in response to a stimulus, eliciting what are known as Event-Related Potentials (ERPs). ERPs can be extracted by averaging across those repetitions 
to cancel out 
inconsistent activation~\cite{Luck_2014}. 

During VR sickness studies, continuous EEG reflects multiple concurrent processes, 
including processing of the visual environment, optic flow, and activity associated with discomfort. Optic flow can induce \textit{vection}, illusory sensations of self-motion, which is itself an important contributor to cybersickness~\cite{Keshavarz_Riecke_Hettinger_Campos_2015}. 
Auditory stimuli presented during visual VR experiences provide repeated, precisely time-locked events against this background of ongoing activity. As presentation of the same auditory stimulus engages broadly similar sensory and cognitive processing across trials, the resulting ERPs provide a common reference point for examining differences in brain activity across periods in which participants did and did not experience discomfort~\cite{Luck_2014}. 

During ERP data analysis, data is segmented from the continuous raw EEG signal beginning from typically 100 to 200 ms prior to onset and ending one to two seconds after the stimulus. This time window  of EEG data around the stimulus onset for a single stimulus (trial) is called an epoch. Changes in the signal at the trial and condition (sets of trials within an experimental manipulation) level can reveal novel and consequential insights into the neural and cognitive processes engaged by the stimulus~\cite{Donchin_1981,Luck_2014,Woodman_2010}.

Throughout more than half a century of research, reliable ERP waveform features (components) 
have been mapped to cognitive states and resources. For example, the P3b component (also called the P3 or P300), is a positive-going potential that is typically maximal over centro-parietal electrodes beginning as early as 300 ms after an attended, task-relevant stimulus
~\cite{Polich_2007,Ghani_Signal_Niazi_Taylor_2020,Verleger2020,Vanbilsen_Kotz_Rosso_Leman_Triccas_Feys_Moumdjian_2023,Ladouce2019}. 
The amplitude of the P3b has consistently been shown to be influenced by task difficulty, cognitive load, and attention,  
and has been used in myriad studies to investigate attentional allocation
~\cite{Vanbilsen_Kotz_Rosso_Leman_Triccas_Feys_Moumdjian_2023,Ghani_Signal_Niazi_Taylor_2020,Polich_2007}.
One way to elicit a P3b is with an auditory oddball paradigm, wherein participants listen to a series of similar sounds (the standards). Randomly and infrequently, one of the sounds will be different (the oddballs), and participants must note these or make a response when they hear an oddball~\cite{Polich_2007}. 
In previous work, we showed that changes in the auditory P3b were found to be associated with the intensity of cybersickness experienced~\cite{Mimnaugh2023}. However, there are likely other patterns present in the ERP data that reflect sickness and discomfort, including in response to the presentation of the standards, which elicit much less P3b activity. Uncovering these previously unknown discomfort-related patterns was the motivation for this work.

\subsection{Cybersickness and Discomfort}

When a person uses a VR 
head-mounted display, they sometimes develop feelings of discomfort, called cybersickness~\cite{Chang_Kim_Yoo_2020}. Cybersickness symptoms include nausea, dizziness, eyestrain, headache, and fatigue, among others~\cite{Caserman_Garcia-Agundez_GámezZerban_Göbel_2021,Chang_Kim_Yoo_2020,Saredakis_Szpak_Birckhead_Keage_Rizzo_Loetscher_2020}. If the sensations arise from using VR specifically, this discomfort can be referred to as VR sickness~\cite{lavalle2023virtual}. 
Theories on the causes for cybersickness include 
a conflict between the expectation of particular patterns of incoming sensory information and the patterns of sensory information experienced in that moment~\cite{Reason_Brand_1975,Reason_1978}, intravestibular imbalances~\cite{Previc_2018}, vestibulo-autonomic responses~\cite{Bogle_Benarroch_Sandroni_2022,Cohen_Dai_Yakushin_Cho_2019,Muth_2006}, 
head pose in the virtual environment~\cite{Palmisano_Allison_Kim_2020}, unexpected sensations of self-motion (vection)~\cite{Teixeira_Miellet_Palmisano_2022,Teixeira_Miellet_Palmisano_2024}, 
and others~\cite{Stanney_Lawson_Rokers_Dennison_Fidopiastis_Stoffregen_Weech_Fulvio_2020,lavalle2023virtual}. 

There are a number of contributors to cybersickness. These include hardware, software, environmental, and individual factors~\cite{Biswas_Mukherjee_Bhattacharya_2024,Caserman_Garcia-Agundez_GámezZerban_Göbel_2021,Chang_Kim_Yoo_2020,Tian_Lopes_Boulic_2022,Stanney_Lawson_Rokers_Dennison_Fidopiastis_Stoffregen_Weech_Fulvio_2020,Saredakis_Szpak_Birckhead_Keage_Rizzo_Loetscher_2020}. While not everyone that uses VR experiences discomfort or gets sick, a significant portion of users can develop adverse symptoms~\cite{Stanney_Lawson_Rokers_Dennison_Fidopiastis_Stoffregen_Weech_Fulvio_2020,Garrido_Frías-Hiciano_Moreno-Jiménez_Cruz_García-Batista_Guerra-Peña_Medrano_2022,Szpak_Michalski_Saredakis_Chen_Loetscher_2019}. Although brief, these symptoms can sometimes last from 10 minutes to four hours after exposure in the HMD has ended~\cite{Dużmańska_Strojny_Strojny_2018}. 

There are different ways that cybersickness can be measured, but the most common method is to have research participants report how they are feeling after exposure with the Simulator Sickness Questionnaire (SSQ)~\cite{Kennedy_Lane_Berbaum_Lilienthal_1993}. The SSQ is comprised of 16 symptoms to which participants must give a rating from 0 (none of that symptom) to 3 (severe intensity of that symptom). 
Though several concerns 
about the SSQ have been raised~\cite{Tian_Lopes_Boulic_2022}, it is still widely used~\cite{Chang_Kim_Yoo_2020}, which allows results across studies to be compared. 


\subsection{Cybersickness Recognition from EEG}

EEG datasets are often small as a result of significant requirements in terms of time (EEG studies typically take around three hours per subject), resources (for the system and per-run consumable costs), challenge (subject impacts like vomiting in the case of cybersickness studies), and other limitations (there is difficulty in recruiting a sample with high numbers of people who will get sick, as making the study extremely sickening poses risks in human subject research). As a result, EEG datasets, particularly for cybersickness studies, are quite limited, with one review finding that only 18\% (6 total) of EEG cybersickness studies had more than 30 participants~\cite{Chang_Billinghurst_Yoo_2023}. Moreover, there is substantial variability in EEG data, making it difficult for models to generalize. 

Despite this challenge, other work has explored using machine learning for cybersickness detection from EEG \cite{li2020vr, mawalid2018, Rosanne_Benesch_Kratzig_Pare_Bolt_Falk_2025, Yang_Kasabov_Cakmak_2022}. This usually involves feature extraction before training a classifier. Features extracted are often in the frequency or time-frequency domain relating to power in frequency bands \cite{li2020vr,Rosanne_Benesch_Kratzig_Pare_Bolt_Falk_2025, Yu2010, continuousVRDetect2025,Khoirunnisaa_Pane_Wibawa_Purnomo_2018, nam2001}. A combination of EEG features with other modalities like inertial measurement unit for head movement \cite{continuousVRDetect2025} or biosignals like heart rate, blink rate, and electrocardiogram \cite{nam2001} has also been explored. Some work extracts time domain features like variance, standard deviation, and number of peaks \cite{mawalid2018} or Principal Component Analysis (PCA) components \cite{Yu2010}. Commonly chosen classifiers are k-Nearest Neighbors (kNN) \cite{Yu2010, Khoirunnisaa_Pane_Wibawa_Purnomo_2018}, Support Vector Machine (SVM) \cite{Yu2010,Khoirunnisaa_Pane_Wibawa_Purnomo_2018}, Naive Bayes (NB) \cite{mawalid2018}, Linear Discriminant Analysis (LDA) \cite{Khoirunnisaa_Pane_Wibawa_Purnomo_2018}, Convolutional Neural Network (CNN) \cite{borra2021deep}, Long Short-Term Memory (LSTM) \cite{continuousVRDetect2025}, AdaBoost \cite{li2020vr,mawalid2018}, and more. Another line of work directly interacts with EEG in the time domain without first extracting features, usually with neural networks \cite{borra2021deep, eegnetAutoencoder2020, liu2024exploring}. Reported cybersickness detection accuracies can range around 76\% \cite{continuousVRDetect2025} and 80\% \cite{li2020vr} with train and test sets separated across different subjects. 
However, the time window sizes vary from a few (three to six) seconds \cite{li2020vr, continuousVRDetect2025} to longer windows like 30s \cite{Rosanne_Benesch_Kratzig_Pare_Bolt_Falk_2025}. In contrast, the data used in this work are periods of 800 ms after the sound of each auditory stimulus, a much shorter time window to use for identifying cybersickness discomfort. For evaluation, we note that leave-one-subject-out cross validation is a more rigorous method to determine how well a model can generalize to unseen subject data \cite{continuousVRDetect2025, borra2021deep} as opposed to mixing all subjects' data together for dataset splits. Due to the usually small dataset size, it may be better to repeat experiments on all possible folds for analysis, rather than using a small k for k-folds. As our goal here is to identify interpretable EEG features 
for cybersickness discomfort classification, neural networks with an explainability method is our approach.




\subsection{{Complications in Classifying Cybersickness}}

One of the many complicating factors in elucidating neural activity associated with VR sickness arises from the way cybersickness is quantified with the SSQ. The measure was originally developed for pilot trainees using Navy helicopter flight simulators in the 1990's~\cite{Kennedy_Lane_Berbaum_Lilienthal_1993}, and, accordingly, the list of symptoms selected for inclusion in the SSQ reflects what most commonly arose when using those devices. As current HMDs bear little resemblance to the simulators from over 30 years ago, the symptoms experienced now may not be well-captured by the SSQ questionnaire~\cite{Tian_Lopes_Boulic_2022}. Another complicating factor from the SSQ is precision in data labeling. VR users rate their symptoms after exposure has completed, meaning the data is given a cybersickness score as a batch although sickness intensity could have fluctuated during exposure. This motivates the need for real-time detection of cybersickness and related discomfort as it is occurring. 


A novel contribution of this work is our exploration of brain activity associated with one cybersickness symptom, general discomfort, in ERP data.  The 16 symptom severities included in the SSQ total weighted score span a variety of visual, physiological, and cognitive reactions that may each produce \textit{independent} activation which passes through divergent areas of the brain~\cite{Chang_Billinghurst_Yoo_2023}. This may explain why our prior work with deep learning on ERP cybersickness data labeled with SSQ scores was unable to achieve high classification accuracy~\cite{Aboud_2023}. If we use labels from the full SSQ or SSQ subscales, we are asking the models to parse multiple different streams of brain activity in a very small dataset, which could dilute our ability to identify sickness-relevant patterns. Here, by classifying data labeled with discomfort only, we can focus the models on patterns of brain activity related to an unpleasant aspect of cybersickness: discomfort. 


\subsection{EEG Analysis with ML Model Interpretability}

Explainable artificial intelligence to understand ML model decision-making 
can benefit EEG analysis by identifying which features were most useful to push the model's decision in the right direction. An ablation-style analysis can be done by modifying each part of the model and observing changes in accuracy metrics \cite{interpretableMLEEG2021}. Since CNNs are often used in EEG analysis, Gradient-weighted Class Activation Maps (GradCAM), which extract activations in the feature map from a target convolutional layer, are a common choice \cite{yan2021novel, shimizu2024feature}, often for EEGNet \cite{Lawhern_2018}. Saliency maps are another option, which outline features from the input that most contribute to the correct decision \cite{borra2021deep}. Integrated gradients that calculate the attribution of input features to model prediction is another choice \cite{liu2024exploring}. With the introduction of transformers, especially vision transformer (ViT), attention maps reveal how different parts of an image (or EEG data shaped similarly to an image) influence the ViT's decision \cite{shimizu2024feature}.
While these features can be useful, as has been noted 
\cite{Yang_Kasabov_Cakmak_2022}, care should be exercised in the interpretation of salient features identified by ML models. These features may be correlated with the focus of the classification, but other more highly correlated biomarkers may be overlooked by the models. More assessment is needed to verify the generalizability of the features, for example, with statistical tests and additional datasets. However, the ML models do point to a direction in which further study and investigation can be done.

\section{Results}\label{sec:results}

We share results from running our open-source framework that trains deep learning models and extracts model interpretability maps [code will be released on acceptance]. The framework was used to classify discomfort given a 0.8 second window of ERP data and extract which spatio-temporal features of the input data were most useful for the model's correct classification decisions. Here, we report the findings from three different neural networks: a convolutional neural network EEGNet \cite{Lawhern_2018}, a convolutional transformer EEG Conformer \cite{9991178}, and a modified ImageNet pre-trained vision transformer EEG ViT \cite{yang2023vit2eeg}. The classification is evaluated in two phases: 1) leave-one-subject-out (LOSO) cross validations to evaluate how well the models' learned patterns generalize to unseen participants and 2) per-subject calibration to evaluate how well the models can learn patterns more specific to a participant, due to high inter-subject variability in EEG. This evaluation is performed over two datasets. The results for the first ERPVR dataset are shared in Section \ref{sec:results-erpvr}. Results for the second VectionVR dataset are in Section \ref{sec:vection-data-results}. To interpret which data features contributed to the model for classification, we use integrated gradients \cite{pmlr-v70-sundararajan17a} which assign an importance score to each feature in the input data. For EEGNet we also use GradCAM \cite{jacobgilpytorchcam}, which shows the activations in the model's feature map that had more focus. More on model interpretability is in Section \ref{Sec:IGs} and the Supplemental Material Section F. 

\begin{table}[htbp!]
\centering
\begin{tabular}{lllllllllll}
\hline
M & D & S & O & E & DB & DC & NB & NC & MB & MC \\
\hline
EEG & Std & Und & ON & Alp & 71.2 $\pm$ 7.0 & 83.6 $\pm$ 3.9 & 61.1 $\pm$ 2.3 & 67.9 $\pm$ 1.1 & 51.2 $\pm$ 0.7 & 51.1 $\pm$ 0.7 \\
EEG & Std & Und & ON & Spa & 73.0 $\pm$ 1.1 & 81.8 $\pm$ 1.8 & 58.9 $\pm$ 3.0 & 63.7 $\pm$ 1.9 & 52.3 $\pm$ 0.4 & 53.0 $\pm$ 1.7 \\
EEG & Std & Over & ON & Alp & \textbf{76.9 $\pm$ 2.0} & 84.8 $\pm$ 1.8 & 60.8 $\pm$ 2.6 & 65.9 $\pm$ 2.4 & 47.4 $\pm$ 2.7 & 47.1 $\pm$ 4.7 \\
EEG & Std & Over & ON & Spa & 71.2 $\pm$ 1.5 & 82.3 $\pm$ 4.3 & 58.0 $\pm$ 1.5 & 62.8 $\pm$ 3.1 & 50.8 $\pm$ 2.0 & 49.4 $\pm$ 4.6 \\
EEG & Std & Und & OFF & Alp & 47.2 $\pm$ 6.7 & 77.7 $\pm$ 6.5 & 76.4 $\pm$ 0.4 & 67.7 $\pm$ 4.4 & 51.7 $\pm$ 1.6 & 52.5 $\pm$ 1.7 \\
EEG & Std & Und & OFF & Spa & 48.8 $\pm$ 1.1 & 82.1 $\pm$ 3.7 & 75.2 $\pm$ 1.2 & 65.6 $\pm$ 4.6 & 51.7 $\pm$ 0.7 & 53.6 $\pm$ 0.7 \\
EEG & Std & Over & OFF & Alp & 49.4 $\pm$ 2.8 & 76.2 $\pm$ 2.0 & 76.9 $\pm$ 0.6 & 68.4 $\pm$ 5.2 & 51.7 $\pm$ 0.5 & 52.5 $\pm$ 1.4 \\
EEG & Std & Over & OFF & Spa & 50.8 $\pm$ 3.0 & 79.0 $\pm$ 3.0 & 76.0 $\pm$ 2.1 & 68.6 $\pm$ 0.8 & 51.4 $\pm$ 1.7 & 52.7 $\pm$ 2.1 \\
EEG & Odd & Und & ON & Alp & 75.8 $\pm$ 4.5 & 79.0 $\pm$ 5.8 & 57.7 $\pm$ 1.9 & 63.0 $\pm$ 0.3 & 52.0 $\pm$ 0.5 & 51.0 $\pm$ 0.5 \\
EEG & Odd & Und & ON & Spa & 73.9 $\pm$ 0.9 & 78.2 $\pm$ 1.4 & 56.7 $\pm$ 3.1 & 61.9 $\pm$ 4.0 & 51.9 $\pm$ 0.3 & 52.1 $\pm$ 1.8 \\
EEG & Odd & Over & ON & Alp & 73.7 $\pm$ 5.1 & 77.2 $\pm$ 3.2 & 58.6 $\pm$ 3.9 & 63.8 $\pm$ 3.5 & 50.9 $\pm$ 1.1 & 50.6 $\pm$ 1.5 \\
EEG & Odd & Over & ON & Spa & 71.6 $\pm$ 3.1 & 81.1 $\pm$ 1.7 & 52.9 $\pm$ 1.9 & 58.2 $\pm$ 2.3 & 50.7 $\pm$ 1.0 & 51.0 $\pm$ 2.3 \\
EEG & Odd & Und & OFF & Alp & 45.3 $\pm$ 3.0 & 62.8 $\pm$ 6.2 & 72.5 $\pm$ 3.0 & 70.2 $\pm$ 4.1 & 50.8 $\pm$ 1.1 & 51.9 $\pm$ 2.8 \\
EEG & Odd & Und & OFF & Spa & 50.0 $\pm$ 3.2 & 67.8 $\pm$ 5.7 & 73.4 $\pm$ 5.2 & 68.9 $\pm$ 4.4 & 50.6 $\pm$ 0.7 & 51.4 $\pm$ 0.2 \\
EEG & Odd & Over & OFF & Alp & 48.2 $\pm$ 1.9 & 68.2 $\pm$ 2.9 & 73.7 $\pm$ 4.5 & 68.9 $\pm$ 3.3 & 51.7 $\pm$ 0.8 & 51.8 $\pm$ 1.7 \\
EEG & Odd & Over & OFF & Spa & 53.2 $\pm$ 3.6 & 67.6 $\pm$ 5.0 & 77.3 $\pm$ 2.9 & 71.2 $\pm$ 5.3 & 51.7 $\pm$ 1.1 & 51.9 $\pm$ 0.7 \\
\hline
Con & Std & Und & ON & Alp & 51.9 $\pm$ 8.3 & 78.6 $\pm$ 4.6 & 76.7 $\pm$ 3.7 & \textbf{89.3 $\pm$ 2.7} & 52.5 $\pm$ 0.9 & 56.7 $\pm$ 2.9 \\
Con & Std & Und & ON & Spa & 57.3 $\pm$ 4.9 & 79.9 $\pm$ 3.1 & 75.7 $\pm$ 3.1 & \textbf{89.3 $\pm$ 2.6} & 53.8 $\pm$ 0.8 & 56.8 $\pm$ 0.7 \\
Con & Std & Over & ON & Alp & 50.1 $\pm$ 9.6 & 80.9 $\pm$ 3.6 & 76.7 $\pm$ 1.2 & 87.2 $\pm$ 2.2 & 51.5 $\pm$ 2.2 & 56.3 $\pm$ 2.6 \\
Con & Std & Over & ON & Spa & 49.7 $\pm$ 3.6 & 82.8 $\pm$ 2.8 & 73.8 $\pm$ 2.7 & 88.8 $\pm$ 1.4 & 51.7 $\pm$ 1.7 & 54.0 $\pm$ 2.2 \\
Con & Std & Und & OFF & Alp & 33.5 $\pm$ 6.7 & 84.6 $\pm$ 0.7 & \textbf{82.6 $\pm$ 0.4} & 85.9 $\pm$ 4.1 & 53.4 $\pm$ 1.2 & 57.7 $\pm$ 0.2 \\
Con & Std & Und & OFF & Spa & 33.2 $\pm$ 1.9 & 83.0 $\pm$ 3.7 & 81.8 $\pm$ 1.3 & 87.9 $\pm$ 4.8 & 53.2 $\pm$ 1.5 & 56.9 $\pm$ 1.1 \\
Con & Std & Over & OFF & Alp & 37.7 $\pm$ 2.5 & 84.3 $\pm$ 2.9 & 79.9 $\pm$ 0.9 & 86.5 $\pm$ 4.5 & 53.2 $\pm$ 1.8 & 56.7 $\pm$ 1.7 \\
Con & Std & Over & OFF & Spa & 34.5 $\pm$ 5.5 & 84.9 $\pm$ 1.4 & 82.3 $\pm$ 0.8 & 87.3 $\pm$ 3.3 & 54.0 $\pm$ 1.9 & 57.2 $\pm$ 0.6 \\
Con & Odd & Und & ON & Alp & 57.1 $\pm$ 2.7 & 69.7 $\pm$ 2.5 & 68.2 $\pm$ 0.6 & 74.1 $\pm$ 2.0 & 52.8 $\pm$ 1.3 & 52.3 $\pm$ 2.5 \\
Con & Odd & Und & ON & Spa & 59.4 $\pm$ 3.1 & 73.4 $\pm$ 7.4 & 67.6 $\pm$ 3.3 & 75.3 $\pm$ 2.5 & 53.6 $\pm$ 0.4 & 51.7 $\pm$ 1.4 \\
Con & Odd & Over & ON & Alp & 59.4 $\pm$ 1.5 & 69.4 $\pm$ 1.9 & 68.5 $\pm$ 2.1 & 75.6 $\pm$ 2.5 & 51.5 $\pm$ 1.7 & 52.8 $\pm$ 2.4 \\
Con & Odd & Over & ON & Spa & 56.0 $\pm$ 1.8 & 65.8 $\pm$ 5.1 & 71.2 $\pm$ 1.1 & 76.7 $\pm$ 3.5 & 53.2 $\pm$ 0.3 & 51.6 $\pm$ 1.8 \\
Con & Odd & Und & OFF & Alp & 35.0 $\pm$ 2.3 & 68.7 $\pm$ 10.5 & 78.4 $\pm$ 1.3 & 72.8 $\pm$ 2.8 & 53.3 $\pm$ 1.0 & 53.0 $\pm$ 0.9 \\
Con & Odd & Und & OFF & Spa & 30.9 $\pm$ 1.3 & 65.1 $\pm$ 9.2 & 75.0 $\pm$ 3.1 & 75.8 $\pm$ 2.8 & 51.3 $\pm$ 1.4 & 53.4 $\pm$ 1.9 \\
Con & Odd & Over & OFF & Alp & 32.7 $\pm$ 6.2 & 71.6 $\pm$ 3.9 & 76.0 $\pm$ 0.7 & 74.0 $\pm$ 4.3 & 52.2 $\pm$ 0.3 & 53.1 $\pm$ 2.1 \\
Con & Odd & Over & OFF & Spa & 32.2 $\pm$ 1.6 & 67.0 $\pm$ 5.3 & 76.7 $\pm$ 1.7 & 76.2 $\pm$ 0.2 & 50.8 $\pm$ 1.5 & 53.0 $\pm$ 1.3 \\
\hline
ViT & Std & Und & ON & Alp & 74.4 $\pm$ 2.2 & 92.9 $\pm$ 1.9 & 53.7 $\pm$ 5.6 & 86.6 $\pm$ 3.8 & 50.5 $\pm$ 0.7 & 54.0 $\pm$ 3.5 \\
ViT & Std & Und & ON & Spa & 72.6 $\pm$ 4.5 & 91.3 $\pm$ 0.7 & 50.4 $\pm$ 0.8 & 86.2 $\pm$ 1.6 & 51.4 $\pm$ 0.4 & 54.3 $\pm$ 0.5 \\
ViT & Std & Over & ON & Alp & \textbf{76.3 $\pm$ 0.3} & 92.3 $\pm$ 0.3 & 51.4 $\pm$ 0.7 & 87.1 $\pm$ 2.4 & 48.4 $\pm$ 2.6 & 53.1 $\pm$ 1.1 \\
ViT & Std & Over & ON & Spa & 71.2 $\pm$ 2.1 & 91.7 $\pm$ 0.6 & 54.3 $\pm$ 0.7 & 87.0 $\pm$ 4.0 & 51.2 $\pm$ 0.2 & 54.6 $\pm$ 0.9 \\
ViT & Std & Und & OFF & Alp & 51.0 $\pm$ 2.1 & 94.2 $\pm$ 1.6 & 65.6 $\pm$ 2.4 & 87.9 $\pm$ 1.9 & 51.6 $\pm$ 1.4 & 54.5 $\pm$ 1.9 \\
ViT & Std & Und & OFF & Spa & 49.8 $\pm$ 4.8 & 93.4 $\pm$ 2.0 & 66.8 $\pm$ 1.8 & 88.7 $\pm$ 3.2 & 52.8 $\pm$ 0.2 & 56.7 $\pm$ 1.3 \\
ViT & Std & Over & OFF & Alp & 48.7 $\pm$ 4.6 & 94.2 $\pm$ 1.1 & 65.8 $\pm$ 0.6 & 89.0 $\pm$ 1.8 & 51.9 $\pm$ 0.8 & 56.3 $\pm$ 1.0 \\
ViT & Std & Over & OFF & Spa & 45.7 $\pm$ 2.2 & \textbf{95.6 $\pm$ 0.3} & 67.7 $\pm$ 1.3 & \textbf{89.6 $\pm$ 1.8} & 51.6 $\pm$ 0.4 & 56.3 $\pm$ 0.9 \\
\hline
\end{tabular}
\caption{ERPVR dataset balanced accuracy from base LOSO and calibration phases by configuration $\times$ subject group averaged over random seeds. Each cell is run three times with different random seeds and averaged over subjects in the subject group to get one final average for that run. We report (mean $\pm (\max - \min)/2$) over the three final averages.  M is ``Model" for model used: ``EEG" for EEGNet, ``Con" for EEG Conformer, and ``ViT" for the pretrained EEG ViT. D is ``Data type" for Standards or Oddballs. S is ``Sampling" for Undersampling or Oversampling. O is ``Outlier detection" turned on or off. E is ``Electrode order" for alphabetical or spatial. DB is ``Discomfort base", DC is ``Discomfort calibration", NB is ``No Discomfort base", NC is ``No Discomfort calibration", MB is ``Mixed base", and MC is ``Mixed calibration."}
\label{tab:bygroup-all-models-balanced-accuracy-base-calib-mean-half-range-over-seeds-by-configuration-subject-group}
\end{table}

\subsection{ERPVR Dataset Results}\label{sec:results-erpvr}

\subsubsection{ERPVR Model Accuracies}
We show results from 120 experiment runs, varying over different configurations of the pipeline for the ERPVR dataset in Table \ref{tab:bygroup-all-models-balanced-accuracy-base-calib-mean-half-range-over-seeds-by-configuration-subject-group}. We define a sample as ``discomfort" when the participant's General Discomfort rating on the SSQ had a score $> 0$ and ``no discomfort" as a General Discomfort score $= 0$. The configurations were permuted per Figure \ref{fig:graphical-abstract}D for standards vs. oddballs ERP data type, under vs. oversampling, outlier detection on vs. no outlier detection, and electrode order alphabetical vs. spatial. For each model, we aggregate the results by pipeline configuration and by participant groups: subjects who felt discomfort during all trials, subjects who felt no discomfort during all trials, and subjects who felt both (mixed).  On average, we see the participant-specific calibration helps boost balanced accuracy for all three groups in Table \ref{tab:bygroup-all-models-balanced-accuracy-base-calib-mean-half-range-over-seeds-by-configuration-subject-group}. 

For the all trials discomfort participant group, for participant-specific calibration, ViT was able to achieve the highest balanced accuracy around 95.6\% with oversampling, no outlier detection, and spatial electrode ordering. For base LOSO to evaluate generalization to unseen participants, EEGNet performed the best with balanced accuracy around 76.9\% using Standard trials EEG data, oversampling, outlier detection, and alphabetical electrode ordering. ViT follows closely with balanced accuracy around 76.3\% in base LOSO with Standard trials data, oversampling, outlier detection, and alphabetical electrode ordering. For the all trials no discomfort subjects, for subject-specific calibration, ViT also achieves the highest balanced accuracy around 89.6\% with oversampling, no outlier detection, and spatial electrode ordering. Conformer follows closely around 89.3\% with Standard trials data, undersampling, outlier detection, and spatial electrode ordering. For base LOSO, Conformer performed best at 82.6\%. However, the mixed label subjects were more difficult for the models to classify, with the best average performance around 57.7\% balanced accuracy with Conformer using Standard trials, undersampling, and no outlier detection. This may be due to intra-subject label differences, and although calibration still does help improve balanced accuracy, it is less for mixed label subjects compared to the single-label subjects, as seen in Table \ref{tab:bygroup-all-models-balanced-accuracy-base-calib-mean-half-range-over-seeds-by-configuration-subject-group}.


Even though mixed label subjects in general had lower balanced accuracies, there are a few where the model was able to achieve high balanced accuracy. They are especially interesting because these subjects indicate that some patterns learned by the models were able to classify and separate discomfort within that subject. One example is subject 15 (16.1\% of trials are discomfort) with EEGNet, Standard trials, outlier detection, and oversampling, which achieved 81.0\% balanced accuracy after calibration. Another example is subject 10 (83.5\% of trials are discomfort) with Conformer, Standard trials, outlier detection, and undersampling, which achieved 75.1\% balanced accuracy after calibration.

\subsubsection{ERPVR Integrated Gradients}
\begin{figure*}[htbp!]
\begin{subfigure}{\textwidth}
    \centering
    \includegraphics[width=\linewidth]{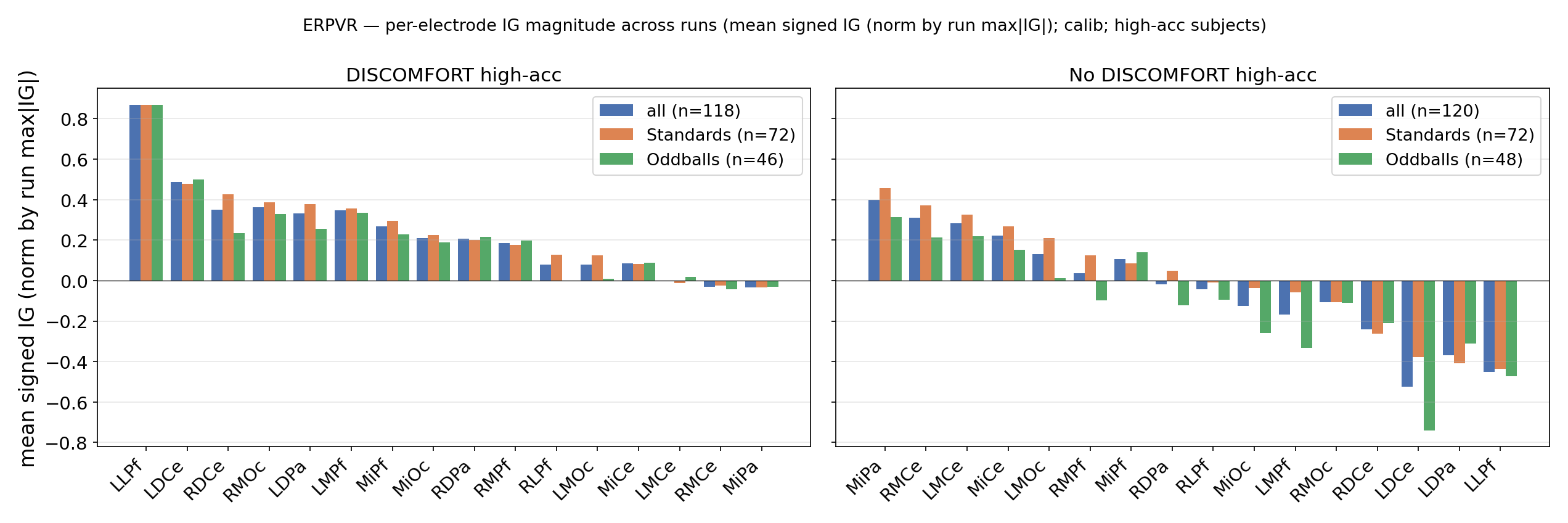}
    \caption{ERPVR integrated gradients for each electrode summed over time, then normalized by the max summed IG magnitude per run, and averaged over all runs, using Standard trials only, or using Oddball trials only, after calibration. N is the number of experiment runs for each group. ``All" combines all model runs, both with Standard trials and with Oddball trials. 
    The ordering of electrodes shown is based on the importance calculations for the standards only. Numbers less than zero mean that electrode was negatively contributing for classification.}
    \label{fig:erpvr_continuous_ig_sum}
\end{subfigure}
\begin{subfigure}{\textwidth}
    \centering
    \includegraphics[width=\linewidth]{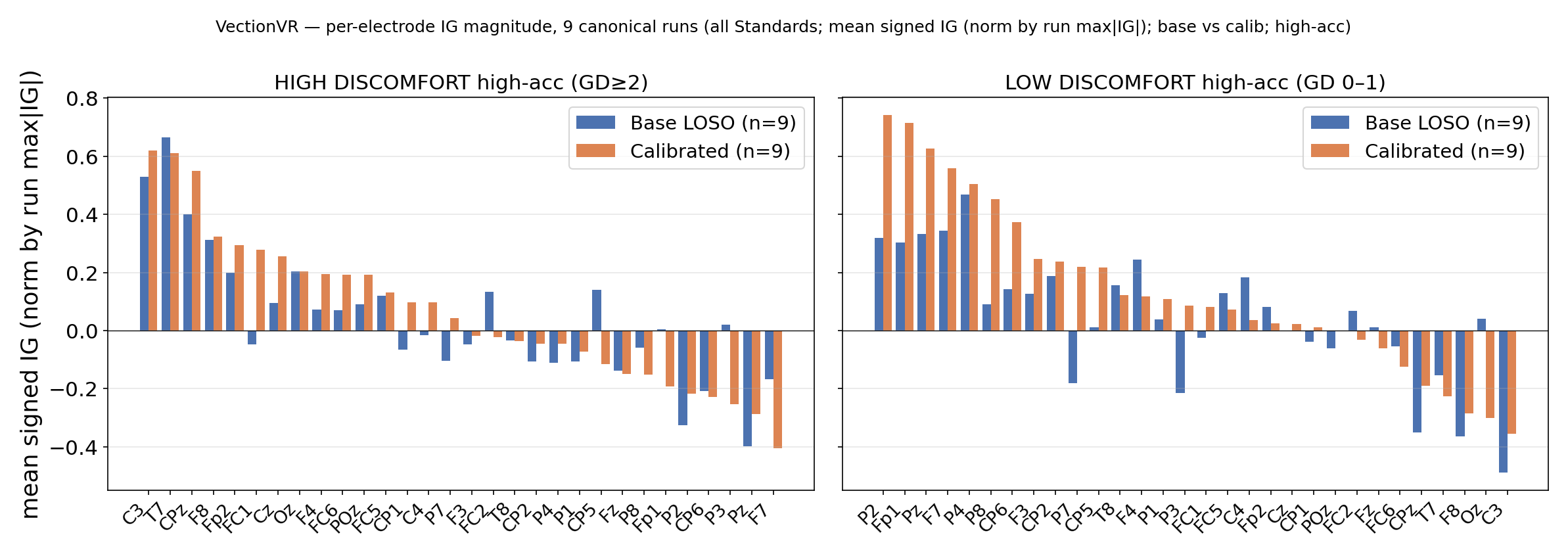}
    \caption{VectionVR integrated gradients for each electrode summed over time then normalized by the max summed IG magnitude per run, and averaged over all runs for base LOSO vs. after calibration. These are Standard trial experiment runs only. High Discomfort is when ``General Discomfort" (GD) is reported to be 2 or greater in the SSQ. Low Discomfort when ``General Discomfort" is reported to be 0 or 1 in the SSQ. The ordering of electrodes shown is based on the importance calculations for calibrated.}
    \label{fig:vectionvr-ig-normmax}
\end{subfigure}
\caption{Integrated gradients for each electrode summed over time, then normalized by the max summed IG magnitude per run, and averaged over all runs. Top plot (a) is from ERPVR runs and bottom plot (b) is from VectionVR runs.}
    \label{fig:continuous-ig-normmax}
\end{figure*}

To determine which electrodes were most useful for discomfort classification, the integrated gradient (IG) for each electrode channel is summed over the time dimension. Then for each run, the summed IG is normalized by the max magnitude summed IG and averaged over the runs (see Figure \ref{fig:erpvr_continuous_ig_sum}). The first plot shows results from runs where there are participants that experienced discomfort which achieved high balanced accuracy over 75\%. The second plot is from runs where no discomfort subjects achieved high balanced accuracy. We also plot the frequency of an electrode reaching top-5 of the summed IG for a run (indicating that the electrode was one of the top five contribution electrodes for model classification) in Supplemental Figure B2. 

\begin{figure*}[t]
    \centering
    \vspace{-0.1cm}
  \begin{subfigure}{0.5\textwidth}
    \centering
    \includegraphics[width=\linewidth]{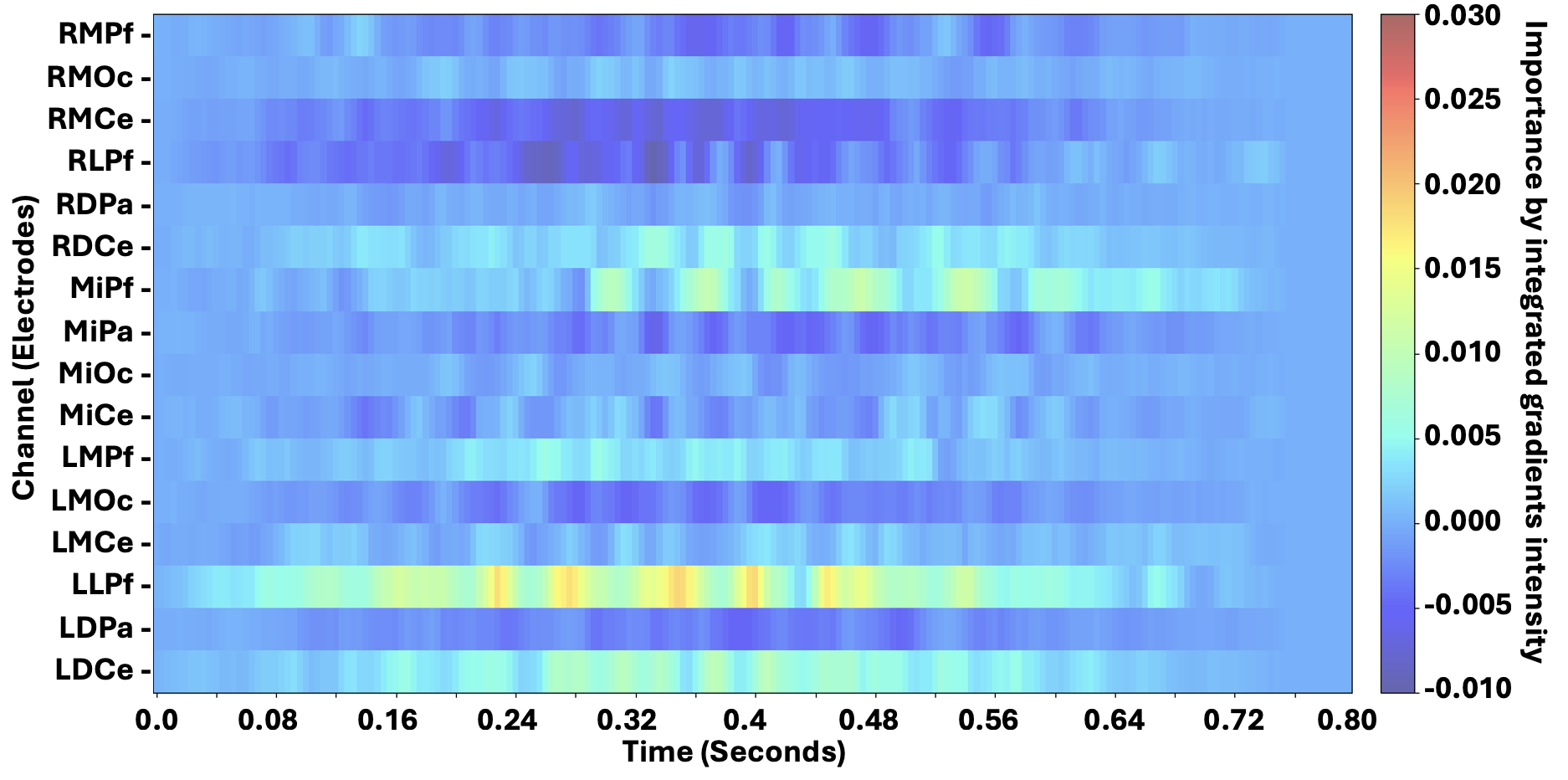} \vspace{-0.55cm}
    \caption{EEG Conformer calibrated IGs}
    \label{fig:conformer-rand-seed-4-calibrated}
  \end{subfigure}\hfill
  \begin{subfigure}{0.5\textwidth}
    \centering
    \vspace{-0.1cm}
    \includegraphics[width=\linewidth]{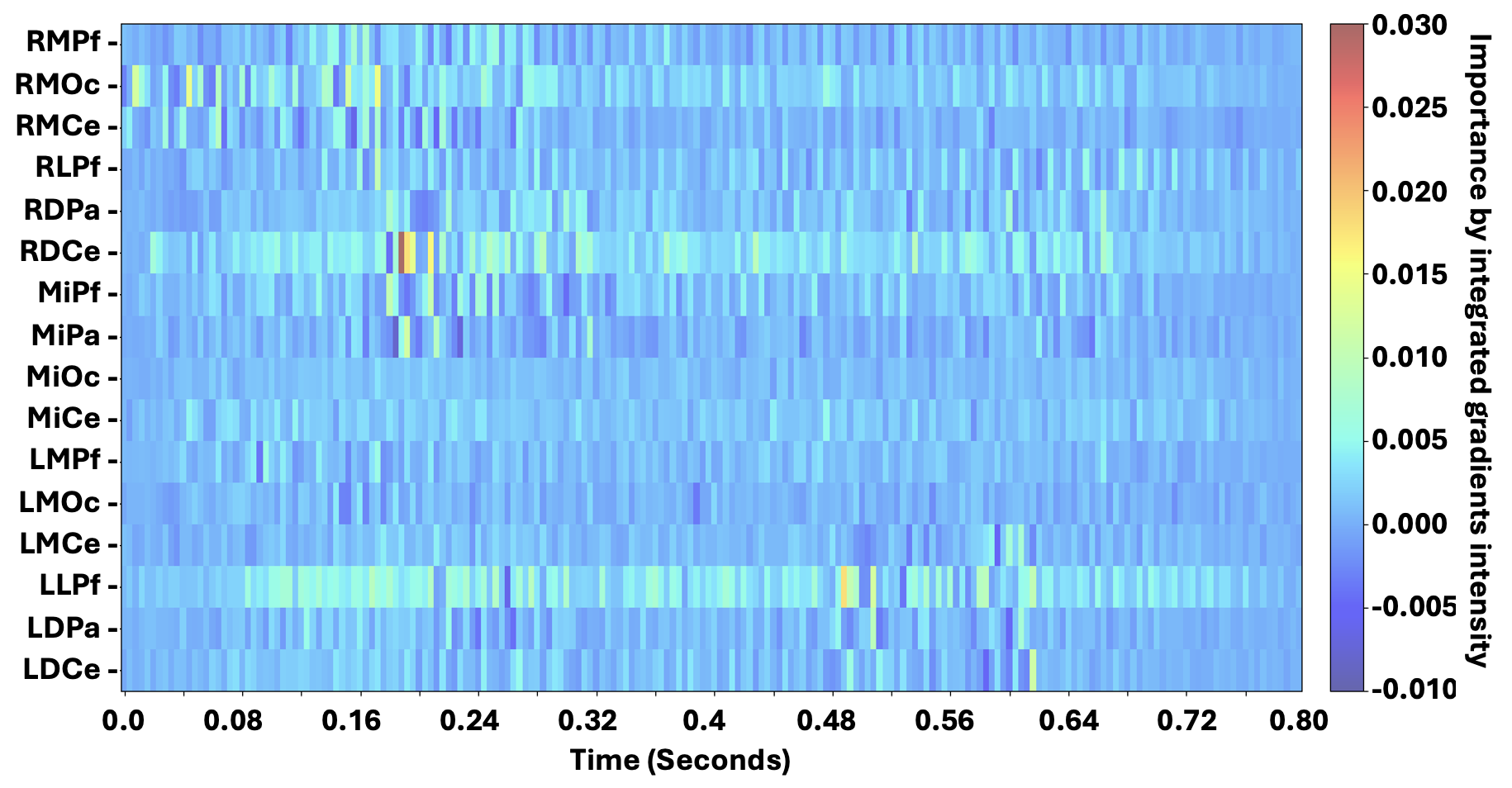} \vspace{-0.55cm}
    \caption{Pre-trained EEG ViT calibrated IGs}
    \label{fig:pretrained-vit-rand-seed-4-calibrated}
  \end{subfigure}

  \vspace{0.7em}

  \begin{subfigure}{0.5\textwidth}
    \centering
    \vspace{-0.1cm}
    \includegraphics[width=\linewidth]{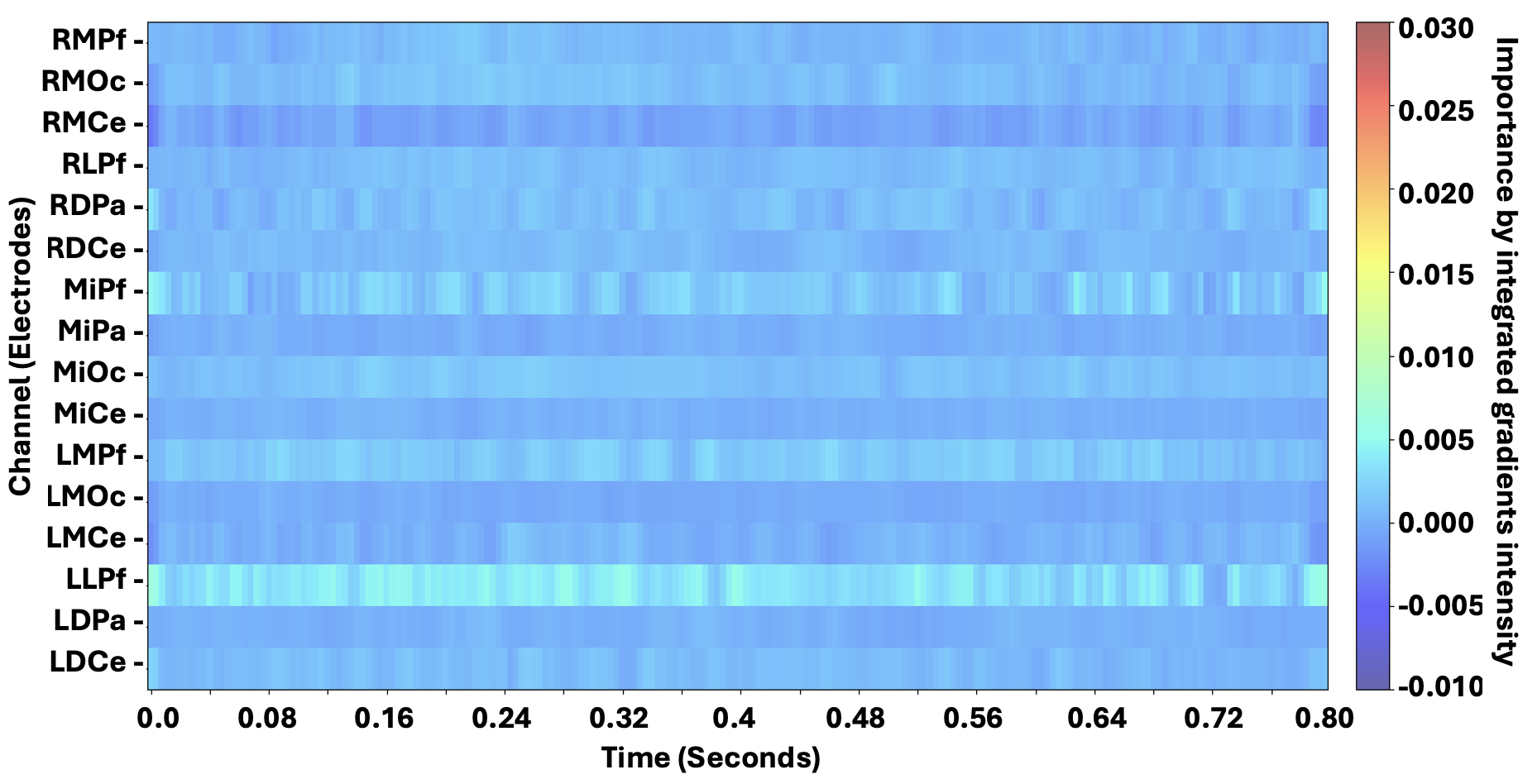} \vspace{-0.55cm}
    
    \caption{EEGNet calibrated IGs}
    \label{fig:eegnet-rand-seed-4-calibrated-ig}
  \end{subfigure}\hfill
  \begin{subfigure}{0.5\textwidth}
    \centering
    \vspace{-0.1cm}
    \includegraphics[width=\linewidth]{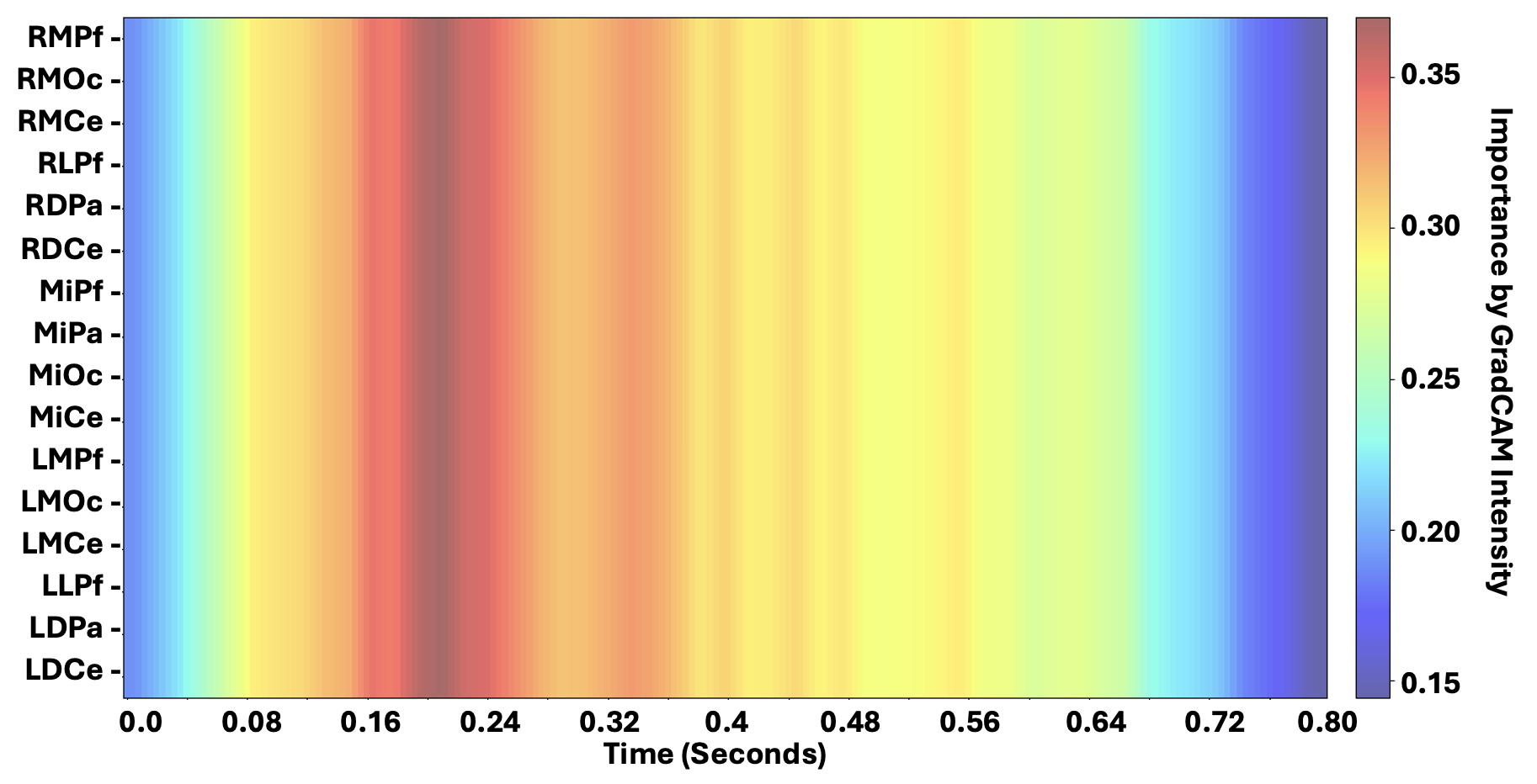} \vspace{-0.55cm}
    \caption{EEGNet calibrated GradCAM}
    \label{fig:eegnet-rand-seed-4-calibrated-gradcam}
  \end{subfigure}
    \caption{ERPVR Integrated Gradients (IGs) averaged over discomfort data from subjects with balanced accuracy~$>75\%$. Warmer colors indicate features have more positive contribution for model classification. These plots are from one of three random seeds.}
    \label{fig:heatmaps-rand-seed-4}
    \vspace{-0.55cm}
\end{figure*}

For each run configuration, the integrated gradients (IGs) map for all models after calibration show the IGs averaged over discomfort data from high balanced accuracy ($>75\%$) subjects. Figure \ref{fig:heatmaps-rand-seed-4} shows the results for Standard trials, outlier detection, oversampling, and alphabetical electrode ordering. Results for more runs, including trials when the models were trained in the spatial electrode ordering, 
are provided in the Supplemental Material. 
In terms of the most relevant features for the models, from the IG calculated across multiple runs, a prominent observation is that some electrodes consistently had high integrated gradient positive attributions. LLPf had the highest integrated gradients for all models and runs, usually followed by LDCe and RDCe. RMOc and LDPa were also generally helpful for the models with a higher average sum of IG over the runs. In the time domain, in EEG Conformer's IG map in Figure \ref{fig:heatmaps-rand-seed-4}, for LLPf there were time steps highlighted between 240 to 440 ms, while EEGNet's GradCAM highlighted an earlier 160 to 240 ms window. Pre-trained EEG ViT highlighted around 160 to 200 ms for LLPf and RDCe.

Comparing the model interpretability maps from mixed label subjects that had balanced accuracy $>$ 75\%, we may identify which features helped the model discriminate between discomfort and no discomfort trials within-subject. The integrated gradients highlight similar electrodes to ones highlighted by all discomfort subjects' IG maps for contributing most to the model classifying discomfort: LDCe, LLPf, LMPf, RDCe, and MiPf. This may indicate that some of the models' learned patterns can carry over to within-subject discomfort classification, as well as between-subject discomfort classification. The integrated gradients maps are in Supplemental Figure B1. 

\subsubsection{ERPVR Configurations}

Based on the metrics increase post-calibration, it seems having some way to bridge the distribution shift between train set and test subjects, especially when working with a small dataset of complex and noisy data like EEG, is crucial for tailoring a classifier to an individual for cybersickness discomfort detection. Since the transformers performed better than EEGNet after calibration, this may indicate that the sequence modeling done by the transformers with attention in the encoder blocks is important for cybersickness discomfort detection in EEG data, which EEGNet as a CNN only, lacks.

For ERPVR, the initial ordering of channels was alphabetical, which did not impact performance for models that learned spatial filters across all electrodes. However, EEG ViT's patch embedding is over four channels, so we re-ordered the channels to group electrodes based on their location spatially on the head and ran all models. We found the metrics were similar and the channels marked as important by IG remained consistent, though EEG ViT did see some changes in attribution. The electrode ordering did have a bigger effect for the VectionVR data runs however, discussed in section \ref{sec:vection-data-results}. 

One configuration that performed well after calibration for discomfort classification for ERPVR was using Standard trials, oversampling, no outlier detection, and spatial ordering. With the ViT model, this achieved the highest average balanced accuracy for discomfort subjects with 95.6\%. EEG Conformer also reached its best average performance with 84.9\%. However, an interesting contrast is that this same configuration had worse performance in base leave-one-subject-out compared to other configurations. This configuration also performed well (compared to other configurations) for mixed label subjects for all three models. For no discomfort subjects, this same configuration also performed the best for ViT with 89.6\% average balanced accuracy. However for EEG Conformer and EEGNet, they performed better with Standard trials, undersampling, and outlier detection with Conformer reaching around 89.3\%.


\subsubsection{ERPVR Statistical Analyses of Configuration Types}

We used a Wilcoxon signed-rank test to test whether the different configurations for ERPVR runs had any significant differences, with detailed numbers reported in Supplemental Material Table B2. 
Since we ran multiple tests, we used Bonferroni Correction, setting $\alpha = \frac{0.05}{N} = 0.01$ for $N=5$ statistical tests performed. We compared the pre-calibration base leave-one-subject-out phase and post-calibration and each of the four configurations: Standards vs. Oddballs data, outlier detection vs. no outlier detection, over vs. undersampling, and alphabetical electrode order vs. spatial order. The Wilcoxon tests were performed by comparing runs where all configurations, except for the configuration being tested, are kept the same. Then, within the matched pair (e.g. Oversampling / Standards / Outlier detection / Alphabetical electrode order/ EEGNet / seed 3 vs. Undersampling / Standards / Outlier detection / Alphabetical electrode order / EEGNet / seed 3 is one example of a pair if the configuration being tested is sampling type), we compute the difference $\Delta$(\%) between each subject's balanced accuracy in the two runs. The delta is computed by the first configuration - second configuration for each model run that satisfies the pairing. Then, this is averaged for each subject across all the paired runs, so we have one averaged delta per subject representing how much the configuration being tested affected the balanced accuracy for that subject. These deltas are then compared. 

We did not find any significant differences between the configurations for outlier detection, sampling, data type, or channel order. However, there was a meaningful difference between calibration vs. base LOSO, 
with calibration positively increasing the balanced accuracy for participants. 
We also evaluated differences between each configuration choice with Wilcoxon signed-rank tests over balanced accuracies from the experiment runs for each of the three main subject groups in ERPVR: pure discomfort, mixed, and no discomfort subjects. These results are reported in Supplemental Material Section B.1. 

\begin{table}[b]
\centering
\begin{tabular}{lllllll}
\hline
Model & HD base & HD calib & LD base & LD calib & Overall base & Overall calib \\
\hline
EEGNet & 49.9 $\pm$ 6.1 & 76.1 $\pm$ 4.6 & 52.9 $\pm$ 2.1 & 73.1 $\pm$ 3.7 & 51.6 $\pm$ 1.4 & 74.4 $\pm$ 1.4 \\
Conformer & 68.2 $\pm$ 5.3 & 88.6 $\pm$ 1.6 & 33.6 $\pm$ 1.8 & 82.6 $\pm$ 2.0 & 48.0 $\pm$ 2.0 & 85.1 $\pm$ 0.8 \\
ViT & 47.0 $\pm$ 4.1 & \textbf{89.9 $\pm$ 0.8} & 50.5 $\pm$ 3.4 & \textbf{92.6 $\pm$ 1.4} & 49.0 $\pm$ 3.3 & \textbf{91.5 $\pm$ 0.6} \\
\hline
\end{tabular}
\caption{VectionVR dataset accuracy averaged over random seeds. Each cell is run three times with different random seeds and averaged over subjects in the subject group to get one final average for that run. We report (mean $\pm (\max - \min)/2$) over the three final averages and per subject group: High discomfort (``HD") or Low discomfort (``LD"). Overall averages over all subjects. These accuracies are for the spatial channel order only.} 
\label{tab:vecgroup-balanced-accuracy-by-subject-group-sick-nonsick-vectionvr}
\end{table} 

\subsection{VectionVR Dataset Results}
\label{sec:vection-data-results}

We ran the main pipeline for VectionVR data as described in Section \ref{sec:vectionvr-main-pipeline} on a second dataset, VectionVR, with results in Table \ref{tab:vecgroup-balanced-accuracy-by-subject-group-sick-nonsick-vectionvr}. Due to the imbalance of the ``General Discomfort" SSQ label in the VectionVR data, we separated the data into high discomfort (SSQ score for ``General Discomfort" was 2 or 3) and low discomfort (SSQ score for ``General Discomfort" was 1 or 0).


\begin{figure*}[t]
    \centering
    \vspace{-0.1cm}
  \begin{subfigure}{0.5\textwidth}
    \centering
    \includegraphics[width=\linewidth]{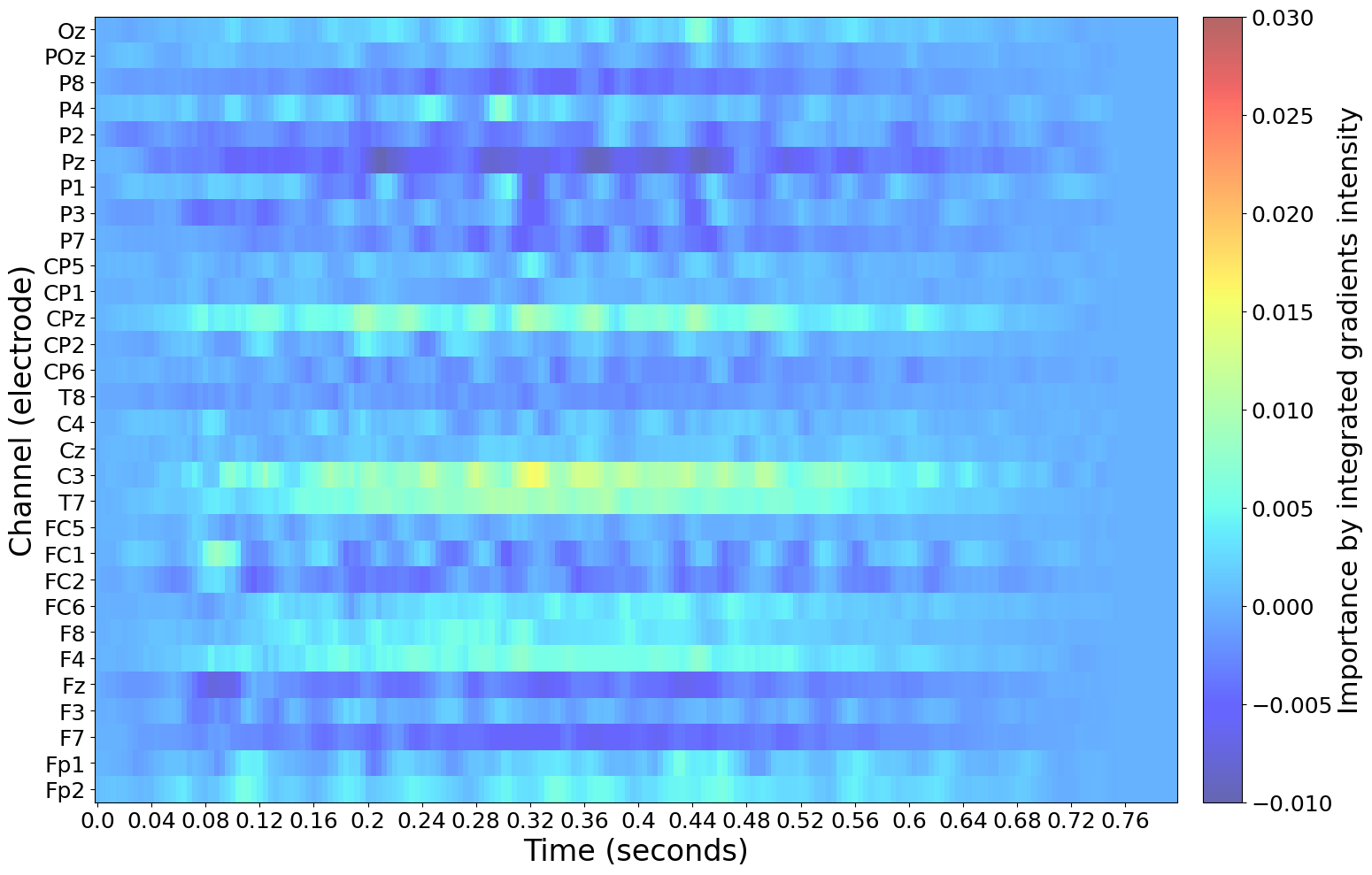} \vspace{-0.55cm}
    \caption{EEG Conformer calibrated IGs}
    \label{fig:vectionvr-conformer-rand-seed-4-calibrated}
  \end{subfigure}\hfill
  \begin{subfigure}{0.5\textwidth}
    \centering
    \vspace{-0.1cm}
    \includegraphics[width=\linewidth]{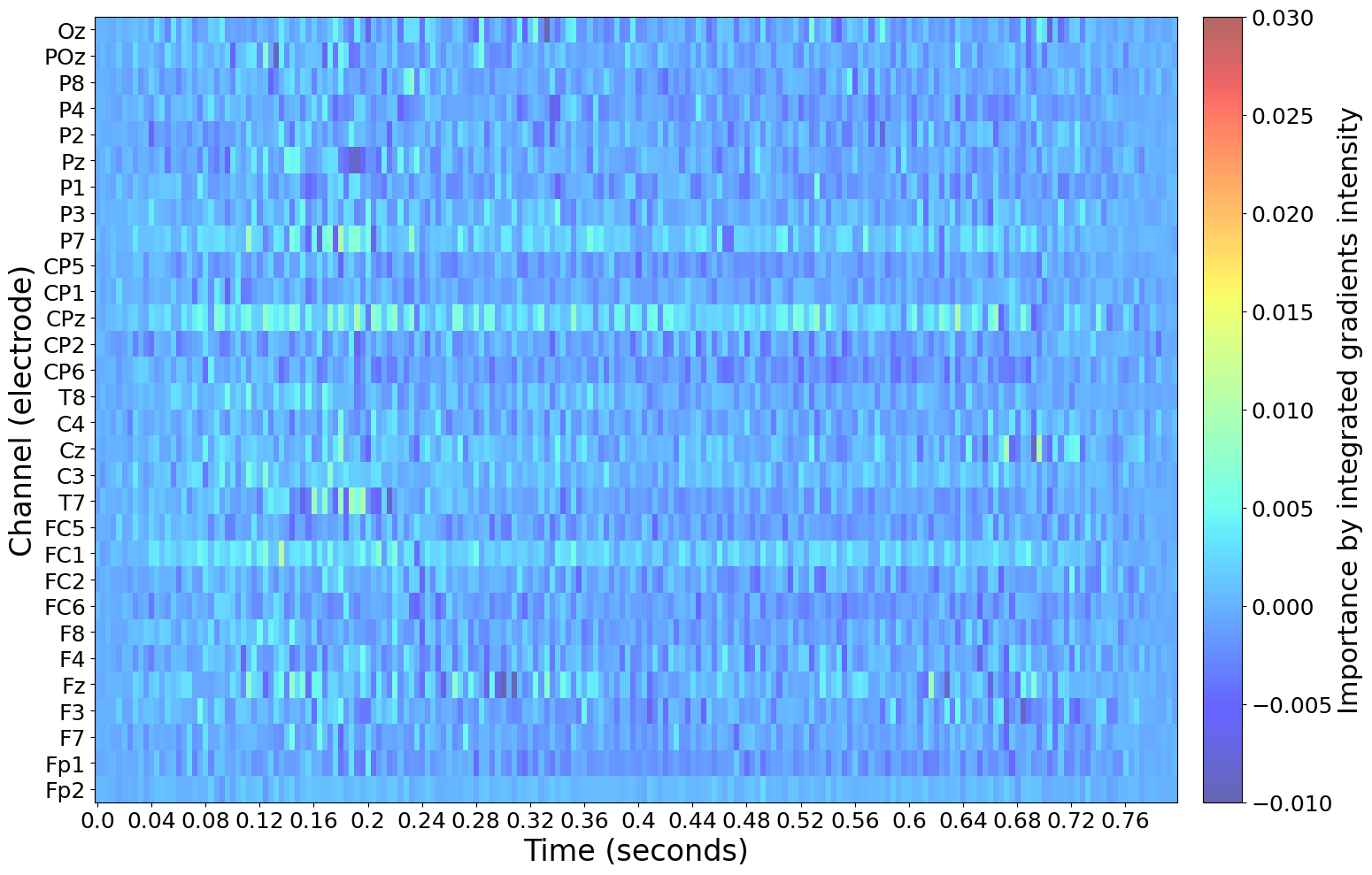} \vspace{-0.55cm}
    \caption{Pre-trained EEG ViT calibrated IGs}
    \label{fig:vectionvr-pretrained-vit-rand-seed-4-calibrated}
  \end{subfigure}

  \vspace{0.7em}

  \begin{subfigure}{0.5\textwidth}
    \centering
    \vspace{-0.1cm}
    \includegraphics[width=\linewidth]{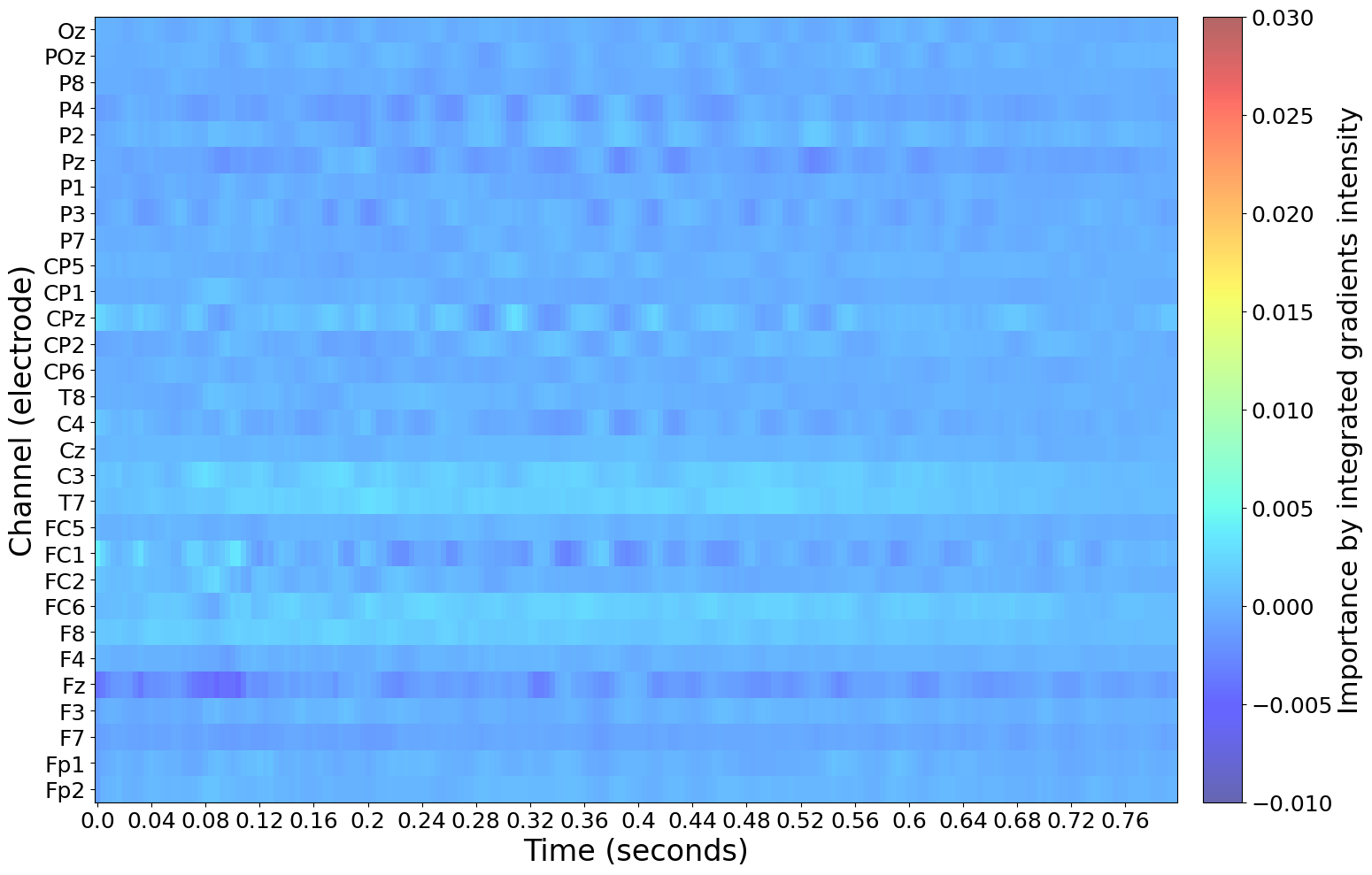}
    
    \caption{EEGNet calibrated IGs}
    \label{fig:vectionvr-eegnet-rand-seed-4-calibrated-ig}
  \end{subfigure}\hfill
  \begin{subfigure}{0.5\textwidth}
    \centering
    \vspace{-0.1cm}
    \includegraphics[width=\linewidth]{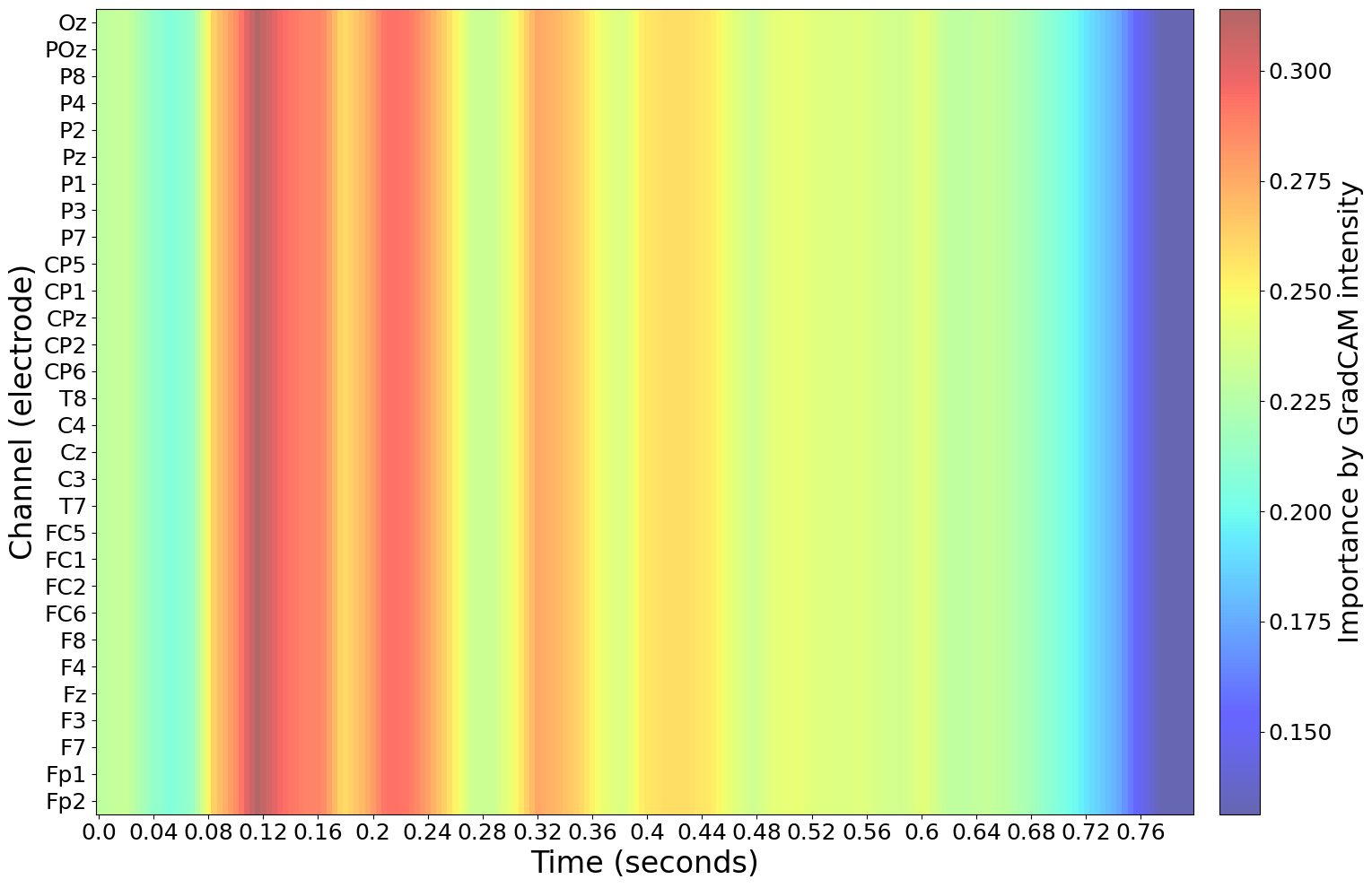}
    \caption{EEGNet calibrated GradCAM}
    \label{fig:vectionvr-eegnet-rand-seed-4-calibrated-gradcam}
  \end{subfigure}
    \vspace{-0.6cm} \caption{VectionVR integrated gradients averaged over high discomfort data from subjects with balanced accuracy~$>75\%$ (from one run of the three seeds). Warmer colors indicate features have more positive contribution for model classification.}
    \label{fig:vectionvr-heatmaps-rand-seed-4}
    \vspace{-0.55cm}
\end{figure*}

\subsubsection{VectionVR Model Accuracies}

We show results over nine experiment runs in Table \ref{tab:vecgroup-balanced-accuracy-by-subject-group-sick-nonsick-vectionvr}, three random seeds for each of the three models, using the pipeline configuration as described in Section \ref{sec:vectionvr-main-pipeline}. We see that the base LOSO on average has much worse accuracy than for ERPVR runs, indicating that the VectionVR dataset was harder for the models to learn to generalize to unseen subjects. One possible reason could be due to the coarser resolution of the labels: ERPVR had four block-level SSQ labels while VectionVR had one label for all eight blocks. Calibration did help pull up the accuracy quite substantially for per-subject classification. Since we did not have mixed label subjects in VectionVR, we cannot evaluate within-subject high to low discomfort classification. And because there are only single label subjects, balanced accuracy collapses to regular accuracy.

From these runs, Imagenet pre-trained ViT performed the best after calibration with around 90\% accuracy for high discomfort subjects, 92.6\% for low discomfort subjects, and 91.5\% overall. This indicates the calibration can help the model learn whether a subject is high or low discomfort, but not within-subject classification because of label resolution. For base leave-one-subject-out, EEG Conformer performed best for high discomfort while EEGNet performed best for low discomfort.

Some exploratory runs were used to test the effects of different configurations on VectionVR data. Outlier detection was tested at varying levels of strictness, and it was found that keeping outlier detection off (balanced accuracy averaged over all subjects in the run with base LOSO is around 52.6\%) performed slightly better than with outlier detection on (49.9\%). For sampling, oversampling (balanced accuracy around 50.9\%) was found to be better than undersampling (around 40.6\%). For electrode ordering, which specifically matters for the pre-trained ViT because the patch embedding groups the channels together depending on the input order, spatial ordering was better at around 46.2\% average balanced accuracy compared to alphabetical around 40.5\%.

\subsubsection{VectionVR Integrated Gradients}

Similar to the analysis for ERPVR runs, we sum the integrated gradient of each electrode channel over the time dimension for the VectionVR data. Then for each run, the summed IG is normalized by the max magnitude summed IG and averaged over the runs (see Figure \ref{fig:vectionvr-ig-normmax}). The left plot shows results from runs where there are high discomfort subjects that achieved accuracy over 75\%, and the right plot for runs where there are low discomfort subjects that achieved accuracy over 75\%. We also plot a histogram of the frequency of an electrode reaching top-5 summed IG for a run, indicating the electrode was a top-5 contributing electrode for model classification of discomfort in VectionVR, in Supplemental Figure B3. 

Figure \ref{fig:vectionvr-heatmaps-rand-seed-4} shows the integrated gradients map for all models after calibration, where the IGs are averaged over high discomfort data from high balanced accuracy ($>$ 75\%) subjects. This result is from Standard trials, oversampling, no outlier detection, and spatial electrode ordering. From these results, there are some electrodes that were consistently useful and positively contributed to high discomfort classification across the models. C3 and CPz were consistently useful for all three models. T7 contributed very positively for EEG Conformer and EEGNet, but not as much to pre-trained ViT. Similarly, F8 is useful for EEG Conformer and EEGNet, but relatively neutral for ViT. ViT highlights FC1 and P7 as useful electrodes for high discomfort classification, whereas they were pretty neutral for EEGNet and Conformer. This is also reflected in Figure \ref{fig:vectionvr-ig-normmax}. We observe C3, T7, and CPz are the three most contributing spatial features over the runs. 

In the time domain, EEGNet's GradCAM in Figure \ref{fig:vectionvr-heatmaps-rand-seed-4} highlighted 80 to 240ms, which overlaps with the time window highlighted by EEGNet in ERPVR runs (from 160 to 200ms). EEG Conformer's IG map highlighted a longer time window of 120 to 560ms. Pre-trained ViT seems to also highlight an early time window of around 80 to 280ms. The three models seem to overlap around the 120 to 240ms time window. This may indicate that there may be some patterns in the ERP data around this time window, especially since the data is time-locked, that help the models with discomfort classification. The windows may be slightly earlier for VectionVR compared to ERPVR as a result of differences in stimulus timing. For ERPVR, there was a small amount of latency with the auditory stimulus that was corrected with an auxiliary channel for time-locking during VectionVR data collection. Thus, the timing may be a bit more exact in the VectionVR data. More analysis plots in Supplemental Material Section B. 

\subsubsection{VectionVR Configurations}

Similar to the ERPVR results, post-calibration helped increase metrics for VectionVR, reinforcing the importance of bridging the distribution shift between train set and test subjects for cybersickness discomfort detection for each subject. 
Unlike ERPVR, where the electrode order between alphabetical and spatial did not affect metrics too much, for VectionVR we found that spatial electrode ordering performed much better for base LOSO for the pre-trained ViT. This may be because of montage naming difference between the two datasets. For ERPVR, alphabetical order sorts the electrodes into areas on the left and right sides of the head, but for VectionVR, which uses the 10-20 montage, alphabetical ordering sorts them roughly by region. Since ViT's convolution is over a subset of electrodes at once, it is more likely to be affected by electrode channel ordering. Details about the model architectures in Supplemental Material Section E. 

\section{Discussion}\label{sec12}

In this article, we present a comprehensive evaluation of brain data collected while participants were experiencing cybersickness. We wanted to identify which temporal and spatial features were most helpful for classifying cybersickness-related discomfort in epoched ERP data collected from an auditory oddball task while participants viewed visual stimuli in VR. 
From datasets that were collected in different countries, with different VR environments, different VR headsets, different primary tasks, different EEG systems, and different intensities of cybersickness, we found similar patterns of important electrodes (left frontal, lateral central, and right parietal) and time periods (roughly around 80 to 260 ms post-stimulus). These patterns were highlighted by three different deep learning models across multiple random seeds and across multiple permutations of machine learning pipeline configurations. By training and evaluating the models on two completely separate datasets, we could assess whether patterns identified consistently in one dataset generalized to the other. 

\subsection{Important Spatial Features}

Discomfort labels for the EEG data came from one item (general discomfort) of the SSQ questionnaire. The SSQ asks participants to rate at that moment whether their symptoms are none (0), slight (1), moderate (2), or severe (3). As a result of individual variability in cybersickness responses to the two different VR stimuli, the participant data split generally into three groups across the two (ERPVR and VectionVR) datasets: people who experienced no cybersickness-related discomfort, people who experienced mild-to-moderate levels of cybersickness discomfort, and people who experienced moderate-to-severe levels of cybersickness discomfort. For the no discomfort group in ERPVR, all ratings of general discomfort were 0. For the discomfort group in ERPVR, the ratings were  primarily 1's and 2's, with an average score of 1.098. For the VectionVR data, the low discomfort group had three people with a score of 0, and the rest a score of 1, with an average of 0.857. For the high discomfort group, the scores were all 2's and 3's with an average of 2.133. Thus, the overlap between the two datasets happened in the discomfort ERPVR group and the low discomfort VectionVR group. 

For participants who did not experience any cybersickness-related discomfort, centro-parietal regions (MiPa, LMCe, RMCe, MiCe, and LMOc) were the most useful features for the model. For participants that experienced intermediate levels of cybersickness discomfort (in both datasets), a distributed pattern was found across left frontal (LLPf, Fp1, F7) and right parietal (P2, P4, P8, RMOc) areas. Bilateral central (LDCe, RDCe) areas were helpful as well. For participants that experienced the highest levels of cybersickness discomfort, central and left centerline (C3, T7, CPz) areas were again tagged. Right frontal areas (F8, Fp2), to a lesser extent, were also useful. 

Interesting patterns of specific important areas were highlighted despite the different layouts (montages) of electrodes for the two studies. For example, LDCe and C3 (and a little further out T7) along the left side of the head in central areas above the ears were tagged in both the ERPVR and VectionVR datasets as being useful for the group with a greater level of discomfort. LLPf and Fp1 with F7 (which are on prefrontal / frontal left areas of the head) as well as RMOc and P2 with Pz and P4 (which are all on right occipito-parietal areas of the head) were collectively determined as useful for the group that experienced discomfort in ERPVR and the low discomfort group in VectionVR. Overall, ERPVR generally elicited lower levels of discomfort than VectionVR, so this overlap may reflect participants experiencing broadly similar (intermediate) levels of discomfort across the two datasets. Another interesting pattern involved the MiPa and CPz electrode locations which are relatively close to each other. For the ERPVR dataset, MiPa was rated as the most important spatial feature for classification of the samples when people felt no discomfort. In the VectionVR dataset, CPz was the third most helpful feature for the samples with high discomfort. This similar location being important for both high and no discomfort could hypothetically be from high amplitudes in one case and low amplitudes in another, or the presence of signal in one case and absence of signal in the other. We cannot determine what exactly the model was picking up on during classification decisions. However, electrodes in centro-parietal regions have been found as important features in other EEG cybersickness and frequency analysis studies~\cite{yang2022machine}, pain studies~\cite{Rockholt2023} and motion sickness studies~\cite{Li2022}.

The distributed pattern of informative scalp sites may also be expected given contemporary understanding of brain function. Complex perceptual, affective, and cognitive states arise through coordinated activity across distributed neural systems, with large-scale functional networks providing an organizing architecture that can be dynamically modulated by current task and state demands ~\cite{Park2013,Sporns2013,bressler2010large}. Cybersickness-related discomfort likewise likely reflects the combined activity of multiple sensory, autonomic, interoceptive, and affective areas. From this perspective, EEG is a particularly useful tool for investigating cybersickness because it samples electrophysiological activity simultaneously across the scalp, providing sensitivity to distributed networks and their rapidly changing patterns of neural activity~\cite{o2018dynamics}. Thus, the spatially distributed features identified by the models may reflect contributions from a broader network of brain regions involved in the experience of discomfort.

For ERPVR, across 120 model runs, with and without calibration, with and without outlier detection, with undersampling and oversampling, with Oddballs or Standards, and with two different channel orders, LLPf was consistently the channel with the highest integrated gradients, meaning that the left frontal scalp region helped the most in cybersickness discomfort classification. This result harmonizes with previous research and narrows in on an important component of cybersickness: the feeling of discomfort. Yang, Kasabov, and Cakmak~\cite{Yang_Kasabov_Cakmak_2023} also found that activity in a left frontal electrode site above the inferior frontal gyrus was an important feature in predicting cybersickness. Other studies have found beta power in left frontal regions to be important as well~\cite{Khoirunnisaa_Pane_Wibawa_Purnomo_2018,Oh_Son_2022}. Thus, we postulate that tagging the left frontal region as a feature may be useful for (online) cybersickness discomfort detection.   


It is important to note that signals from activity in different brain areas are distorted as they pass through the skull and meninges and propagate across the head. Additionally, the amplitudes that are picked up from an individual electrode can reflect the combination of activity from numerous neuronal ensembles, including those at some distance, making it difficult to map between (two-dimensional) scalp space and (three-dimensional) brain space (the EEG inverse problem~\cite{Luck_2014}). This makes it very challenging to determine where in the brain the signals we record with EEG are coming from. It is also difficult to determine what it was about the patterns in the original signal that the models were picking up on after convolution. As such, care should be exercised in translating from ML model features to characteristics of underlying brain activity. 

For this reason, we cannot interpret with certainty the neural or functional significance of the pattern in amplitudes 
that were here determined as an important feature by our models. At the same time, it is interesting to note what is known about the activity in left frontal and prefrontal areas from fMRI research (which provides substantially greater spatial specificity) regarding pain and sickness processing. Prior work has pointed to the prefrontal and orbitofrontal cortices as areas involved in regulation of autonomic outflow~\cite{Sclocco_Kim_Garcia_Sheehan_Beissner_Bianchi_Cerutti_Kuo_Barbieri_Napadow_2016}, and nausea induced from vection stimulation during fMRI has been shown to result in activation in frontal areas~\cite{Napadow_Sheehan_Kim_Lacount_Park_Kaptchuk_Rosen_Kuo_2013}. Moreover, activity in the left dorsolateral prefrontal cortex (DLPFC) and left ventrolateral prefrontal cortex were significantly correlated with autonomic reactions (change in skin conductance) during recovery from vection-induced nausea~\cite{Sclocco_Kim_Garcia_Sheehan_Beissner_Bianchi_Cerutti_Kuo_Barbieri_Napadow_2016}, further implicating vestibulo-autonomic responses in cybersickness~\cite{Bogle_Benarroch_Sandroni_2022, Cohen_Dai_Yakushin_Cho_2019,Muth_2006}.

Furthermore, the consequential role of the DLPFC in pain regulation and suppression, especially for headache, has been well-established~\cite{Apkarian_Bushnell_Treede_Zubieta_2005,Ong_Stohler_Herr_2019,Seminowicz_Moayedi_2017}. Abnormalities in the DLPFC have been found in patients with chronic back pain~\cite{Ong_Stohler_Herr_2019} and migraines~\cite{Zhou_Wang_Luo_Fitzgerald_Cash_Fitzgibbon_Che_2024}, and the relationship between migraines~\cite{Seminowicz_Moayedi_2017,Zhou_Wang_Luo_Fitzgerald_Cash_Fitzgibbon_Che_2024} and cybersickness has been investigated previously~\cite{Carvalho_Mehnert_Basedau_Luedtke_May_2021}. Interestingly, transcranial magnetic stimulation (TMS) to the left DLPFC specifically has been demonstrated as an effective treatment for thermal pain~\cite{Brighina_De_Tommaso_Giglia_Scalia_Cosentino_Puma_Panetta_Giglia_Fierro_2011} and migraine~\cite{Zhou_Wang_Luo_Fitzgerald_Cash_Fitzgibbon_Che_2024}. In view of this, it would be interesting to explore if TMS to the left DLPFC could reduce discomfort from cybersickness.

\subsection{Important Time Features}

Amplitudes recorded at the scalp are comprised of mixtures of activity generated by multiple neuronal populations over periods ranging from a few to hundreds of milliseconds. As a result, amplitudes at any specific time point or scalp location are highly correlated with activity at temporally and spatially adjacent points. Features identified as important by the models would therefore be expected to extend over ranges of time and space rather than appear at isolated points. 
In EEGNet’s GradCAM maps, importance in the ERPVR dataset increased beginning around 100 ms, peaked around 200 ms, and declined by approximately 300 ms. In VectionVR, the corresponding window began around 80 ms and extended to approximately 250 ms, with a peak around 110 ms. Although the precise timing shifted somewhat across seeds, the maps generally showed similar gradients. These patterns are broadly consistent with the timing of relatively early auditory sensory processing. Cortical auditory ERP components can be observed beginning within the first several tens of milliseconds after stimulus onset, with prominent sensory responses appearing over approximately the first 200 ms~\cite{Luck_2014}. These early responses are relatively tightly time-locked to the eliciting stimulus, whereas later activity increasingly reflects attentional, mnemonic, and other processes whose timing may vary across trials and individuals. One possibility, therefore, is that the relatively stereotyped timing of early auditory processing provides a temporally aligned signal in which differences associated with participants’ ongoing discomfort state are particularly accessible to the model.

The use of auditory event-related potentials during visual VR tasks was one contribution of this work, as most previous ML approaches to EEG-based cybersickness detection have used continuous EEG data~\cite{Chang_Billinghurst_Yoo_2023,Yang_Kasabov_Cakmak_2022}. Repeated auditory events provide periods of neural processing that are precisely aligned in time and relatively constrained by a common eliciting stimulus. This consistency may help models identify variation associated with cybersickness-related discomfort that would be more difficult to isolate in unconstrained continuous EEG. Consistent with this possibility, EEGNet’s GradCAM maps in both datasets emphasized relatively early portions of the auditory-evoked response, despite substantial differences in the primary VR tasks performed in the two studies. In one case, participants were viewing a simulated museum and were asked to pay attention to the artwork and statues, about which they were quizzed at the end of the study. In the other experiment, participants were viewing rotating black and white lines and indicating when they felt sensations of self motion. In both cases, while participants did those primary tasks in VR, their secondary task was to listen to a series of beeps and respond by silently counting (ERPVR) or pressing a button (VectionVR) when they heard an infrequent random beep that sounded different from the others (the auditory oddball paradigm, see Section~\ref{EEG-background-sec}). The next question naturally becomes whether an auditory ERP task would be necessary to find similar features or use these features online. For practical purposes, it would be easier to not have VR users completing any task. If repeated time-locked auditory events are important for model sensitivity, a passive auditory paradigm, such as one designed to elicit the mismatch negativity (MMN;~\cite{Sussman2014}), in which the person can ignore sounds playing in the background, could provide such probes without requiring an overt task. Alternatively, the spatial patterns identified here could guide analyses of continuous EEG to determine whether related discomfort-sensitive activity can be detected without any external probe.

\subsection{Framework and Models}

EEG data are multi-dimensional, complex, and variable across individuals, and EEG datasets are often small because of logistical constraints on data collection. Thus, we needed to develop a framework that would allow us to use different types of models, handle participants with imbalanced label data, calibrate the data by individual to overcome the issue of variability, and determine which features were most helpful for the models. Our framework, ``Neural Network Interpretability with Leave-One-Subject-Out and subject-specific Calibration" (NNILC), is open-source and can be used by other researchers to classify ERP data with imbalanced labels and map out the importance of features to models. The method presented here utilizes CNNs and transformer models, calibration per data source, and interpretability maps to provide a visual representation of what the models focused on during classification. NNILC is designed to be modular to allow evaluation at different phases of training and handle classification label imbalance, especially in the per-subject calibration. The modular design of the two phases, the base training with LOSO evaluation and later the per-subject calibration phase, is so the models can be evaluated on two different abilities: 1) generalization to new unseen subject data with the LOSO phase and 2) subject-specific calibration to learn patterns more specific to that subject. Neural networks were chosen to allow this two phase split of training from scratch (for compact models) and fine-tuning to a specific individual. NNILC is also model-agnostic; any neural network with classification capability can be loaded into this framework. 

Using a sample of ERP brain activity collected while 29 participants were exposed to cybersickness-inducing stimuli in VR for ERPVR and 36 participants who were exposed to vection and cybersickness-inducing stimuli in VR for VectionVR, we compared the highest ranked features between models of increasing complexity by trainable parameters: a CNN (EEGNet), a convolutional transformer (EEG Conformer), and a pre-trained vision transformer (EEG ViT). We found that certain areas of the head (left frontal and right parietal) best helped predict cybersickness consistently. 
These features provide testable hypotheses about cybersickness-related patterns in EEG data and could be used to improve cybersickness classification in the future.


From the metrics, for ERPVR, calibration generally seemed to improve the models' classification performance per subject, though 
subjects with both labels were harder for the models. This may be the result of a larger distribution shift between these subjects and the others in the train set, where the majority had only one label. Reducing $L2_{SP}$ regularization may allow models to calibrate more to these subjects, though there is the trade-off of possibly losing helpful patterns learned from other subjects and overfitting. Alternatively, training the model on only one subject's data or supplementing with more data could help. However, logistical constraints limit both options. The amount of EEG data that can practically be collected from a single participant is typically limited for machine-learning applications. Also, since brain recordings are sensitive and potentially identifiable, they are not always made publicly available. 

A review of 33 studies on brain activity during cybersickness found associated changes across broad areas of the head and in every frequency band~\cite{Chang_Billinghurst_Yoo_2023}. This heterogeneous evidence made it difficult to predict which features might best help classify cybersickness. Here, we extracted features only from participants who experienced cybersickness-related discomfort and for whom the models achieved high classification accuracy. Though the ERPVR data collection study used a stimulus that caused cybersickness in a different sample previously, experimenters could not control how much discomfort participants experienced. Thus, from 29 participants in ERPVR, only 14 felt discomfort, and high accuracy classification was only achieved for a subset of those. Thus, our features from ERPVR are from a small number of people: five for Conformer, five for EEGNet, and six for pre-trained EEG ViT. Although this does decrease the amount of contributing data, we have greater confidence that the identified features actually reflect something about cybersickness discomfort and not other patterns present in the data. This may explain why a fewer number of features were especially important here versus in other work. Indeed, when we compared the most useful features between high accuracy discomfort subjects and all subjects where more participants did not feel discomfort, the IGs both within multiple runs of the same neural network and between different neural networks were different. 
However for VectionVR, more subjects were able to achieve high accuracy for high discomfort classification by the models. Unlike ERPVR where only a handful of subjects felt discomfort, most of VectionVR's subjects felt some level of discomfort. There were 15 high discomfort subjects and all 15 were able to achieve good accuracy after calibration: 9 for EEGNet, all 15 for Conformer, and all 15 for pre-trained EEG ViT.


\section{Limitations}

Though we have shown that our method can handle this highly variable data, the datasets are small for ML, even if that is expected for EEG. The number of people that experienced cybersickness discomfort in the ERPVR dataset was limited because individual differences in susceptibility meant that a random sample of participants would have significant variety in how much cybersickness or discomfort they experience during a study~\cite{Tian_Lopes_Boulic_2022}. Though there were a small number of subjects who experienced cybersickness discomfort and had high accuracy, the question of cross-dataset generalizability is addressed with the second dataset, and we provide our code so other EEG cybersickness datasets can be compared. Another limitation is that across both montages there were few occipital electrodes, so we do not know if those areas of the head would have recorded brain data helpful in cybersickness discomfort classification. Also, the labels for the datasets are quite coarse since the SSQ was given to subjects after a block of trials (about every two minutes for four periods) in ERPVR, and only once at the end of eight blocks for VectionVR, though this is standard for cybersickness measurement~\cite{Chang_Billinghurst_Yoo_2023}. The scoring options for discomfort in the SSQ are also only 0 - 3, which does not allow for fine-grained analysis based on the intensity of sickness. Because feature interpretation focused on participants for whom classification accuracy was $>75\%$, the features identified here may not fully characterize the neural patterns present in participants who were more difficult for the models to classify.

\section{Conclusion}\label{sec13}

In this paper, we identify features of EEG data that deep learning models consistently find most useful when classifying cybersickness-related discomfort in two datasets.
By using NNILC, our model-agnostic framework to train neural networks, calibrating them by individual, and extracting interpretability maps of the models, we identified the features in the input EEG data that each neural network found useful. 
We demonstrate use of this framework with a CNN and transformers which classified whether or not a person was experiencing cybersickness discomfort given a short window of brain activity data. 
Across 120 ERPVR experiment runs varying trial type, sampling, outlier detection, and electrode order, a subset of scalp locations consistently contributed strongly to accurate discomfort classification. Related spatial and temporal patterns were also identified in the independent VectionVR dataset despite differences in participants, VR stimuli, EEG systems, and electrode montages. These recurring features provide testable candidates for future work, including studies examining whether similar patterns can support discomfort detection in continuous EEG without an auditory ERP paradigm.

\section{Methods}\label{sec11}

\begin{figure*}
    \includegraphics[width=\linewidth]{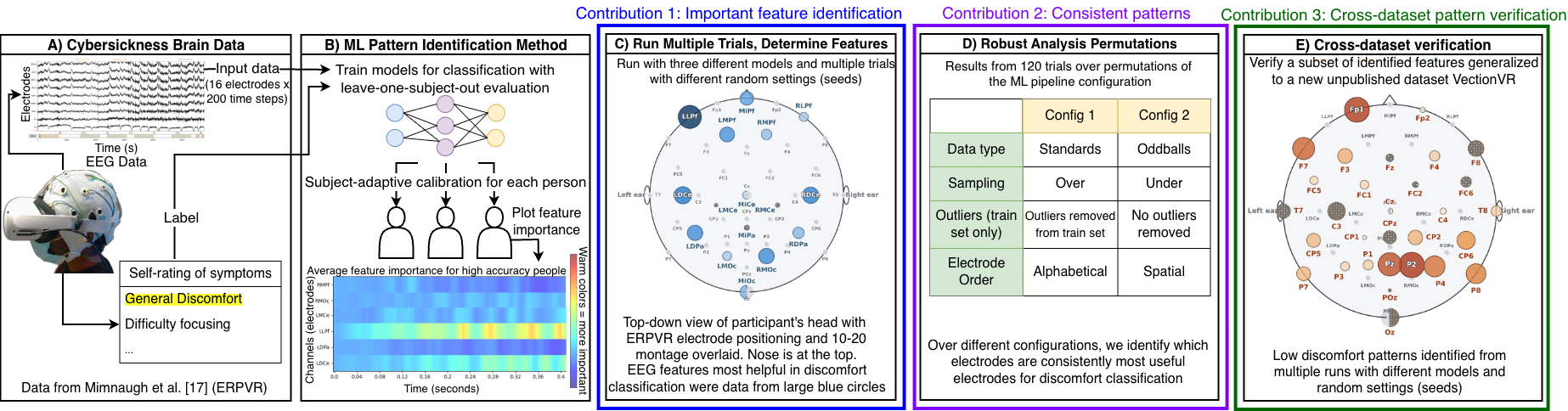}
    \caption{Methods overview. A) ERPVR data from Mimnaugh et al.~\cite{Mimnaugh2023} was brain activity recorded with EEG 
    while participants experienced cybersickness. 
    B) Discomfort 
    rating classification with leave-one-subject-out evaluation to measure model generalizability across subjects before per-subject calibration to evaluate subject-specific classification performance 
  (further details in Figure~\ref{fig:main-method}). To ensure feature specificity, for people with discomfort classification accuracy $>$75\%, extracted feature interpretation maps from each model. C) Over multiple experiments, a subset of EEG electrodes consistently contributed the most information to the models for cybersickness-related discomfort classification. D) To verify the patterns, ablation experiments were run with all permutations of different configurations of the pipeline. E) Cross-dataset verification of identified patterns on a new, unpublished dataset (VectionVR).} 
    \label{fig:graphical-abstract}
\end{figure*}







\subsection{ERPVR Dataset Description}\label{sec:Dataset}

This dataset is from Mimnaugh et al.~\cite{Mimnaugh2023}. The research study was conducted at the University of Illinois Urbana-Champaign and received approval from the IRB ($\#22519$). Participants consented to data sharing and future use of their data for new research. The data were from 29 participants (17 female and 12 male, mean age 23.1). 

{\textbf{\underline{ERPVR Study design.}}} As described in the Supplemental Material Section K 
and original manuscript, ERPVR cybersickness data was collected using a dual-task auditory oddball paradigm.
Participants passively viewed two tours through a virtual museum, each twice, in an Oculus Quest 2 headset (Figure \ref{fig:graphical-abstract}A) while their brain activity was recorded. The two tours through the museum differed based on their amount of rotation and forward speed, and were shown in a previous study to evoke either mild or moderate amounts of VR sickness~\cite{Becerra_Suomalainen_Lozano_Mimnaugh_Murrieta-Cid_LaValle_2020}. During the sickening museum tours, participants were instructed their primary task was to pay attention to the virtual museum because they would be quizzed on it later. They were also instructed to simultaneously listen to a series of beeps and keep a mental count of the infrequent (oddball) beeps (the P3b secondary task). After a non-sickening VR immersion condition (baseline) and immediately after each two minute tour through the VR museum, participants filled out an SSQ questionnaire rating their cybersickness symptoms during the museum tour or at the current moment. All participant SSQ scores and other demographic and questionnaire responses are available from~\cite{Mimnaugh2023} online at: \url{osf.io/v9fst}.

{\textbf{\underline{ERPVR EEG data.}}} Sixteen scalp electrodes, three electrooculogram, and two reference (left and right mastoids) were used for data collection. 
The full electrode montage was arranged in an approximately equidistant configuration to ensure uniform spatial sampling of the scalp potential field ~\cite{Oostenveld_Praamstra_2001} (see~\cite{Ganis_Kutas_Sereno_1996} for a comparison to the 10-20 layout). A broadly distributed subset (16 of 32) electrodes was used to shorten data collection time (under three hours).  For these analyses, ERP windows (epochs associated with each trial) of 800 ms after the participant heard each beep were extracted. 

We separately ran experiments with the standard and oddball trials because there is significantly different cognitive processing in the ERP window related to making a response after the oddballs. As the dataset was already small and the ratio of standards to oddballs was about 70\% to 30\% per block (9,583 standards and 3,641 oddball total), we selected the standard trials for full analyses using all deep learning models. Selecting the standards provided the benefit that we could find patterns in the data not associated with the P3b, which we already knew to be impacted by cybersickness. Regardless of whether or not a participant responded to the oddball, cognitive processing of cybersickness-related discomfort (if experienced) was happening during both the oddball and standard trials. The oddball trials were run with only the more compact models that can handle the smaller dataset size. 





{\textbf{\underline{ERPVR data preprocessing.}}} \label{ERPVRpreproc}
Following guidance on ERP data preprocessing~\cite{Luck_2014}, raw EEG data were downsampled to 250 Hz, re-referenced to the average of both mastoids, cleaned of eye artifacts using independent component analysis, filtered, cut into periods of 800 ms after the auditory stimulus, baselined to the 200 ms before the stimulus, and then removed of any trials where voltages exceeded $\pm100$ microvolts from artifacts~\cite{Mimnaugh2023}. Per recommendations for free viewing EEG experiments~\cite{Dimigen_2020} and optimal filter settings for the P3~\cite{Zhang_Garrett_Luck_2024}, the EEG data was filtered with second-order IIR Butterworth filter with a 12 dB/octave roll-off and a 0.2 to 30 Hz half-amplitude cutoff. 

{\textbf{\underline{ERPVR samples.}}} There were approximately 330 Standard and 126 Oddball trials per person with 200 samples per trial in the time domain. Binary classification labels were based on each user's self-reported intensity for one VR sickness symptom (general discomfort from the SSQ). 
If participants reported any level of discomfort (score $>0)$, then the sample was labeled as ``discomfort," else it was ``no discomfort." Binary classification was chosen instead of four-class because in the dataset only a subset of people felt discomfort (14 of 29), leaving few examples per class where the score $>$ 0. Therefore, the samples could not be further segmented into varying levels of discomfort. 



\subsection{VectionVR Dataset Description}

The VectionVR dataset was collected at the John Paul II Catholic University of Lublin. Ethical approval for the project was obtained from the Research Ethics Committee of the Institute of Psychology (\#~KEBN\_25/2024). Participants consented to future use of their data for new research. The data are from 36 of 40 participants (19 females and 17 males, mean age 22.7, min = 19, max = 39). Three of the 40 people were excluded (for atypical stereovision; an equipment issue; and vomiting), and one person did not consent to future use of their data.

{\textbf{\underline{VectionVR study design.}}} Similar to ERPVR, this study was designed with a dual-task auditory oddball paradigm. However, there were several significant differences. This study was designed to determine how a factor that contributes to cybersickness, vection, impacted attention as compared to cybersickness. To elicit vection and cybersickness in participants, the subjects viewed alternating black and white lines which spun either clockwise or counterclockwise for each approximately two-minute block. The stimuli appeared similar to~\cite{Strozak2019a}, except one outer ring of lines spun in one direction, and a smaller inner ring of lines spun in the opposite direction. Eight blocks were viewed in a VR HMD, the Meta Quest 3, and another eight blocks on a large screen (the screen trials were not included here). While viewing the lines, participants indicated if they were feeling vection by pressing on a joystick in their dominant hand, and responded to the oddball auditory stimulus with a button in their non-dominant hand. At the end of eight blocks, the participants completed an SSQ questionnaire. Thus, all eight blocks had only one discomfort rating for each subject. 

{\textbf{\underline{VectionVR EEG data.}}} A Brain Vision actiCHamp plus amplifier and actiCAP active electrodes with a standard 10-20 montage were used to collect EEG data. Thirty scalp electrodes and two electrodes removed from the EEG cap and affixed to the mastoids (TP9 and TP10) were used. The montage was modified from a 10-20 layout to provide a greater density of electrodes around centro-parietal areas. Data were recorded at 1000 Hz and impedances were kept under 10 k$\Omega$. Windows (epochs) of EEG data for 800 ms after the participant heard the beep were extracted for analysis. Only standards were analyzed (as discussed in Section~\ref{EEG-background-sec}).

{\textbf{\underline{VectionVR data preprocessing.}}} The VectionVR EEG data were preprocessed in EEGLAB~\cite{delorme2004eeglab} with the same preprocessing pipeline as the ERPVR data (Section~\ref{ERPVRpreproc}, see~\cite{Mimnaugh2023} for a more detailed description of the processing). Data were recorded online with a Cz reference and re-referenced to the average of the mastoids offline. The data were downsampled to 250 Hz and filtered with a Butterworth bandpass filter (second order IIR, 0.2 to 30 Hz half-amplitude cutoff, 12 dB/octave roll-off). They were cleaned of artifacts using the same ICA method as ERPVR~\cite{Dimigen_2020} and epoched from -200 ms pre-stimulus to 800 ms post-stimulus, with pre-stimulus baselining. A $\pm$100 $\mu$V simple voltage threshold was used to remove epochs with artifacts in channels POz, CP1, Pz, P3, P1, P2, P4, CP2, CPz, and Cz (a similar area as ERPVR) across the one second window. For the models, screen trials and the 200 ms pre-stimulus period were removed.

{\textbf{\underline{VectionVR samples.}}} There were approximately 336 Standard and 144 Oddball trials per person with 200 samples per trial in the time domain. EEG data were recorded on 30 channels (electrodes), giving the input data a size of 30 x 200. Binary classification labels were based on each user's self-reported intensity for the same VR sickness symptom (general discomfort from the SSQ). However, unlike for ERPVR data, only three out of 36 subjects reported a score of 0. To handle this imbalance, we instead separate between low discomfort (score $\leq$ 1) and high discomfort (score $>$ 1). This gave us 15 out of 36 high discomfort subjects and 21 low discomfort subjects. Since the SSQ was only collected once at the end of the block of HMD trials, there is only one SSQ label for each subject and not per-block like in ERPVR. So for VectionVR, there are only single-label subjects and no mixed label subjects. In total, there were 12,096 standard and 5,184 oddball trials. \\

\subsection{Data Analysis and Augmentation}\label{sec:data_analysis}

Machine learning (ML) models learn patterns within the train set to generalize to the validation set and unseen test set, so if the train set is skewed or biased in some way, the model could pick up that bias and learn something not inherent in the data. Since the ERPVR binary classification labels were very skewed (3:1 for no discomfort vs discomfort based on the SSQ symptom general discomfort), we analyzed the data through plots and unsupervised learning (in Supplemental Material Section D)
, and explored augmentation methods to alleviate this issue. This included balancing the labels through sampling (both over and undersampling), generating synthetic data, and identifying outliers in the data. More on data augmentation in Supplemental Material Section C. 
{\textbf{\underline{Z-score.}}} 
Input data size is a $(C,T)$ matrix by channel (electrode) $C$ and by time $T$. We first flattened the data to size $C*T$ before computing the z-score, defined as $\mathbf{z}_i = \frac{\mathbf{x}_i - \mu}{\sigma}$ where $\mathbf{x}_i \in \mathbb{R}^{C * T}$ is the flattened vector being normalized, $\mu \in \mathbb{R}^{C * T}$ is the mean, and $\sigma \in \mathbb{R}^{C * T}$ is the standard deviation, calculated differently to serve two distinct purposes with different computations.
\begin{enumerate}
    \item Per-feature normalization: mean and standard deviation were computed per feature across the train set regardless of label. Each of the 3200 channel-time features for ERPVR and 6000 features for VectionVR has its own mean and standard deviation, thus channels are not forced to share a common distribution. This is necessary for neural network training stability. Without it, EEG’s varying amplitudes would cause gradient instability and bias the model toward larger absolute amplitude features regardless of discomfort content.
    \item Outlier detection: mean and standard deviation were computed separately per label per feature from the train set. Trials where any feature exceeds z=3 are removed. This prevents extreme discomfort-related patterns from being flagged as outliers against the no discomfort distribution. Z-score $>= 3$ is used to delineate outliers in other work \cite{nolan2010faster, artist}. We apply it at finer granularity per feature and validate by comparing strict, lenient, and no removal. Validation and test sets are never modified, so these steps cannot inflate evaluation metrics.
\end{enumerate}

{\textbf{\underline{Outlier detection and removal.}}} 
%
%
Adding outlier removal improved performance compared to no outlier removal for subjects with discomfort-related sickness labels for ERPVR, analysis in Supplemental Table B3.
Since removing trials where any feature exceeds z=3 was quite strict, we also tested a milder version that identified outliers if over half the features exceeded three standard deviations. However, test performance dropped for ERPVR, especially on subjects with more discomfort data, compared to the strict version. For VectionVR, no outlier detection performed better than with outlier detection.

{\textbf{\underline{Sampling.}}} To balance the class labels, we tried variations of under and oversampling. For ERPVR, for undersampling, we randomly picked majority no discomfort data to match the number of discomfort. For oversampling, we randomly replicated minority discomfort data until the classes were balanced. For VectionVR, it was also high discomfort as the minority class so oversampling would sample from there, and low discomfort is the majority class so undersampling would drop from that. These approaches were not ideal, however, since with undersampling we lose data, whereas with oversampling repeating the same data does not provide new information to the models. 
Generating synthetic data for the minority class was investigated but found to perform worse than random sampling for neural network models. 

\subsection{Main Data Processing Pipeline for ERPVR dataset}\label{sec:erpvr-main-pipeline}

The primary data processing pipeline for ERPVR data classification of discomfort was: 1) individual stimuli outlier detection and removal, 2) normalization, and 3) random oversampling. The strict outlier detection was run first on the train set only, then z-score normalization on the remaining data. Validation, test, and calibration sets were normalized with the same average $\mu$ and standard deviation $\sigma$ obtained from the train set. Finally, the train set was augmented using random oversampling, implemented with Pytorch's \cite{paszke2019pytorchimperativestylehighperformance} weighted random sampler by giving higher weights to the minority class data. To check for consistency however, we vary the random seed so the sampler picks different samples. We separate the standards and oddball ERPs because the cognitive processing is different between the two.

\subsection{Main Data Processing Pipeline for VectionVR dataset}\label{sec:vectionvr-main-pipeline}

The primary data processing pipeline for VectionVR data classification of high vs. low discomfort was: 1) no outlier detection, 2) normalization, and 3) random oversampling. Spatial electrode channel ordering does better than alphabetical for VectionVR's 10-20 montage, especially for the ImageNet pre-trained Vision Transformer model (more in Section \ref{sec:dl-patterns}). The normalization and random oversampling use the same method as in ERPVR, z-score normalization by average $\mu$ and standard deviation $\sigma$ from train set and Pytorch's weighted random sampler.

\label{sec:traditional-ml}

\subsection{Deep Learning for Pattern Recognition}\label{sec:dl-patterns}

Since unsupervised learning by clustering the EEG data without labels was unable to identify meaningful groupings, and traditional machine learning models struggled to capture the complexities of the EEG data in the time domain (more in Supplemental Material Section D)
, deep learning models were considered. Convolution seemed promising to extract spatially local features in the signal, so EEGNet \cite{Lawhern_2018} was chosen as it was designed for EEG data, including ERPs, and compact enough to work with small datasets. Since EEG has a time dimension component, models that take sequential information into account were also considered, specifically transformers. EEG Conformer was chosen as it is a compact convolutional transformer designed to encapsulate local and global features in EEG data \cite{9991178}. An ImageNet pre-trained EEG Vision Transformer (ViT) was chosen to leverage the reasoning patterns it had learned from images to EEG data with fine-tuning \cite{yang2023vit2eeg}. Each input sample is a matrix of electrical potential amplitudes in microvolts with shape EEG electrodes. For ERPVR it is $C=16$ by time $T=800$ms (200 time steps in the data) and for VectionVR it is $C = 30$ by time $T = 800$ms for 200 time steps as well. 
The three models are described by order of increasing complexity.

\subsection{Model Interpretability}  \label{Sec:IGs}

For EEGNet, the final convolution layer was set for GradCAM extraction. The heatmaps show where along the time dimension the activation was strongest for the model to classify discomfort. This is because EEGNet's final convolution layer is a pointwise convolution that mixes all feature maps together. For EEG Conformer and pre-trained EEG ViT, integrated gradients (IG) provides more fine-grained information about the gradient strength by channel and time. While GradCAM interpretation very much depends on which layer the maps were extracted from, IG passes through the whole model to compute the cumulative gradients. Descriptions in Supplemental Material F. 

\subsection{Calibration}

\begin{figure*}
    \centering
    \includegraphics[width=\linewidth]{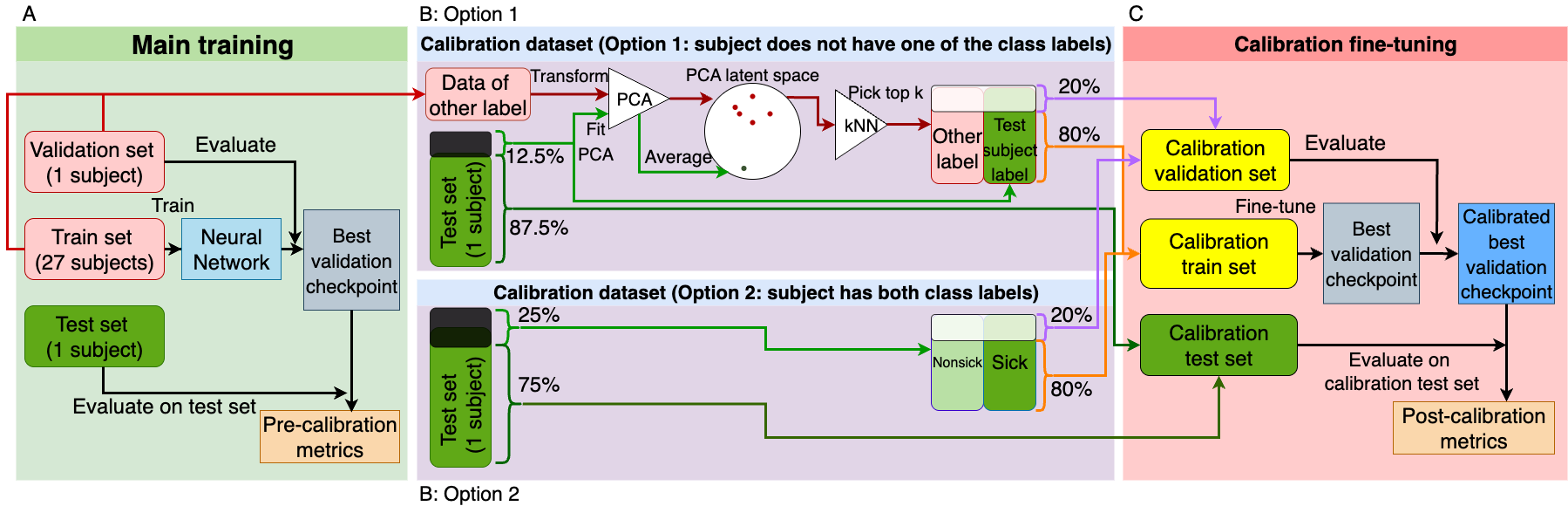} 
    \caption{Neural Network Interpretability with Leave-One-Subject-Out and subject-specific Calibration (NNILC) Framework: 1) main training (A) for leave-one-subject-out cross validation and 2) calibration (B and C) to fine-tune model to the test subject. There are two possible calibration datasets depending on if the test subject data has both labels or not. For main training, the train set goes through outlier detection, z-score normalization, and random oversampling. Best validation checkpoint is fine-tuned in calibration. If test subject has both labels, 25\% of the test subject's data is sampled. If test subject only has one label, 12.5\% of test subject's data is sampled and top k data of the other label from other subjects are taken that are closest to sampled test subject data centroid in whitened PCA space.}
    \label{fig:main-method}
    \vspace{-0.4cm}
\end{figure*}

Due to the high inter-subject variability of the ERP data, it was difficult for the model to generalize to the data of the subject left out. To bridge the domain shift between the train set distribution and the subject left out's distribution, with the goal of improving subject-specific accuracy, we used some data of the left out subject to fine-tune the trained model checkpoint (Figure \ref{fig:main-method}). Many works on EEG classification tasks rely on calibration to specific subjects to improve classification performance, by collecting labeled data of the test subject to build a classifier \cite{shi2025plug, 10.3389/fnins.2021.733546} or fine-tune a model \cite{ding2025eeg}.

{\textbf{\underline{Calibration dataset.}}} We sampled either 12.5 or 25\% from the dataset of the subject left out. If the subject left out had both labels, then 25\% of the data were sampled, where exactly 12.5\% were from each class. However, many subjects only had one label, so for those subjects, 12.5\% of their data was sampled. To handle the imbalance, the same amount of the other class label were taken from the remaining 28 subjects' data for ERPVR or 35 for VectionVR. Instead of random sampling, we took the top $k$-nearest neighbors that were closest to the subject left out's distribution since the goal of the calibration set was to bridge the distribution shift for each individual subject. 


The subject left out's sampled data were used to fit a whitened PCA space of 16 components for both ERPVR and VectionVR. Each sampled data was projected to this space then averaged into a centroid. Next, each of the other label data from the remaining subjects was projected into PCA space and kNN was run to find and retrieve the data that were closest to the centroid by Euclidean distance. $K$ was the same as 12.5\% of the subject left out's data. Together, this formed the calibration dataset.

The remaining test subject's data were used as a test set. The validation set was formed by taking 20\% of each class from the calibration dataset. Since each subject had about 330 data samples for Standards, about 40 for each class label were sampled for calibration for ERPVR. For VectionVR Standards, each subject has 336 samples, about 42 for each class label were sampled for calibration. If the test subject only had one label, then about 40 of the other label was taken from the other subjects for a total of 80 data samples, of which 
approximately 16 were used as validation for ERPVR. For VectionVR, since all subjects only had one label, then about 42 of the other label was taken from the other subjects for a total of 84 data samples, of which approximately 18 were used as validation.


{\textbf{\underline{Calibration fine-tuning.}}} Once the calibration dataset was created, the model with the best validation score from the main training phase was selected to be fine-tuned by the calibration set. The validation metric for this phase was a custom weighted balanced accuracy to identify early stopping from evaluation on the validation set, and was designed to have more weight on the correct classification of discomfort data.

For EEGNet, all layers were unfrozen to fine-tune all weights. For the EEG Conformer and pre-trained EEG ViT, sections of the model were frozen. For EEG Conformer, the first three of four encoder blocks were frozen. For the pre-trained EEG ViT, all 12 encoder blocks remained frozen. L2-starting point ($L2_{SP}$) \cite{xuhong2018explicit} regularization, defined in Equation \ref{eq:l2-sp} where $w_0$ is a vector of model parameters from the best validation checkpoint after the main training phase and $w$ is the current model's unfrozen parameters in the calibration fine-tuning phase and $\beta$ is the regularization strength, was added to prevent the pre-trained weights from drifting too much during fine-tuning. This was crucial to prevent the model checkpoints from overfitting to the small calibration set and ``forgetting" the distribution it had learned during the main training phase. In experiment runs, $\beta = \num{2e-3}$. The L2-starting point regularization is defined as:

\vspace{-0.2cm}
\begin{equation}
    L2_{SP}(w) = \frac{\beta}{2}||w - w_0||_2^2
    \label{eq:l2-sp}
\end{equation}
where $S$ is the test subject's dataset, $b_i = \begin{cases}
    2.0 \text{ if } i \in S \\
    1.0 \text{ otherwise }
\end{cases}$ is the weight for data sample, $i \in N$ for calibration train set size $N$, $y_i$ is the ground truth label, $x_i$ is the classification output logits, and $p_i$ is the softmax probability for the discomfort label 1 Equation \ref{eq:sick-label-prob}, defined as: 

\begin{equation}
    p_i = \frac{e^{x_{i,1}}}{e^{x_{i,0}} +  e^{x_{i,1}}}\;,
    \label{eq:sick-label-prob}
\end{equation} 
and the cross entropy loss (CE) is defined as:
\vspace{-0.3cm}
\begin{equation}
    CE(x, y) = - \sum_{i=1}^N \frac{b_i}{||\textbf{b}||} (y_i \log (p_i) + (1 - y_i) \log (1 - p_i)).
    \label{eq:ce-weighted}
\end{equation}
\vspace{-0.2cm}

We used cross entropy loss (CE) with normalized weight by sample and $L2_{SP}$ regularization to form the calibration phase loss here:

\vspace{-0.1cm}
\begin{equation}
    L(x,y,w) = CE(x,y) + L2_{SP}(w).
    \label{eq:calibration-loss}
\end{equation}
\vspace{-0.2cm}


{\textbf{\underline{Custom weighted balanced accuracy.}}} Since we placed more emphasis on the model's ability to correctly classify discomfort data at the cost of potentially creating more false positives, we created a custom metric to identify an early train stopping point by validation. This custom weighted balanced accuracy (WBA) allows tuning the weight on the recall of the discomfort class (positive label) and no discomfort class (negative label) while keeping the weights normalized. Let $TP$ be true positive, $TN$ true negative, $FP$ false positive, and $FN$ false negative. In experiment runs, $\alpha = 0.7$ for the weighted balanced accuracy, defined as:

\vspace{-0.3cm}
\begin{equation}
    \text{WBA} = \alpha \left ( \frac{TP}{TP + FN} \right ) + (1 - \alpha) \left ( \frac{TN}{TN + FP} \right ).
    \label{eq:custom-weighted-bal}
\end{equation}

\subsection{Experiment Setup and Metrics}



For ERPVR, each model runs each configuration of the pipeline for three trials with different random seeds. For VectionVR, each model runs the main pipeline for three trials with different random seeds. Exact training, calibration, and testing details in Supplemental Material Section G. 

To evaluate our approach in the main training pre-calibration phase, we use ``leave-one-subject-out" cross-validation where one specific subject was left out of the train set to serve as the test set and another subject is randomly chosen to serve as the validation set. The remaining subjects (27 for ERPVR, 34 for VectionVR data) were in the train set. This ensured that the test set was completely unseen by the model at this phase and the evaluation results reflected how well the model could generalize to the unseen subject. For the calibration phase, the validation and test sets were as described in Figure \ref{fig:main-method} where minimal amounts of the test subject's data is seen by the model, with the evaluation reflecting how well the model closed the distribution shift to the test subject. Since the dataset was very imbalanced, the metric we considered is balanced accuracy. Another metric for imbalanced data, F1-score, is reported in Supplemental Material Section A. 




{\textbf{\underline{Balanced accuracy.}}} Balanced accuracy (BA) is the average of the recall for each class. Let $TP$ be true positive, $TN$ true negative, $FP$ false positive, and $FN$ false negative. Balanced accuracy for binary classification \cite{scikit-learn} is: 

\begin{equation}
    \text{BA} = \frac{1}{2} \left(\frac{TP}{TP + FN} + \frac{TN}{TN + FP} \right).
\end{equation}


\subsection{LLM Usage}
AI was used to create the montage images (Figure~\ref{fig:electrode-montages-all}) from our electrode location coordinates and calculated IGs, plot integrated gradients charts, format tables in latex, find related literature, and proofread the final draft.

\backmatter

\bmhead{Supplementary information}

This article does have supplementary files.




\bmhead{Acknowledgements}

The authors wish to thank Ruohan Zhang, Dariusz Zapała, Hanna Zgorzelska, Weronika Mroczka, Markku Suomalainen, Israel Becerra, Rafael Murrieta-Cid, Eliezer Lozano, Larry, Lucy, Yoshi, Abdulsatar Aboud, Michael Mimnaugh, James Motes, and Marco Morales.

\section*{Declarations}




\section{Funding}

This work was supported by a European Research Council
Advanced Grant ILLUSIVE: Foundations of Perception
Engineering, 101020977. 

\section{Data availability}






Framework code and additional analyses will be available on Github. ERPVR participant data, with the exception of raw EEG (for participant privacy protection), is available on OSF (https://osf.io/v9fst). Aggregate EEG data measurements for ERPVR are on Harvard Dataverse (https://doi.org/10.7910/DVN/MENRFT). VectionVR data, with the exception of raw EEG, will be made available on OSF on publication of the paper analyzing its ERPs.

\section{Code availability}

Framework code will be available on Github on acceptance.







\section{Author contribution}


K.M., J.Y., K.F., S.L., T.O., W.Y., P.A., P.S., and N.A. contributed to the conception. K.M., J.Y., E.C., K.F., P.A., P.S., and M.K. contributed to the design. J.Y. and K.M. wrote the paper. J.Y., K.M., K.F., M.K., E.C., and P.A. contributed to the analysis. K.F., K.M., J.Y., E.C., and M.K. contributed to the interpretation of the results. 


\section{Competing interests}

The authors declare no competing interests.


\bibliography{sn-bibliography}


\clearpage

\begin{appendices}
\suppressfloats[t] 

    \section{Additional Results}\label{sec:supplemental-results}
    {\textbf{\underline{F1-score.}}} F1-score is the harmonic mean between precision and recall \cite{scikit-learn} where TP is true positive, FP is false positive, and FN is false negative. F1-score is defined as: \vspace{-0.3cm}

\begin{equation}
    F1 = \frac{2 * TP}{2 * TP + FP + FN}
\end{equation}
    
where the F1-scores for the ERPVR runs are reported in Supplementary Table \ref{tab:bygroup-all-models-f1-base-calib-mean-half-range-over-seeds-by-configuration-subject-group}. For single-label subjects, F1-score becomes degenerate and defaults this to $F1 = \frac{2 * TP}{2 * TP + FN}$. F1-scores not reported for VectionVR due to degeneracy from single-label subjects only.

\begin{table}[htbp]
\centering
\begin{tabular}{lllllllllll}
\hline
M & D & S & O & E & DB & DC & NB & NC & MB & MC \\
\hline
EEG & Std & Und & ON & Alp & 79.2 $\pm$ 4.9 & 90.7 $\pm$ 2.4 & 71.3 $\pm$ 2.2 & 79.1 $\pm$ 1.3 & 45.8 $\pm$ 0.7 & 47.5 $\pm$ 3.6 \\
EEG & Std & Und & ON & Spa & 81.0 $\pm$ 1.0 & 89.6 $\pm$ 1.5 & 69.6 $\pm$ 2.8 & 75.7 $\pm$ 2.3 & 46.4 $\pm$ 0.9 & 47.5 $\pm$ 5.1 \\
EEG & Std & Over & ON & Alp & 84.4 $\pm$ 0.7 & 91.5 $\pm$ 1.2 & 70.9 $\pm$ 1.8 & 77.8 $\pm$ 1.7 & 44.2 $\pm$ 2.6 & 49.0 $\pm$ 3.7 \\
EEG & Std & Over & ON & Spa & 79.1 $\pm$ 1.0 & 89.6 $\pm$ 3.3 & 67.9 $\pm$ 1.5 & 75.6 $\pm$ 2.7 & 45.8 $\pm$ 1.3 & 47.1 $\pm$ 3.3 \\
EEG & Std & Und & OFF & Alp & 60.1 $\pm$ 6.0 & 87.0 $\pm$ 4.3 & 83.8 $\pm$ 0.6 & 79.2 $\pm$ 3.0 & 44.3 $\pm$ 0.9 & 50.2 $\pm$ 1.9 \\
EEG & Std & Und & OFF & Spa & 61.3 $\pm$ 1.7 & 89.9 $\pm$ 2.5 & 82.6 $\pm$ 1.0 & 77.4 $\pm$ 3.4 & 42.7 $\pm$ 1.2 & 48.4 $\pm$ 2.2 \\
EEG & Std & Over & OFF & Alp & 62.7 $\pm$ 2.5 & 86.2 $\pm$ 1.4 & 84.1 $\pm$ 0.9 & 79.8 $\pm$ 3.7 & 43.0 $\pm$ 0.5 & 49.2 $\pm$ 2.9 \\
EEG & Std & Over & OFF & Spa & 63.4 $\pm$ 4.0 & 88.0 $\pm$ 1.9 & 83.7 $\pm$ 2.1 & 80.0 $\pm$ 0.4 & 42.2 $\pm$ 2.4 & 50.6 $\pm$ 3.0 \\
EEG & Odd & Und & ON & Alp & 83.6 $\pm$ 3.4 & 87.0 $\pm$ 4.4 & 68.7 $\pm$ 2.3 & 74.8 $\pm$ 0.3 & 46.8 $\pm$ 2.7 & 49.4 $\pm$ 6.0 \\
EEG & Odd & Und & ON & Spa & 83.3 $\pm$ 0.7 & 87.2 $\pm$ 0.7 & 67.9 $\pm$ 3.1 & 73.6 $\pm$ 4.0 & 44.3 $\pm$ 1.5 & 47.5 $\pm$ 2.5 \\
EEG & Odd & Over & ON & Alp & 81.7 $\pm$ 3.6 & 85.2 $\pm$ 2.9 & 70.1 $\pm$ 3.2 & 75.4 $\pm$ 2.3 & 43.3 $\pm$ 1.1 & 45.0 $\pm$ 4.1 \\
EEG & Odd & Over & ON & Spa & 80.5 $\pm$ 2.0 & 89.2 $\pm$ 1.2 & 64.5 $\pm$ 1.7 & 70.8 $\pm$ 2.7 & 43.4 $\pm$ 1.9 & 44.4 $\pm$ 2.9 \\
EEG & Odd & Und & OFF & Alp & 59.2 $\pm$ 2.0 & 75.5 $\pm$ 5.7 & 80.6 $\pm$ 1.5 & 80.6 $\pm$ 3.5 & 42.7 $\pm$ 1.0 & 47.7 $\pm$ 7.4 \\
EEG & Odd & Und & OFF & Spa & 63.2 $\pm$ 1.7 & 80.2 $\pm$ 4.1 & 82.1 $\pm$ 3.7 & 79.9 $\pm$ 3.2 & 41.9 $\pm$ 0.3 & 46.9 $\pm$ 3.2 \\
EEG & Odd & Over & OFF & Alp & 60.1 $\pm$ 3.2 & 80.7 $\pm$ 2.0 & 81.5 $\pm$ 3.3 & 79.4 $\pm$ 2.6 & 41.7 $\pm$ 0.8 & 45.8 $\pm$ 3.4 \\
EEG & Odd & Over & OFF & Spa & 65.8 $\pm$ 4.2 & 79.7 $\pm$ 4.1 & 84.6 $\pm$ 1.8 & 80.8 $\pm$ 3.5 & 42.9 $\pm$ 2.5 & 47.4 $\pm$ 1.8 \\
\hline
Con & Std & Und & ON & Alp & 63.8 $\pm$ 7.7 & 87.5 $\pm$ 3.1 & 84.8 $\pm$ 3.7 & 94.2 $\pm$ 1.5 & 47.9 $\pm$ 1.3 & 59.8 $\pm$ 2.5 \\
Con & Std & Und & ON & Spa & 68.1 $\pm$ 4.2 & 88.5 $\pm$ 2.1 & 83.9 $\pm$ 3.2 & 94.2 $\pm$ 1.5 & 49.0 $\pm$ 2.8 & 58.9 $\pm$ 0.3 \\
Con & Std & Over & ON & Alp & 62.2 $\pm$ 7.8 & 89.2 $\pm$ 2.2 & 85.0 $\pm$ 1.4 & 92.9 $\pm$ 1.2 & 50.4 $\pm$ 2.5 & 60.2 $\pm$ 2.5 \\
Con & Std & Over & ON & Spa & 62.1 $\pm$ 2.4 & 90.3 $\pm$ 1.7 & 82.2 $\pm$ 2.3 & 93.9 $\pm$ 0.9 & 46.9 $\pm$ 2.3 & 57.1 $\pm$ 3.3 \\
Con & Std & Und & OFF & Alp & 47.0 $\pm$ 7.1 & 91.5 $\pm$ 0.5 & 89.0 $\pm$ 0.7 & 91.9 $\pm$ 2.6 & 47.0 $\pm$ 2.3 & 62.0 $\pm$ 0.8 \\
Con & Std & Und & OFF & Spa & 46.5 $\pm$ 1.5 & 90.5 $\pm$ 2.3 & 88.5 $\pm$ 1.5 & 93.2 $\pm$ 3.0 & 48.0 $\pm$ 2.5 & 59.8 $\pm$ 1.3 \\
Con & Std & Over & OFF & Alp & 51.1 $\pm$ 2.6 & 91.2 $\pm$ 1.9 & 87.2 $\pm$ 1.1 & 92.5 $\pm$ 2.8 & 47.0 $\pm$ 2.4 & 59.7 $\pm$ 2.1 \\
Con & Std & Over & OFF & Spa & 46.9 $\pm$ 4.8 & 91.6 $\pm$ 0.9 & 88.8 $\pm$ 0.7 & 93.0 $\pm$ 2.0 & 46.7 $\pm$ 1.5 & 59.4 $\pm$ 2.5 \\
Con & Odd & Und & ON & Alp & 69.5 $\pm$ 2.7 & 81.3 $\pm$ 2.1 & 78.4 $\pm$ 0.8 & 84.1 $\pm$ 1.7 & 47.5 $\pm$ 1.4 & 55.2 $\pm$ 3.3 \\
Con & Odd & Und & ON & Spa & 71.4 $\pm$ 2.9 & 83.8 $\pm$ 5.5 & 78.5 $\pm$ 2.3 & 85.2 $\pm$ 1.9 & 47.4 $\pm$ 0.8 & 55.5 $\pm$ 1.2 \\
Con & Odd & Over & ON & Alp & 71.0 $\pm$ 1.2 & 80.9 $\pm$ 1.1 & 78.9 $\pm$ 1.3 & 85.5 $\pm$ 1.8 & 46.3 $\pm$ 1.8 & 55.2 $\pm$ 1.4 \\
Con & Odd & Over & ON & Spa & 67.5 $\pm$ 1.5 & 77.8 $\pm$ 4.1 & 81.3 $\pm$ 1.0 & 86.3 $\pm$ 2.4 & 47.3 $\pm$ 2.0 & 54.8 $\pm$ 2.1 \\
Con & Odd & Und & OFF & Alp & 47.7 $\pm$ 2.9 & 79.5 $\pm$ 9.2 & 86.2 $\pm$ 1.5 & 83.2 $\pm$ 2.1 & 44.7 $\pm$ 1.8 & 51.7 $\pm$ 1.4 \\
Con & Odd & Und & OFF & Spa & 43.8 $\pm$ 0.8 & 76.7 $\pm$ 8.7 & 83.9 $\pm$ 1.9 & 85.6 $\pm$ 2.0 & 44.0 $\pm$ 1.1 & 55.1 $\pm$ 1.8 \\
Con & Odd & Over & OFF & Alp & 45.6 $\pm$ 6.5 & 82.5 $\pm$ 3.3 & 84.0 $\pm$ 0.9 & 84.1 $\pm$ 2.7 & 43.3 $\pm$ 2.0 & 54.7 $\pm$ 4.8 \\
Con & Odd & Over & OFF & Spa & 45.8 $\pm$ 3.1 & 79.1 $\pm$ 4.5 & 84.6 $\pm$ 1.5 & 85.5 $\pm$ 0.8 & 42.3 $\pm$ 4.2 & 53.1 $\pm$ 5.1 \\
\hline
ViT & Std & Und & ON & Alp & 83.0 $\pm$ 2.1 & 96.3 $\pm$ 1.0 & 66.7 $\pm$ 4.3 & 92.6 $\pm$ 2.4 & 47.1 $\pm$ 1.8 & 56.8 $\pm$ 1.5 \\
ViT & Std & Und & ON & Spa & 81.7 $\pm$ 3.2 & 95.4 $\pm$ 0.4 & 63.8 $\pm$ 0.9 & 92.4 $\pm$ 1.0 & 48.6 $\pm$ 2.1 & 58.4 $\pm$ 1.1 \\
ViT & Std & Over & ON & Alp & 84.8 $\pm$ 0.6 & 96.0 $\pm$ 0.2 & 64.7 $\pm$ 0.1 & 92.9 $\pm$ 1.4 & 46.8 $\pm$ 3.2 & 57.9 $\pm$ 1.2 \\
ViT & Std & Over & ON & Spa & 80.3 $\pm$ 1.5 & 95.6 $\pm$ 0.4 & 67.0 $\pm$ 0.7 & 92.8 $\pm$ 2.6 & 48.0 $\pm$ 0.8 & 58.3 $\pm$ 0.3 \\
ViT & Std & Und & OFF & Alp & 64.4 $\pm$ 2.5 & 97.0 $\pm$ 0.9 & 77.2 $\pm$ 2.6 & 93.4 $\pm$ 1.1 & 49.5 $\pm$ 2.6 & 58.9 $\pm$ 2.5 \\
ViT & Std & Und & OFF & Spa & 62.8 $\pm$ 4.8 & 96.5 $\pm$ 1.1 & 78.2 $\pm$ 2.0 & 93.8 $\pm$ 2.0 & 50.0 $\pm$ 1.0 & 60.9 $\pm$ 1.0 \\
ViT & Std & Over & OFF & Alp & 62.5 $\pm$ 4.1 & 97.0 $\pm$ 0.6 & 76.9 $\pm$ 1.3 & 94.0 $\pm$ 1.1 & 47.5 $\pm$ 2.1 & 60.2 $\pm$ 1.1 \\
ViT & Std & Over & OFF & Spa & 57.8 $\pm$ 2.2 & 97.7 $\pm$ 0.2 & 78.0 $\pm$ 1.4 & 94.4 $\pm$ 1.2 & 46.6 $\pm$ 0.8 & 61.4 $\pm$ 1.3 \\
\hline
\end{tabular}
\caption{Supplementary Table A1. ERPVR dataset F1-score from base LOSO and calibration phases averaged over random seeds and subject group (mean $\pm (\max - \min)/2$). Results organized by configuration $\times$ subject group. M is “Model” for model used: ``EEG” for EEGNet, ``Con” for EEG Conformer, and ``ViT” for the pretrained EEG ViT. D is ``Data type” for Standards or Oddballs. S is ``Sampling” for Undersampling or Oversampling. O is ``Outlier detection” turned on or off. E is ``Electrode order” for alphabetical or spatial. DB is ``Discomfort base”, DC is ``Discomfort calibration”, NB is ``No Discomfort base”, NC is ``No Discomfort calibration”, MB is ``Mixed base”, and MC is ``Mixed calibration.”}
\label{tab:bygroup-all-models-f1-base-calib-mean-half-range-over-seeds-by-configuration-subject-group}
\end{table}






\section{Additional Analyses}\label{sec:supplemental-additional-analyses}

\begin{figure*}
\begin{subfigure}{\textwidth}
    \centering
    \includegraphics[width=\linewidth]{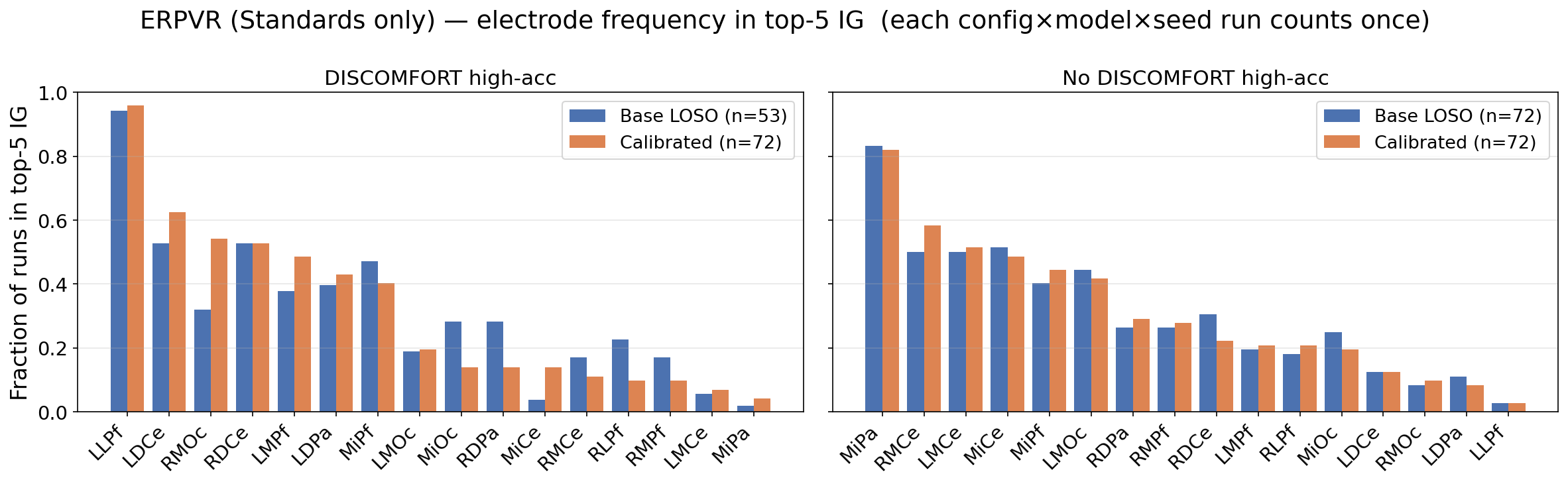}
    \caption{ERPVR fraction of runs using Standards where electrode was in the top-5 integrated gradients for high-accuracy subjects.}
    \label{fig:erpvr_freq_top5_stan}
\end{subfigure}
\begin{subfigure}{\textwidth}
    \centering
    \includegraphics[width=\linewidth]{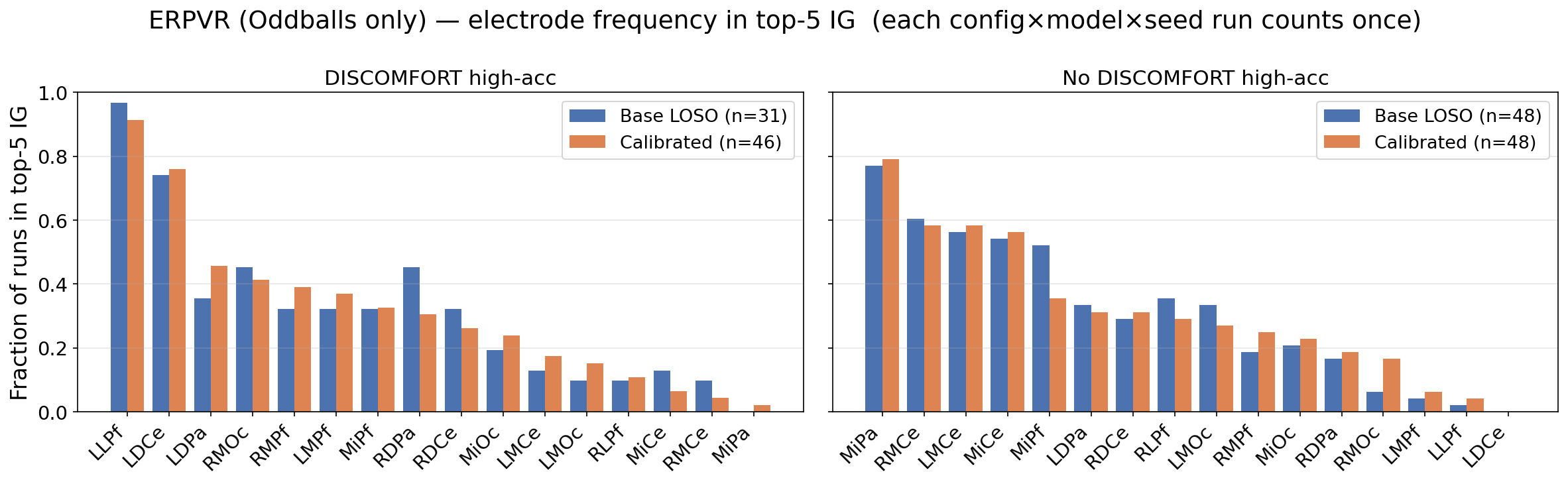}
    \caption{ERPVR fraction of runs using Oddballs where electrode was in the top-5 integrated gradients for high-accuracy subjects.}
    \label{fig:erpvr_freq_top5_odd}
\end{subfigure}
    \caption{Supplementary Figure B2. ERPVR fraction of runs where electrode was in the top-5 integrated gradients for high-accuracy subjects. n is number of experiments where the setting had subjects classified with high accuracy, discomfort subjects on the left and no discomfort right.}
    \label{fig:erpvr_freq_top5}
\end{figure*}

In addition to the normalized IG averaged across ERPVR runs, we also plot the frequency counts of when each electrode was top-5 most useful for an experiment run in Figure \ref{fig:erpvr_freq_top5}. 

\begin{figure*}
    \begin{subfigure}{0.5\textwidth}
        \centering
        \includegraphics[width=\linewidth]{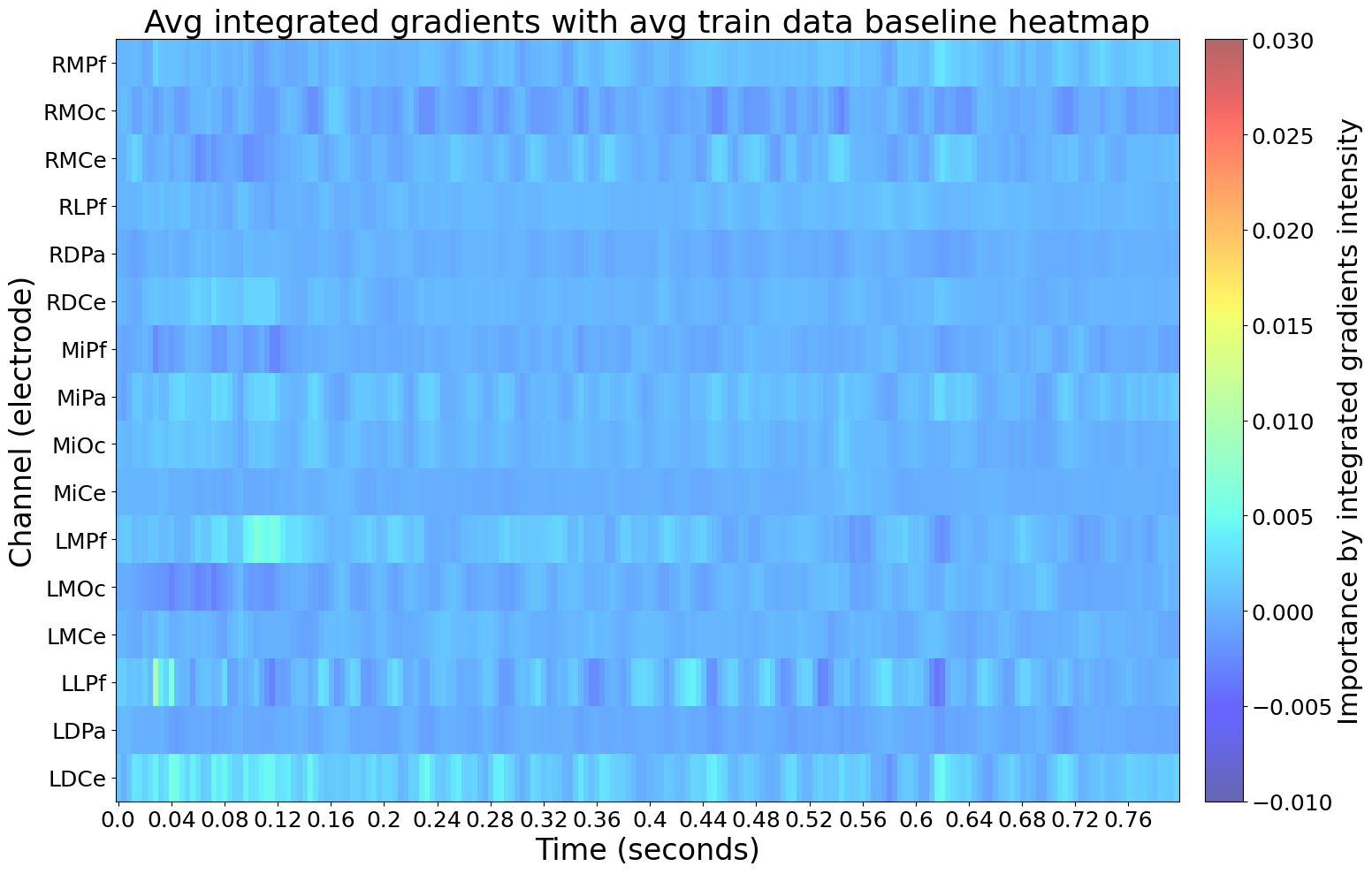}
        \caption{ERPVR Subject 15's IGs averaged over sick data from experiment run when balanced accuracy $ > $ 75\%.}
        \label{fig:erpvr-mixed-sub-15}
    \end{subfigure}
    \begin{subfigure}{0.5\textwidth}
        \centering
        \includegraphics[width=\linewidth]{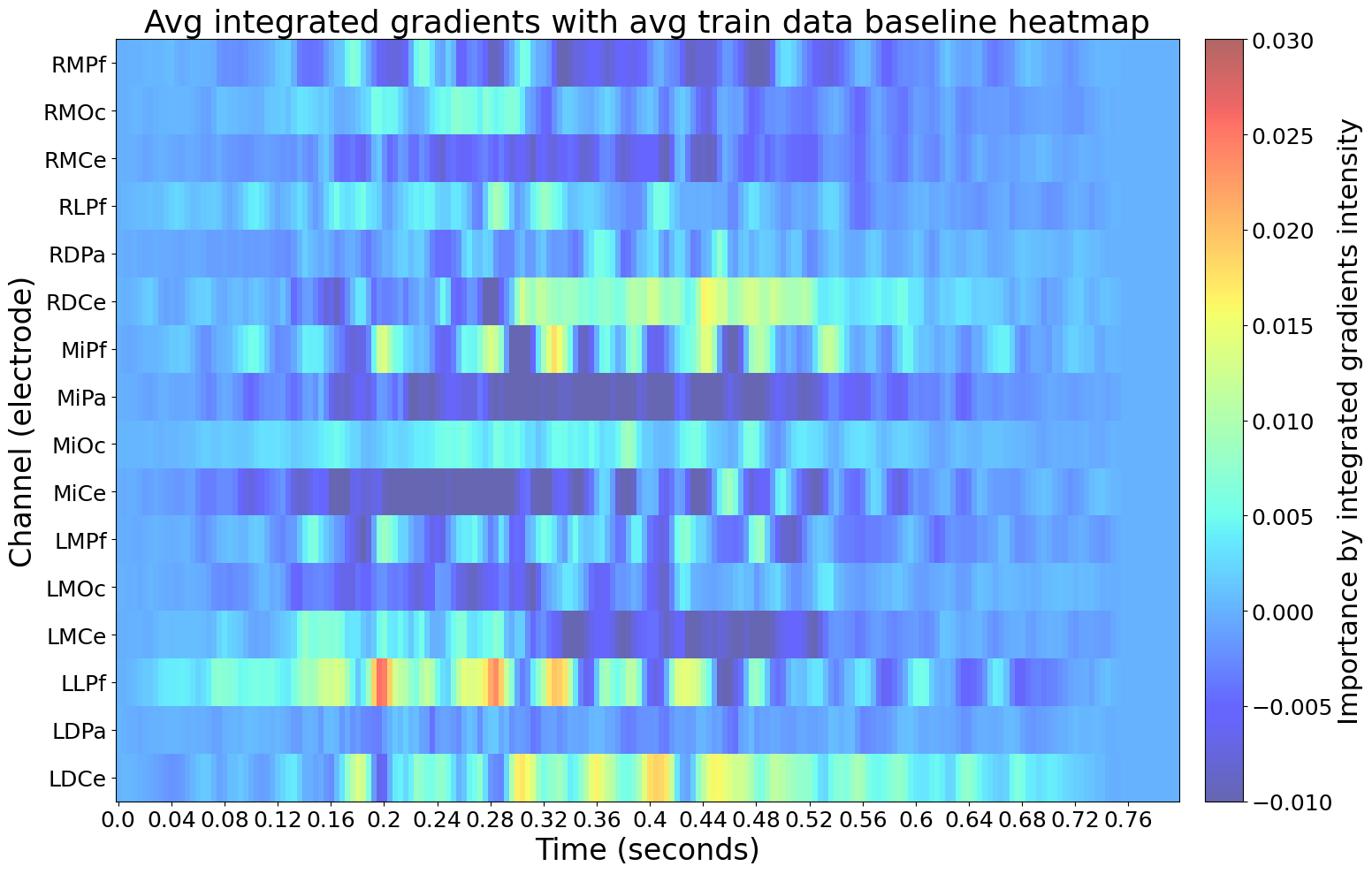}
        \caption{ERPVR Subject 10's IGs averaged over sick data from experiment run when balanced accuracy $ > $ 75\%.}
        \label{fig:erpvr-mixed-sub-10}
    \end{subfigure}
    \caption{Supplementary Figure B1. ERPVR mixed label subjects that reached high balanced accuracy.}
    \label{fig:erpvr-mixed-label-high-acc}
\end{figure*}

\begin{figure*}
    \centering
    \includegraphics[width=\linewidth]{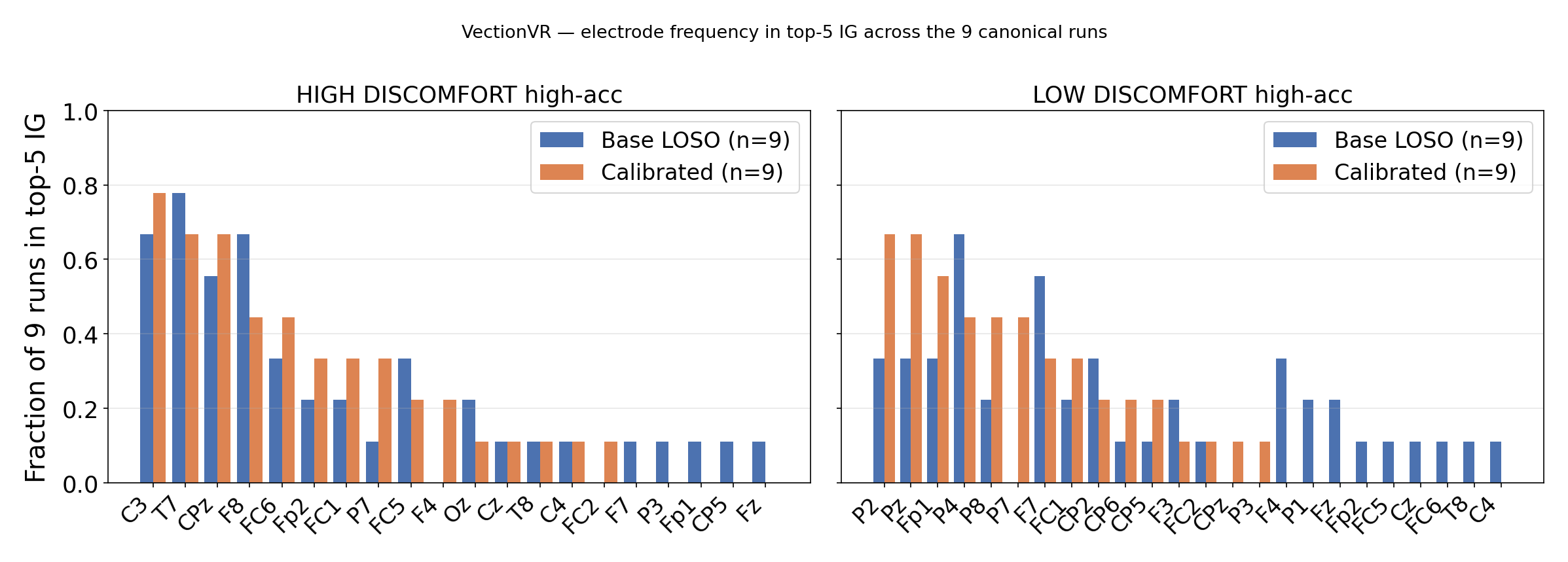}
    \caption{Supplementary Figure B3. VectionVR fraction of runs using Standards where electrode was in the top-5 integrated gradients for high-accuracy subjects. n is number of experiments where the setting had subjects classified with high accuracy, discomfort subjects on the left and no discomfort on the right.}
    \label{fig:electrode_frequency_top5_vectionvr}
\end{figure*}

\begin{figure*}[t]
    \centering
    \vspace{-0.1cm}
  \begin{subfigure}{0.5\textwidth}
    \centering
    \includegraphics[width=\linewidth]{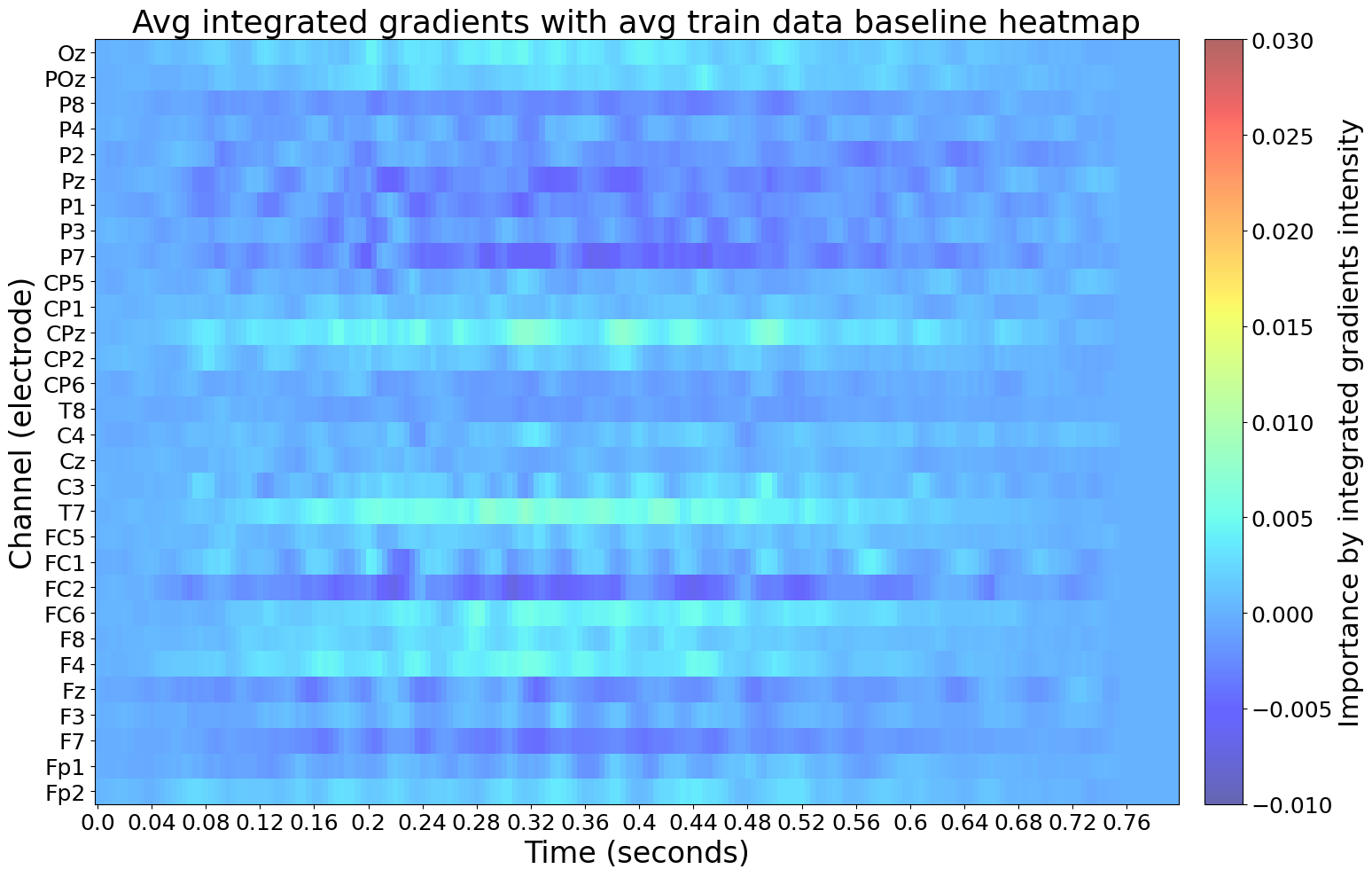} \vspace{-0.55cm}
    \caption{EEG Conformer calibrated IGs}
    \label{fig:vectionvr-conformer-rand-seed-3-calibrated}
  \end{subfigure}\hfill
  \begin{subfigure}{0.5\textwidth}
    \centering
    \vspace{-0.1cm}
    \includegraphics[width=\linewidth]{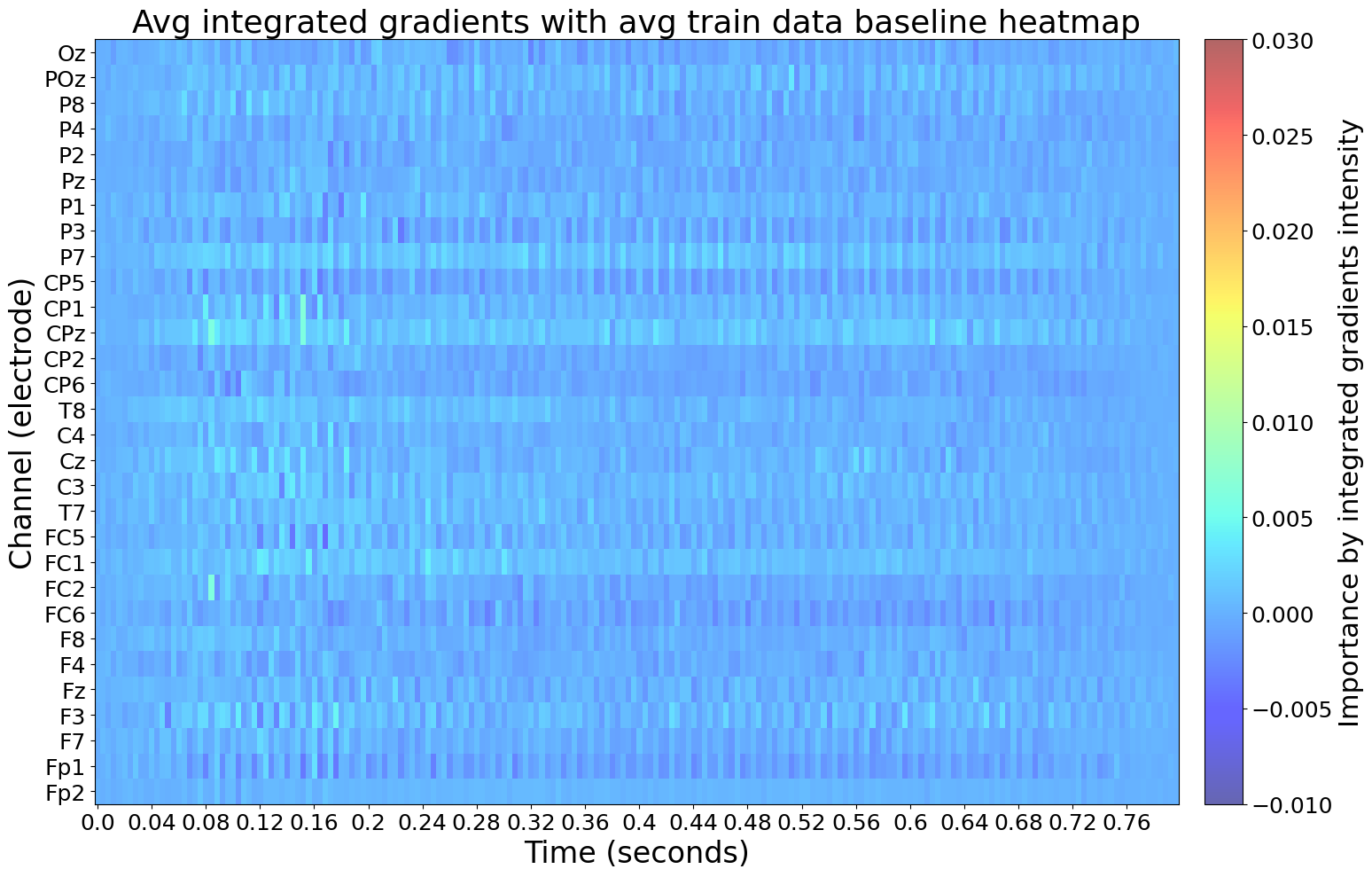} \vspace{-0.55cm}
    \caption{Pre-trained EEG ViT calibrated IGs}
    \label{fig:vectionvr-pretrained-vit-rand-seed-3-calibrated}
  \end{subfigure}

  \vspace{0.7em}

  \begin{subfigure}{0.5\textwidth}
    \centering
    \vspace{-0.1cm}
    \includegraphics[width=\linewidth]{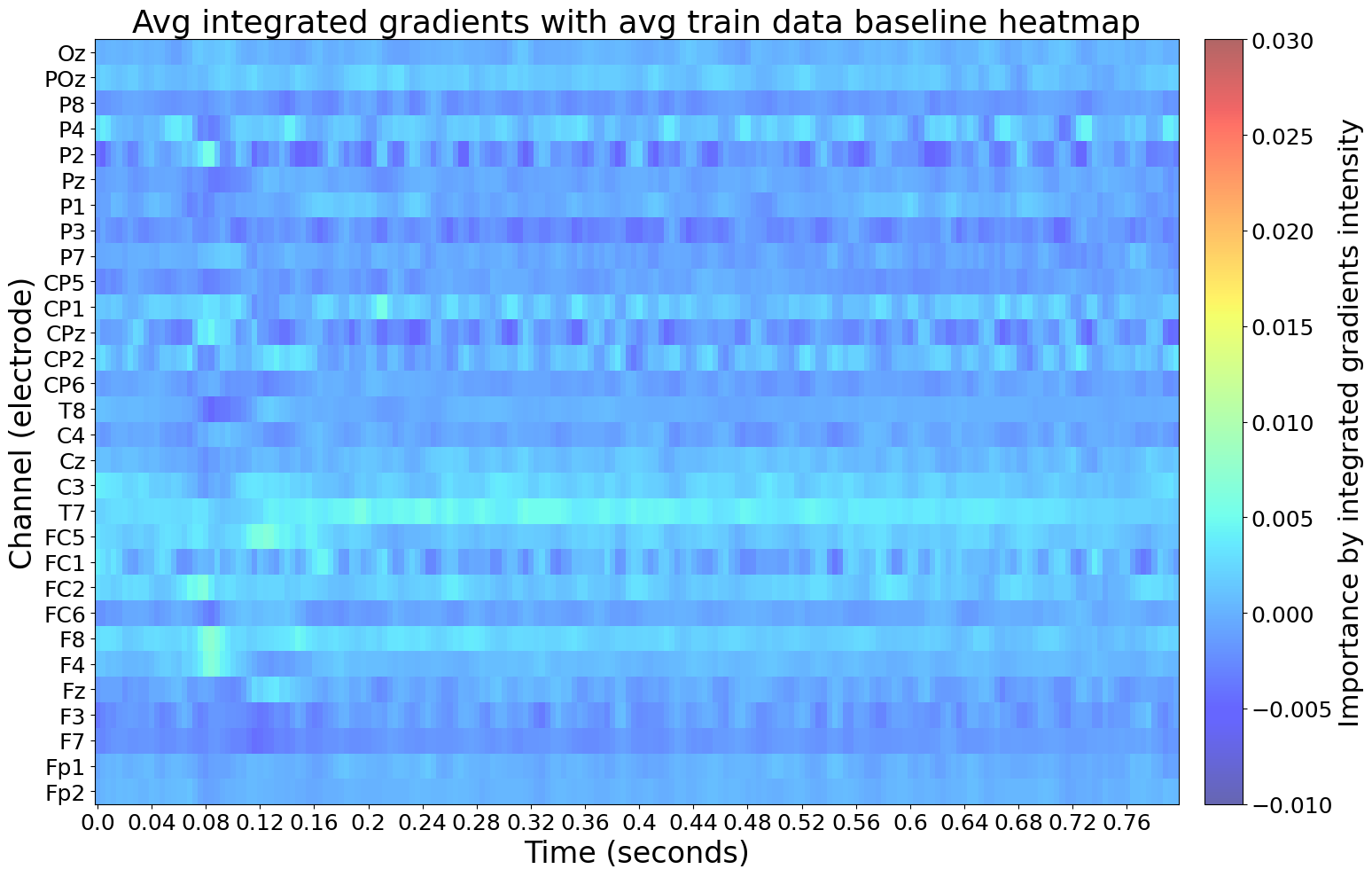} \vspace{-0.55cm}
    
    \caption{EEGNet calibrated IGs}
    \label{fig:vectionvr-eegnet-rand-seed-3-calibrated-ig}
  \end{subfigure}\hfill
  \begin{subfigure}{0.5\textwidth}
    \centering
    \vspace{-0.1cm}
    \includegraphics[width=\linewidth]{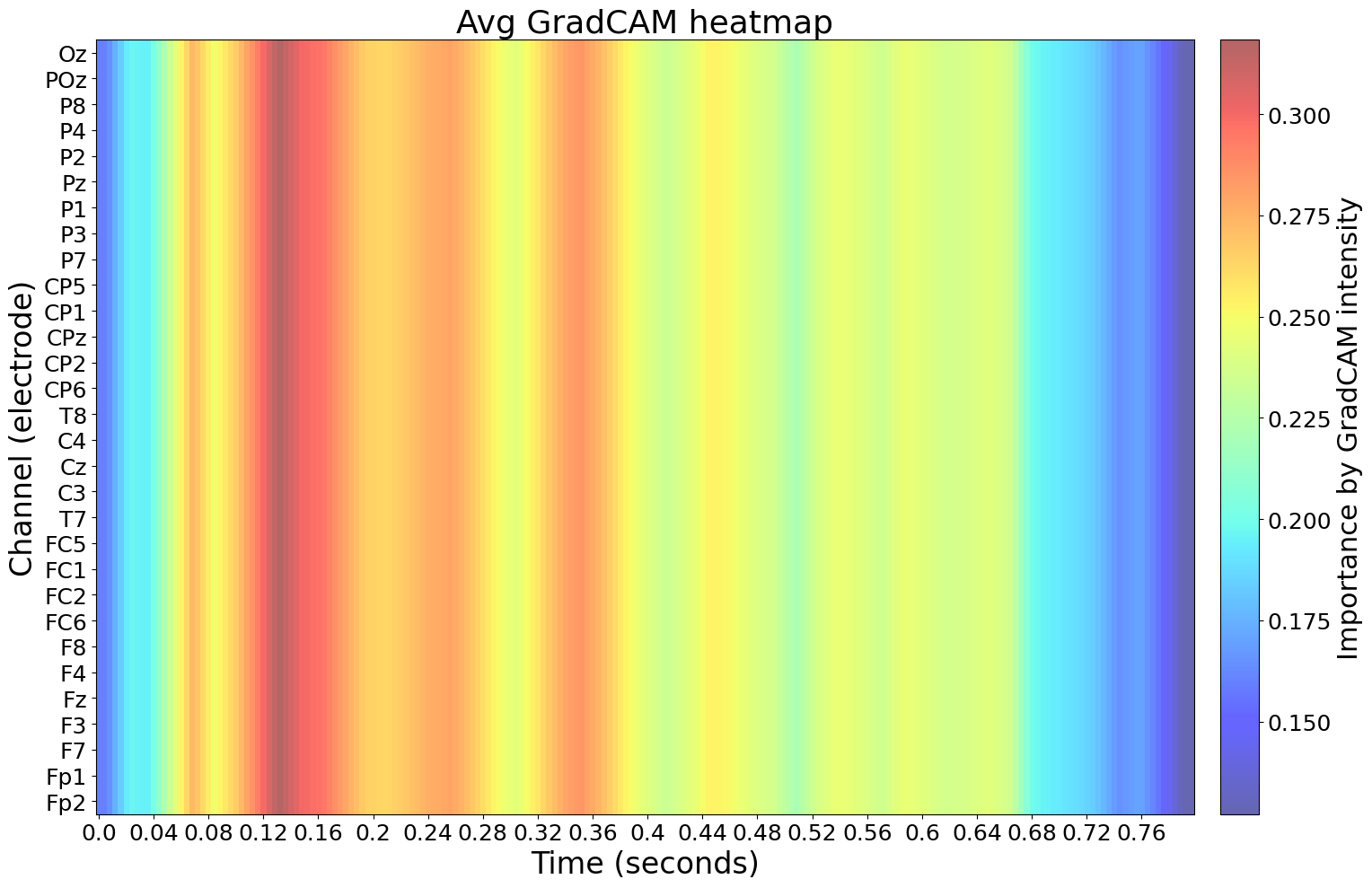} \vspace{-0.55cm}
    \caption{EEGNet calibrated GradCAM}
    \label{fig:vectionvr-eegnet-rand-seed-3-calibrated-gradcam}
  \end{subfigure}
    \vspace{-0.6cm} \caption{Supplementary Figure B4. VectionVR Integrated Gradients (IGs) averaged over discomfort data from subjects with balanced accuracy~$>75\%$ (from seed 3). Warmer colors indicate more positive attribution.}
    \label{fig:vectionvr-heatmaps-rand-seed-3}
    \vspace{-0.55cm}
\end{figure*}

For VectionVR, this is in Figure \ref{fig:electrode_frequency_top5_vectionvr}. We also include the integrated gradient maps of ERPVR mixed label subjects that had balanced accuracy $>$ 75\% in Figure \ref{fig:erpvr-mixed-label-high-acc}. Additional integrated gradient plots for VectionVR run with seed 3 in Figure \ref{fig:vectionvr-pretrained-vit-rand-seed-3-calibrated}.

\begin{figure*}[t]
    \centering
    \vspace{-0.1cm}
  \begin{subfigure}{0.5\textwidth}
    \centering
    \includegraphics[width=\linewidth]{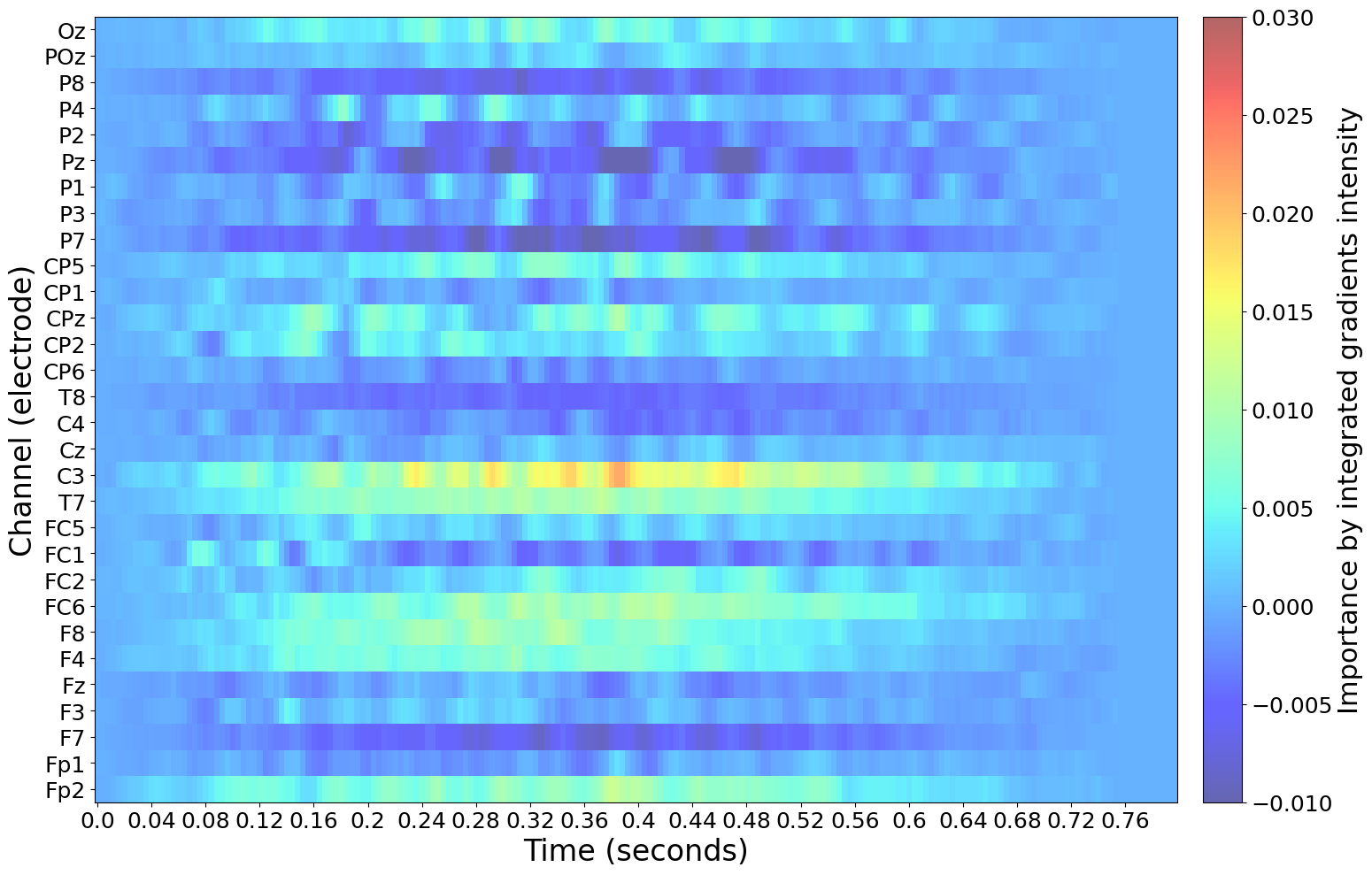} \vspace{-0.55cm}
    \caption{EEG Conformer calibrated IGs}
    \label{fig:vectionvr-conformer-rand-seed-2-calibrated}
  \end{subfigure}\hfill
  \begin{subfigure}{0.5\textwidth}
    \centering
    \vspace{-0.1cm}
    \includegraphics[width=\linewidth]{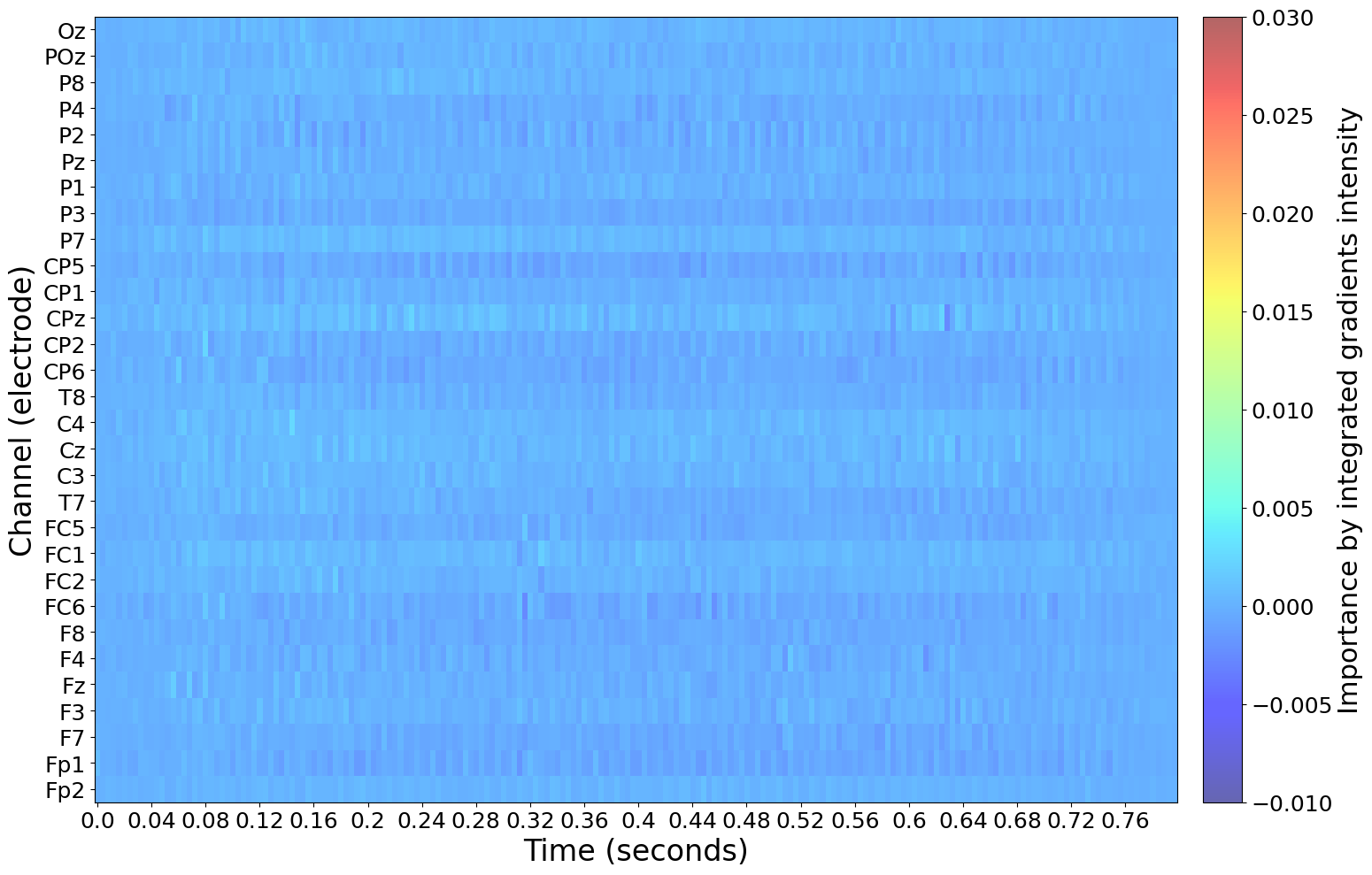} \vspace{-0.55cm}
    \caption{Pre-trained EEG ViT calibrated IGs}
    \label{fig:vectionvr-pretrained-vit-rand-seed-2-calibrated}
  \end{subfigure}

  \vspace{0.7em}

  \begin{subfigure}{0.5\textwidth}
    \centering
    \vspace{-0.1cm}
    \includegraphics[width=\linewidth]{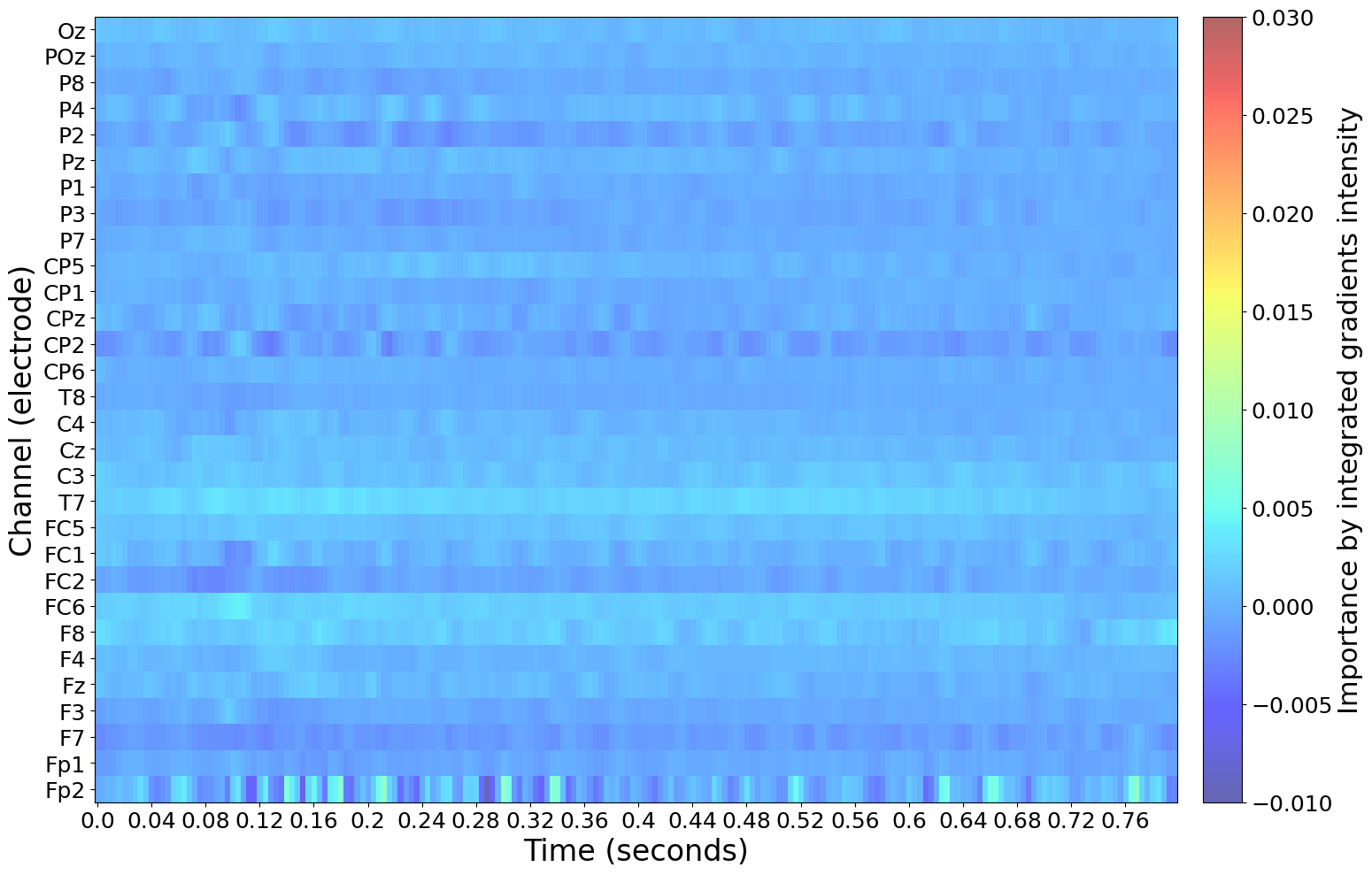} \vspace{-0.55cm}
    
    \caption{EEGNet calibrated IGs}
    \label{fig:vectionvr-eegnet-rand-seed-2-calibrated-ig}
  \end{subfigure}\hfill
  \begin{subfigure}{0.5\textwidth}
    \centering
    \vspace{-0.1cm}
    \includegraphics[width=\linewidth]{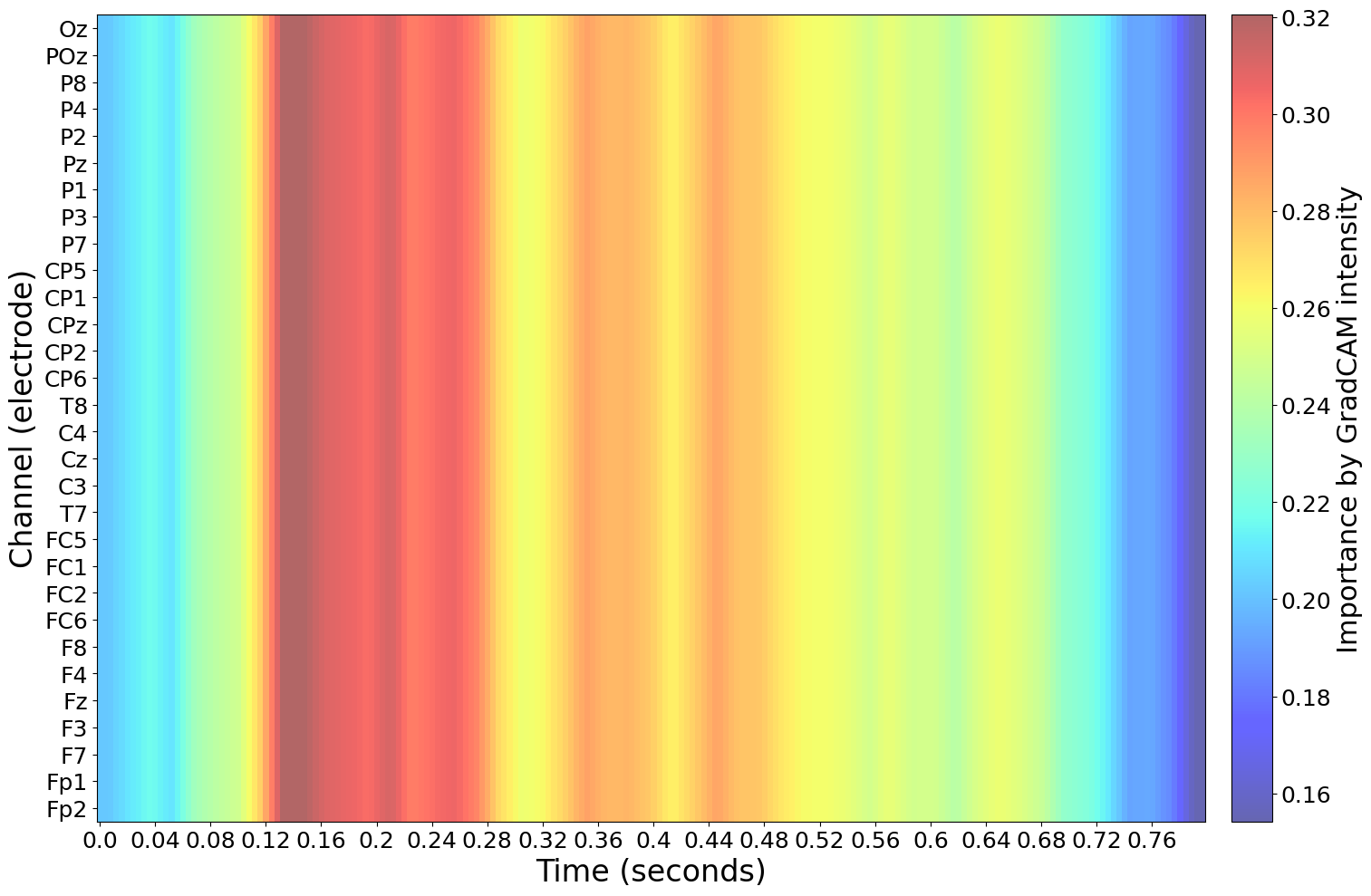} \vspace{-0.55cm}
    \caption{EEGNet calibrated GradCAM}
    \label{fig:vectionvr-eegnet-rand-seed-2-calibrated-gradcam}
  \end{subfigure}
    \vspace{-0.6cm} \caption{Supplementary Figure B5. VectionVR Integrated Gradients (IGs) averaged over discomfort data from subjects with balanced accuracy~$>75\%$ (from seed 2). Warmer colors indicate more positive attribution.}
    \label{fig:vectionvr-heatmaps-rand-seed-2}
    \vspace{-0.55cm}
\end{figure*}

\subsection{ERPVR Statistical Tests and Analyses}\label{sec:supplemental-erpvr-stats-tests}

We report Wilcoxon signed-rank tests in Table \ref{tab:stats-all-subjects-combined-n-29-transformation-effects} over all subjects in ERPVR. We also report Wilcoxon signed-rank tests in Table \ref{tab:grpstats-paired-t-test-and-wilcoxon-signed-rank-tests-by-effect-and-subject-group} by subject group subsets for ERPVR runs: subjects with all discomfort data, subjects with mixed labels, and subjects with all no discomfort data. They are treated as three separate families of five comparisons. Since we run $N=5$ tests, the Bonferroni Correction is $\alpha = \frac{0.05}{5} = 0.01$ for statistical significance.

For the subject group split of the Wilcoxon tests in Table \ref{tab:grpstats-paired-t-test-and-wilcoxon-signed-rank-tests-by-effect-and-subject-group}, for per-subject calibration compared to leave-one-subject-out base, we see that for all three subject groups, calibration led to higher balanced accuracies. 
For outlier detection, we see that for all discomfort subjects, having outlier detection on is helpful for increasing balanced accuracy. However, for all no discomfort subjects, it's in the opposite direction where having outlier detection decreased balanced accuracy. For sampling method, it seems over vs. undersampling had no statistical significance for all three subject groups. For data type of Oddballs vs. Standards, we do see for no discomfort subjects, there is statistical significance for using Standards instead of Oddballs for classification, with Standards leading to increased balanced accuracy. And for electrode channel order, it seems there is no statistical significance for classification.

\begin{table}[htbp]
\centering
\begin{tabular}{lllll}
\hline
Transformation & n & Mean $\Delta$ (\%) & Median $\Delta$ (\%) & Wilcoxon \\
\hline
Outlier detection (OFF - ON) & 29 & +1.74 & +2.84 & W=159, p=0.213 \\
Sampling (over - under) & 29 & -0.12 & -0.17 & W=190, p=0.565 \\
Data type (Oddballs - Standards) & 29 & -2.02 & -2.69 & W=102, p=0.011 \\
Channel order (spatial - alpha) & 29 & -0.22 & -0.29 & W=174, p=0.358 \\
Calibration (calib - base LOSO) & 29 & +9.64 & +3.71 & W=54, p=0.000174 \\ 
\hline
\end{tabular}
\caption{Supplementary Table B2. Wilcoxon signed-rank test on balanced accuracy from ERPVR runs on all subjects (n=29). Each run in the paired test comparison aggregates per-subject balanced accuracy differences $\Delta$, one averaged value per subject over its matched pairs. n is the number of contributing subjects.}
\label{tab:stats-all-subjects-combined-n-29-transformation-effects}
\end{table}

\begin{table}[htbp]
\centering
\begin{tabular}{lllllll}
\hline
Transformation & Pairs & Group & n & Mean $\Delta$ (\%) & Med $\Delta$ (\%) & Wilcoxon\\
\hline
Calibration (Calib - base LOSO) & 120 & AD & 6 & +25.10 & +18.06 & W=0, p=0.031 \\
Calibration (Calib - base LOSO) & 120 & mixed & 8 & +1.67 & +1.72 & W=3, p=0.039 \\
Calibration (Calib - base LOSO) & 120 & ND & 15 & +7.71 & +3.71 & W=26, p=0.055 \\
\hline
Outlier detection (OFF - ON) & 60 & AD & 6 & -23.24 & -22.45 & W=0, p=0.031 \\
Outlier detection (OFF - ON) & 60 & mixed & 8 & +0.56 & +0.15 & W=15, p=0.742 \\
Outlier detection (OFF - ON) & 60 & ND & 15 & +12.35 & +10.85 & W=0, p=6.10e-05 \\
\hline
Sampling (over - under) & 60 & AD & 6 & -0.12 & -0.95 & W=6, p=0.438 \\
Sampling (over - under) & 60 & mixed & 8 & -0.73 & -0.78 & W=14, p=0.641 \\
Sampling (over - under) & 60 & ND & 15 & +0.20 & +0.22 & W=54, p=0.762 \\
\hline
Data type (Oddballs - Standards) & 48 & AD & 6 & +1.12 & +0.95 & W=7, p=0.562 \\
Data type (Oddballs - Standards) & 48 & mixed & 8 & -0.15 & +0.55 & W=15, p=0.742 \\
Data type (Oddballs - Standards) & 48 & ND & 15 & -4.28 & -4.33 & W=6, p=8.54e-04 \\
\hline
Channel order (spatial - alpha) & 60 & AD & 6 & -0.52 & -0.08 & W=7, p=0.562 \\
Channel order (spatial - alpha) & 60 & mixed & 8 & +0.48 & +0.76 & W=13, p=0.547 \\
Channel order (spatial - alpha)& 60 & ND & 15 & -0.48 & -0.82 & W=36, p=0.188 \\
\hline
\end{tabular}
\caption{Supplementary Table B3. Wilcoxon signed-rank tests by effect by configuration and subject group from all ERPVR runs over balanced accuracy. AD means ``all discomfort" and ND means ``no discomfort".}
\label{tab:grpstats-paired-t-test-and-wilcoxon-signed-rank-tests-by-effect-and-subject-group}
\end{table}

\section{Additional details on data analysis and augmentation}\label{sec:supplemental_data_aug}

\subsection{Individual trial-level outlier detection}\label{sec:supplemental_outlier_detection}

With the strict outlier detection and removal method outlined in section 7.3, 
approximately one-third of the nonsick trials and one-third of the sick trials were identified as outliers in ERPVR. Exact numbers varied across random seeds depending on which subjects are left out as validation and test. An example breakdown of the outlier data removal by subject is shown for nonsick in Supplementary Figure \ref{fig:nonsick-outlier-removal} and for sick in Supplementary Figure \ref{fig:sick-outlier-removal}. In this case, approximately one-third of nonsick and sick data were removed from a total of 6,108 nonsick data trials and 2,808 sick data trials.

\begin{figure}[h]
    \centering
    \includegraphics[width=0.7\linewidth]{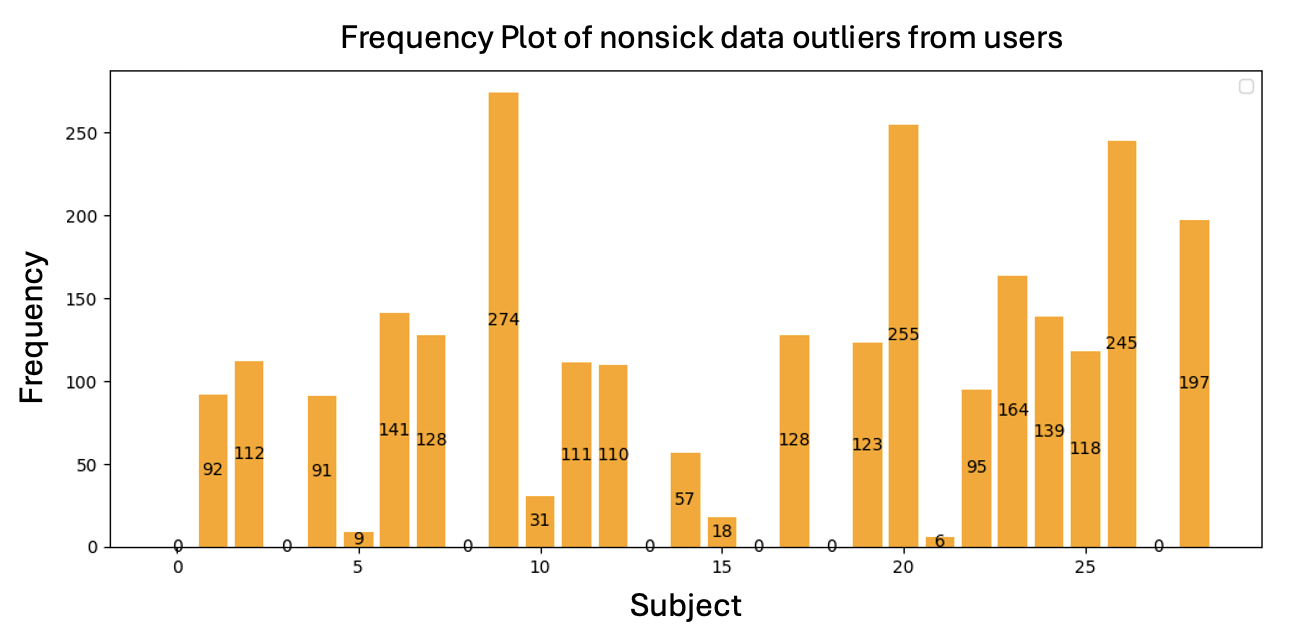}
    \caption{Supplementary Figure C6. Example trial of nonsick data outlier removal frequency plot by subject with subject 0 left out as test and subject 27 as validation}
    \label{fig:nonsick-outlier-removal}
\end{figure}

\begin{figure}[h]
    \centering
    \includegraphics[width=0.7\linewidth]{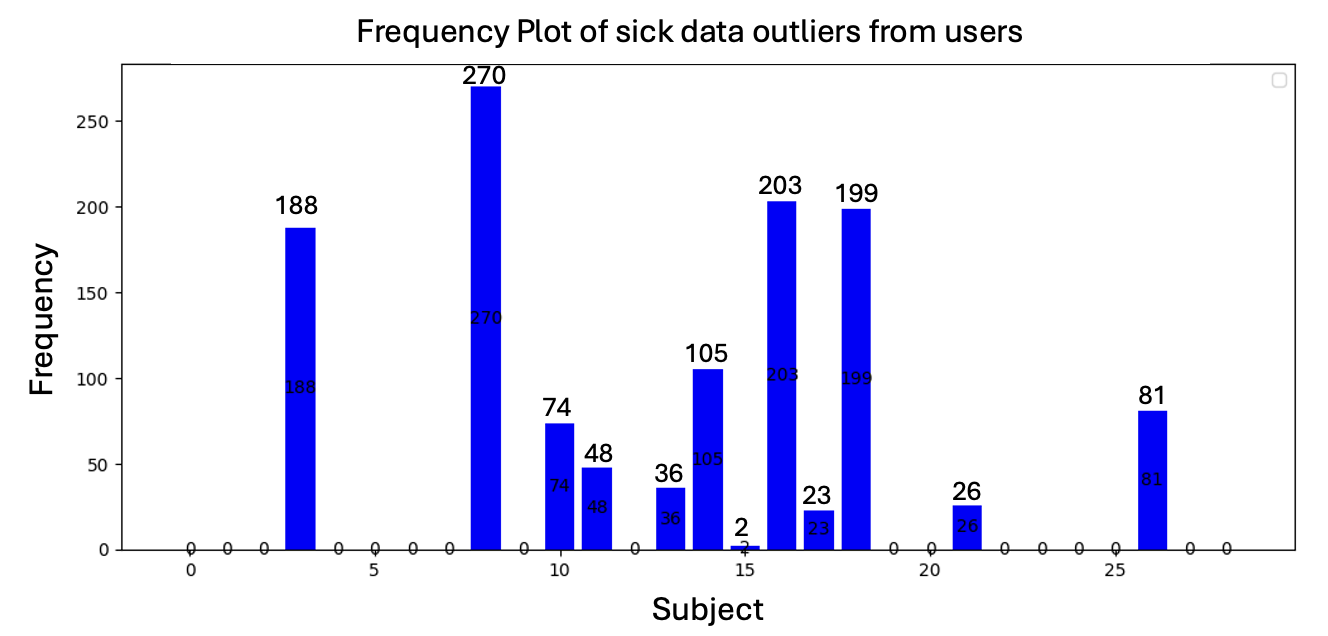}
    \caption{Supplementary Figure C7. Example trial of sick data outlier removal frequency plot by subject with subject 0 left out as test and subject 27 as validation}
    \label{fig:sick-outlier-removal}
\end{figure}

\subsection{Subject-level outlier detection}\label{sec:supplemental_subj_outlier}

\begin{figure}[h]
    \centering   \includegraphics[width=\linewidth]{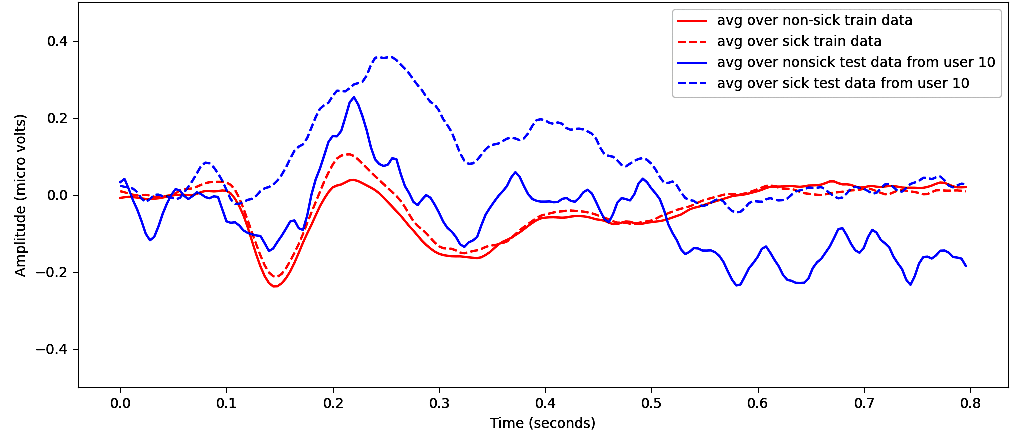}
    \vspace{-0.3cm}
    \caption{Supplementary Figure C8. Leaving out Subject 10 from the remaining subjects.}
    \label{fig:loso-sub-11}
    \vspace{-0.1cm}
\end{figure}

\begin{figure}[h]
    \centering
    \includegraphics[width=\linewidth]{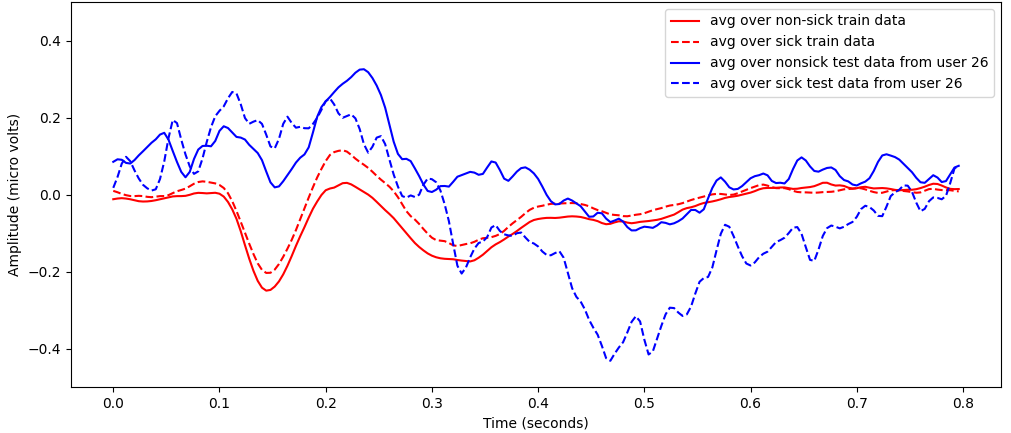}
    \caption{Supplementary Figure C9. Leaving out Subject 26 from the remaining subjects}
    \label{fig:loso-sub-26}
\end{figure}


Since there was high inter-subject variability in EEG data, we visualized how different each subjects' data was compared to others out of the 29 subjects in the ERPVR study. 
For subject outlier detection, we examined the plots from data analysis. Plots were made by averaging across the EEG data between sick and non-sick labels and between the 28 subjects vs. the one subject left out, providing a qualitative view of the data (see Figure \ref{fig:loso-sub-11}). For subjects whose data visually looked significantly different from the average of the remaining 28 subjects, we checked the notes from the cybersickness study for a possible explanation. For example, after observing the variation when Subject 10 was left out (Figure \ref{fig:loso-sub-11}), we discovered from the notes that they felt extremely sick to the point of nearly throwing up, whereas the other participants did not feel such an intense level of sickness. Another method used was observing if there were subjects that all models performed poorly on. For example, all models in this work did not evaluate well on Subject 26 prior to calibration (Figure \ref{fig:loso-sub-26}). As mentioned in the study notes, Subject 26 had very noisy data from jaw clench muscle contractions. 
Although there were individuals that deviated quite a bit from the average of the remaining subjects, we did not remove any from the data because these distribution shifts were representative of how different people can and will have EEG variability.


\subsection{Synthetic data for class imbalance}\label{sec:supplemental_smote}
A common method for generating synthetic data is Synthetic Minority Over-sampling Technique (SMOTE) \cite{Chawla_2002}. We explored several variants \cite{smote-comparison} using the smote-variants library \cite{smote-variants}. 
Since SMOTE uses kNN to interpolate synthetic data, it does not work well for high dimensional and noisy data like EEG. In earlier test runs, we observed that neural networks like EEGNet performed better with random oversampling. Thus, SMOTE was not used for the deep learning pipeline. However we did find 
Safe-level SMOTE, which generates synthetic data by using ``safe" samples identified by kNN, worked well for our data after it has been projected to PCA space to reduce dimensions. The principal components, both real and synthetic, were passed into traditional non-neural network models for classification.

\section{ML for Pattern Recognition}\label{sec:supplemental_ML_for_patterns}

We considered both supervised and unsupervised learning methods to extract patterns from the ERPVR data. With supervised learning, both traditional and neural networks were explored. CNN and vision transformers were found to perform the best across the LOSO cross-validations, so we focused on those models for interpretation.

\begin{figure}[t]
    \centering
    \vspace{-0.1cm}
\includegraphics[width=\linewidth]{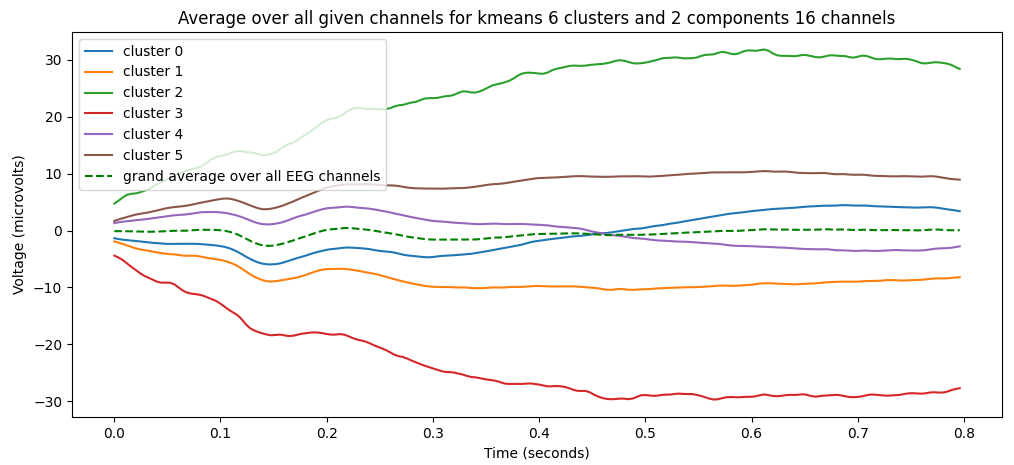}
    \caption{Supplementary Figure D10. K-means centroids of 6 clusters and average over ERPVR data.}
    \label{fig:kmeans-6-clusters}
    \vspace{-0.4cm}
\end{figure}

\begin{figure}
    \centering   \includegraphics[width=\linewidth]{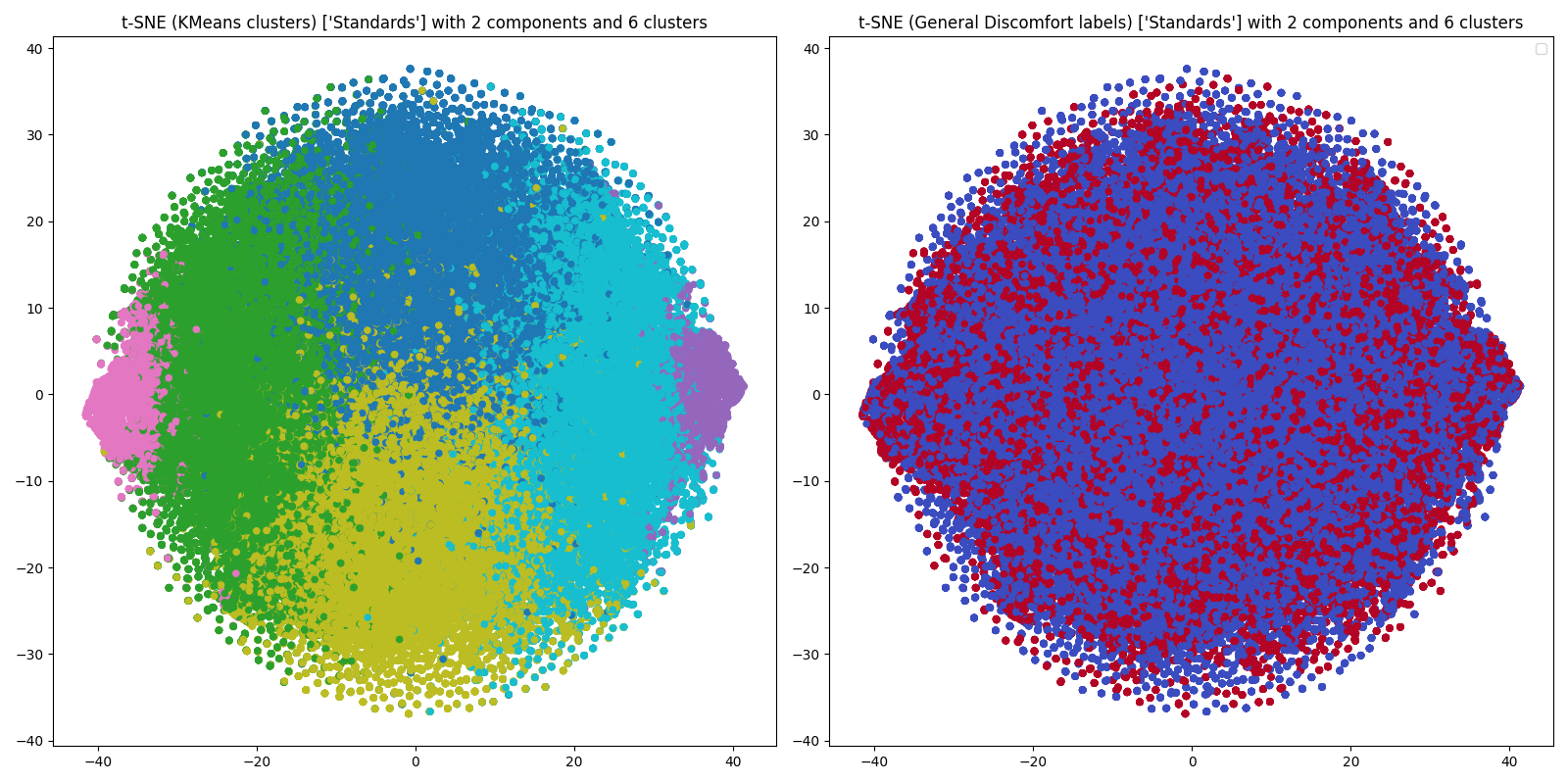}
    \vspace{-0.3cm}
    \caption{Supplementary Figure D11. t-SNE for 2 components of ERPVR data. Left: data points from the same cluster have the same color. Right: data points are colored based on sick and non-sick label.}
    \label{fig:tsne-2}
    \vspace{-0.4cm}
\end{figure}

\subsection{Unsupervised learning}
Clustering was run to see if any patterns could be identified from individual channel EEG without any labels in the ERPVR data. We used k-means clustering, where 6 to 7 clusters were found to be the ideal number based on silhouette score \cite{ROUSSEEUW198753}. 
Silhouette score measures inter and intra-cluster distance for each data sample. Each cluster centroid is plotted in Figure~\ref{fig:kmeans-6-clusters}. 
To visualize the high dimensional data, we employed T-distributed Stochastic Neighbor Embedding (t-SNE) \cite{JMLR:v9:vandermaaten08a} for two components, seen in Figure~\ref{fig:tsne-2}. From the right plot, we can observe that separation between sick and non-sick labeled data is difficult. When decomposing data in each cluster, each subject and electrode had some data in each cluster, indicating that the EEG data were difficult to classify across subjects and electrodes.

\subsection{Supervised learning with traditional ML models}\label{sec:supplemental_trad_ML}

\begin{table*}[h]
    \centering
    \begin{tabular}{| c c c c c c c c c c c |}  
   \hline 
kNN & AdaBoost & RF & LDA & QDA & B NB & G NB & M NB & SVM & GP & Relu MLP\\ [0.1em] 
 \hline 
0.5823 & 0.6510 & 0.5566 & 0.5031 & 0.5628 & 0.5066 & 0.5473 & 0.4785 & 0.5983 & 0.5612 & 0.6129\\ 
 \hline
\end{tabular}
    \caption{Supplementary Table D4. ERPVR average balanced accuracy for traditional ML models across all LOSO cross validations with outlier removal, normalization, PCA, and safe-level SMOTE using the SSQ general discomfort symptom intensity as labels. Results are reported for k=2, 10-tree AdaBoost and random forest (RF) with log loss split criterion, Bernoulli Naive Bayes (B NB), Gaussian Naive Bayes (G NB), Multinomial Naive Bayes (M NB), Singular Value Decomposition LDA, quadratic discriminant analysis (QDA), polynomial kernel SVM, and Radial-Basis Function kernel Gaussian Process (GP).}
    \label{tab:traditional-ml-acc-results}
    \vspace{-0.4cm}
\end{table*}

We tried to find underlying patterns present in the data using the binary sickness labels by comparing traditional machine learning model performance based on their inductive biases. Though not as powerful as neural networks in capturing complex patterns, traditional models have the advantage of requiring less train data and being more interpretable. We evaluated multiple traditional models with LOSO cross validation. The results averaged over all LOSO-folds are shown in Table~\ref{tab:traditional-ml-acc-results}. Scikit-learn \cite{scikit-learn} was used for implementation. The input data was a flattened array of number of electrodes (channels) $C=16$ by time $T$ of size $C*T$. Our pipeline removed outliers and normalized with z-score before running PCA for 48 components and safe-level SMOTE.

\section{Deep Learning Model Descriptions}\label{sec:supplemental-dl-model-descrip}

{\textbf{\underline{EEGNet.}}} EEGNet \cite{Lawhern_2018} is a two block CNN that has both spatial depthwise convolutions across the channels and temporal convolutions along the time dimension. In the first block is a temporal convolution with kernel size $(1, 64)$ and stride 1 along the time dimension, followed by a spatial convolution with kernel size $(C, 1)$ and stride 1 to learn a spatial filter across all electrodes, where $C$ is the number of electrodes. The second block has a separable (depthwise) convolution followed by a pointwise convolution to separate learning how to summarize individual feature maps in time (the depthwise convolution) with how to optimally combine the feature maps (the pointwise convolution) \cite{Lawhern_2018}. Finally, a fully-connected layer outputs classification logits.


{\textbf{\underline{EEG Conformer.}}} EEG Conformer \cite{9991178} is a convolutional transformer with three parts: a convolution block to create feature maps and form tokens, a transformer to model relationships within the token sequence, and a classifier block. The convolution block convolves over the EEG signal similarly to EEGNet's temporal and spatial convolution, with temporal kernel $(1, 25)$ and stride 1, and spatial kernel $(C, 1)$ with stride 1. The learned filters are pooled and rearranged into a sequence of tokens to be passed to the transformer, composed of four encoder blocks. Each encoder block employs multi-head self-attention to learn longer-term temporal features in the token sequence. The self-attention focuses on the relationship between tokens in sequence while multiple heads can represent and capture different kinds of relationships simultaneously. In this work, we use 8 heads and 40 hidden size. The classifier block is composed of three fully-connected layers, outputting classification logits.

{\textbf{\underline{Pre-trained EEG ViT.}}} Vision Transformers extract information from an image by decomposing it into a sequence of patch embeddings and learning relationships between the patches with the transformer blocks. We modified EEG ViT \cite{yang2023vit2eeg} for our task. This model has a patch embedding convolution block, a pre-trained transformer composed of 12 encoder blocks with 12 heads, and a classifier block. It was pre-trained on the ImageNet dataset 
so it learned to extract patterns from images. Since it was specifically trained on $224 \times 224$ size images, decomposed into patches of size $16 \times 16$ and has 196 patch tokens in the patch embedding sequence along with the class token for 197 tokens with 768 hidden size \cite{dosovitskiy2021imageworth16x16words}, modifications were made to match those dimensions given our EEG input data. 
Our modified model for ERPVR starts with a temporal convolution of $(1, 35)$ with stride 3 and creates patch embeddings with kernel size (4, 1) to output 224 tokens of $4 \times 1$ size patches and the class token for 225 total tokens. For VectionVR, the patch embeddings are created with kernel size (6, 1) to output 280 tokens of $6 \times 1$ size patches and the class token for 281 total tokens. The token sequence is passed to the pre-trained ViT backbone and two-layer fully-connected classifier block, outputting classification logits.

\section{Model Interpretability}
\label{sec:supplemental-interpretability}
{\textbf{\underline{Integrated gradients.}}} To visualize which input features the neural network models found helpful (or not) for the model's classification of the target class label, integrated gradients \cite{pmlr-v70-sundararajan17a} were extracted using the Captum library \cite{kokhlikyan2020captumunifiedgenericmodel}. Integrated gradients (IG) is an axiomatic model interpretability algorithm that assigns an importance score to each input feature by approximating the integral of gradients of the model’s output with respect to the inputs along a straight line path from a chosen baseline to the input \cite{kokhlikyan2020captumunifiedgenericmodel} by accumulating gradients along the path. IG satisfies the completeness axiom that states attributions add up to the difference between the output of the model $F$ at the input $x$ and the baseline $x^0$ \cite{pmlr-v70-sundararajan17a}. It provides a signed value for each input feature, positive attribution for helping the model classify to target label and negative for the reverse. In our method, the baseline was the grand average over all train data (close to zero due to normalization).



{\textbf{\underline{GradCAM.}}} Gradient-weighted class activation maps (GradCAM) were extracted with the Pytorch CAM methods library \cite{jacobgilpytorchcam}. GradCAM shows which activations in the feature map had more focus for classification of a target label by taking the gradient with respect to the chosen convolution layer's feature map. 


{\textbf{\underline{Interpretability for each model.}}} For EEGNet, the final convolution layer was set for GradCAM extraction. The heatmaps show where along the time dimension the activation was strongest for the model to classify sickness. This is because EEGNet's final convolution layer is a pointwise convolution that mixes all feature maps together. For EEG Conformer and pre-trained EEG ViT, IG provides more fine-grained information about the gradient strength by channel and time. While GradCAM interpretation very much depends on which layer the maps were extracted from, IG passes through the whole model to compute the cumulative gradients.

\section{Experiment Setup Details}
\label{sec:supplemental_exp_setup}

For the main training phase, we used an Adam optimizer with learning rate and weight decay of $1\times 10^{-4}$. A multi-step learning rate scheduler was used for EEGNet and EEG Conformer, with milestones at 10,000 and 20,000 steps. For the pre-trained EEG ViT, cosine scheduler with a warmup of 1,000 steps was used, as done by the original model \cite{dosovitskiy2021imageworth16x16words}. The batch size was 64. Dropout was 0.25 for both EEGNet and EEG Conformer while pre-trained EEG ViT was 0.1. Cross entropy loss was used (focal loss was tested with the class label imbalance but with random oversampling to balance each batch, cross entropy was found to perform better). For EEGNet and EEG Conformer, the models were trained from scratch. For the pre-trained EEG ViT, fine-tuning was run where the 12 encoder blocks were frozen while the initial convolution and batchnorm, patch embedding, and final classifier layers were unfrozen. For VectionVR runs, pre-trained EEG ViT also had positional embeddings unfrozen. For the transformer models, gradient clipping was applied to stabilize training. Balanced accuracy was the metric used in validation to determine early stopping point. For each stimulus in ERPVR, 800 ms from 16 EEG electrodes (channels) were used as input data to the models, a matrix of shape $C=16$ by $T=200$ time steps. For VectionVR, it is 800 ms from 30 EEG electrodes (channels) used as input data to the models, a matrix of shape $C = 30$ by $T = 200$ time steps.

Next was calibration, where the best validation checkpoint for each model for each subject left out was fine-tuned on the smaller calibration dataset with a constant learning rate of $1\times 10^{-5}$. Cross entropy loss was weighted per sample (eq. 3),
where data from the test subject had double the weight of the data from the others for subjects who only had one label. A custom weighted balanced accuracy (eq. 5)
was used to determine early stopping points.

Some EEGNet and EEG Conformer experiments were run with an Nvidia RTX 4060 GPU, but most were run on an Nvidia RTX 4090. The pre-trained EEG ViT experiments were run on an Nvidia RTX 4090. All are run on Ubuntu Linux computers.

\section{Additional plots}

\begin{figure*}[h]
    \centering
  \begin{subfigure}{0.5\textwidth}
    \centering
    \includegraphics[width=\linewidth]{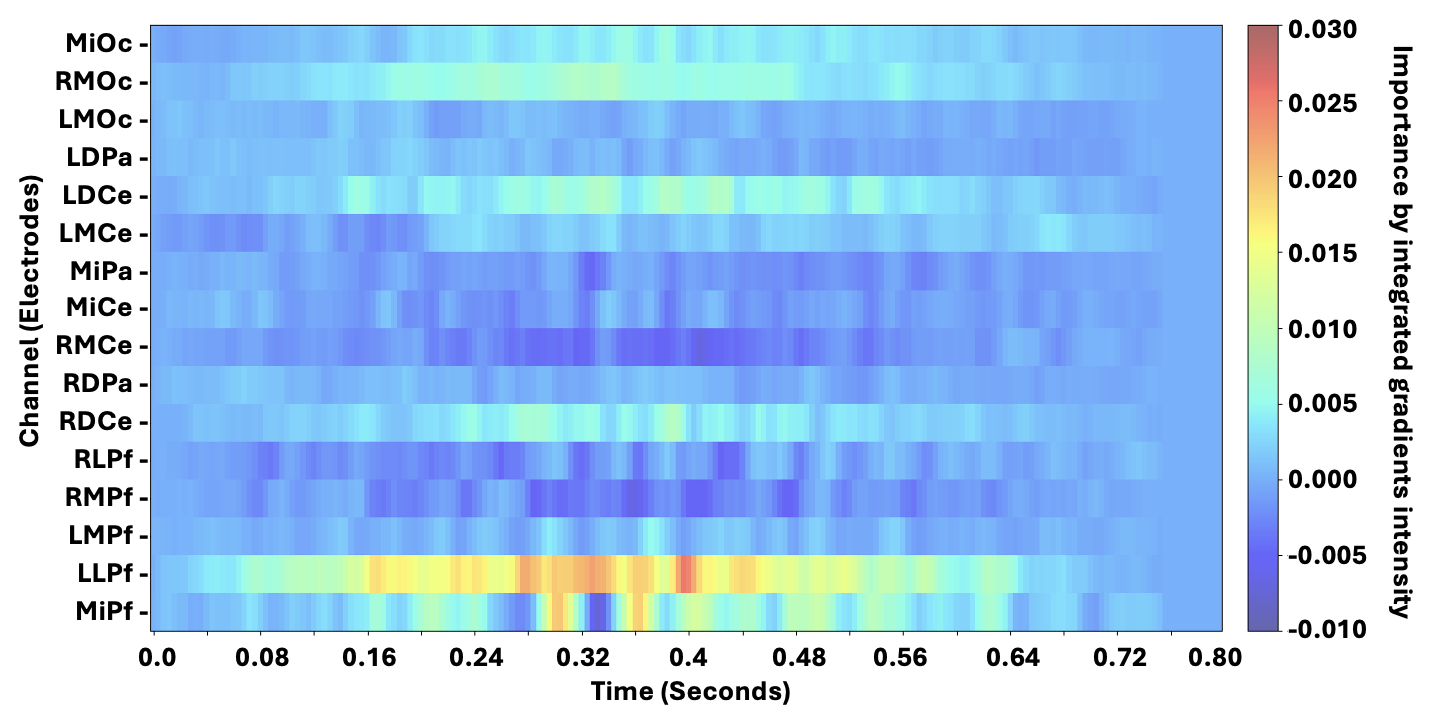} \vspace{-0.4cm}
    \caption{EEG Conformer calibrated integrated gradients}
    \label{fig:snake-order-conformer-rand-seed-4-calibrated}
  \end{subfigure}\hfill
  \begin{subfigure}{0.5\textwidth}
    \centering
    \includegraphics[width=\linewidth]{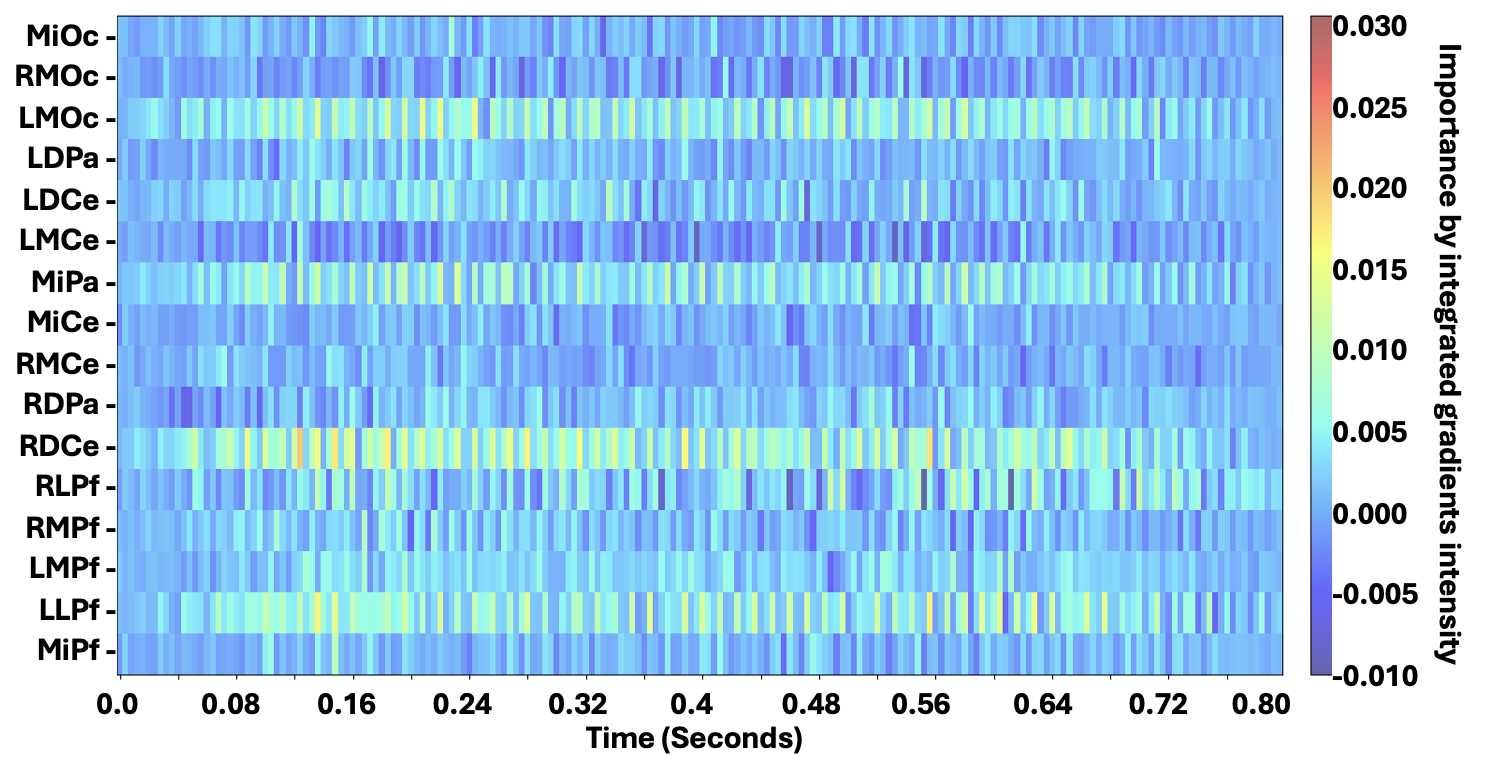} \vspace{-0.4cm}
    \caption{Pre-trained EEG ViT calibrated integrated gradients}
    \label{fig:snake-order-pretrained-vit-rand-seed-4-calibrated}
  \end{subfigure}

  \vspace{0.75em}

  \begin{subfigure}{0.5\textwidth}
    \centering
    \includegraphics[width=\linewidth]{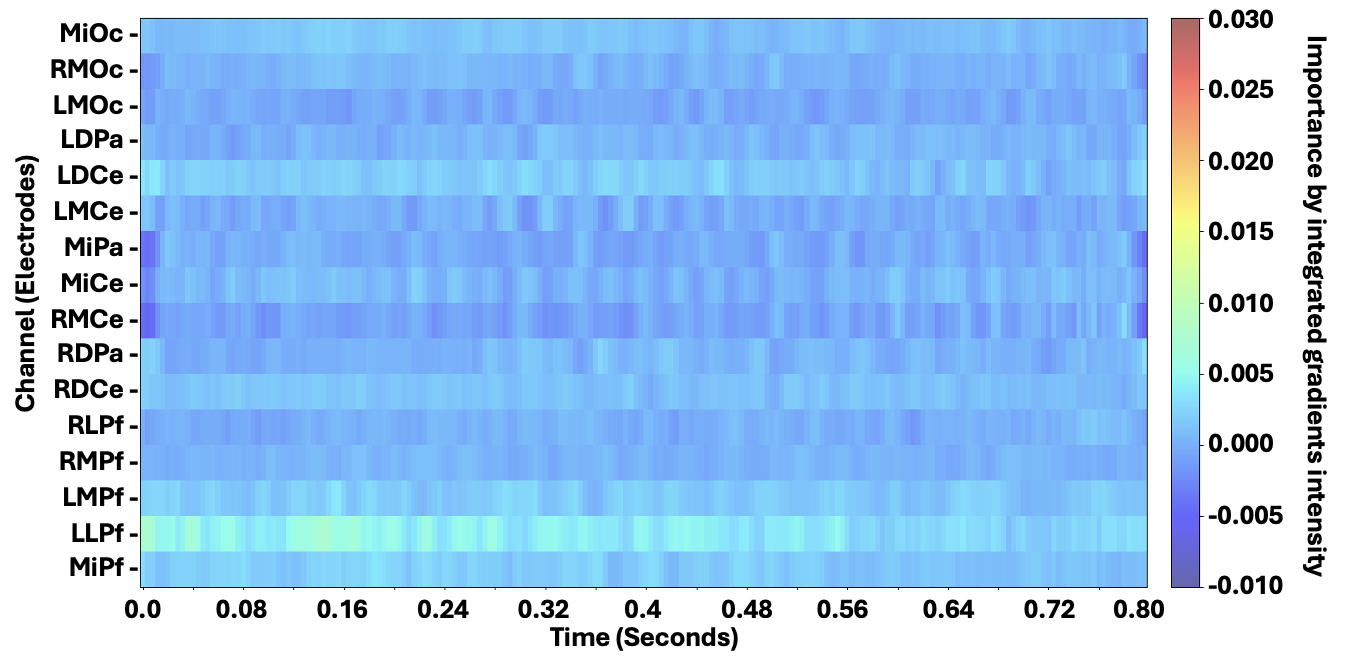} \vspace{-0.4cm}
    \caption{EEGNet calibrated integrated gradients}
    \label{fig:snake-order-eegnet-rand-seed-4-calibrated-ig}
  \end{subfigure}\hfill
  \begin{subfigure}{0.5\textwidth}
    \centering
    \includegraphics[width=\linewidth]{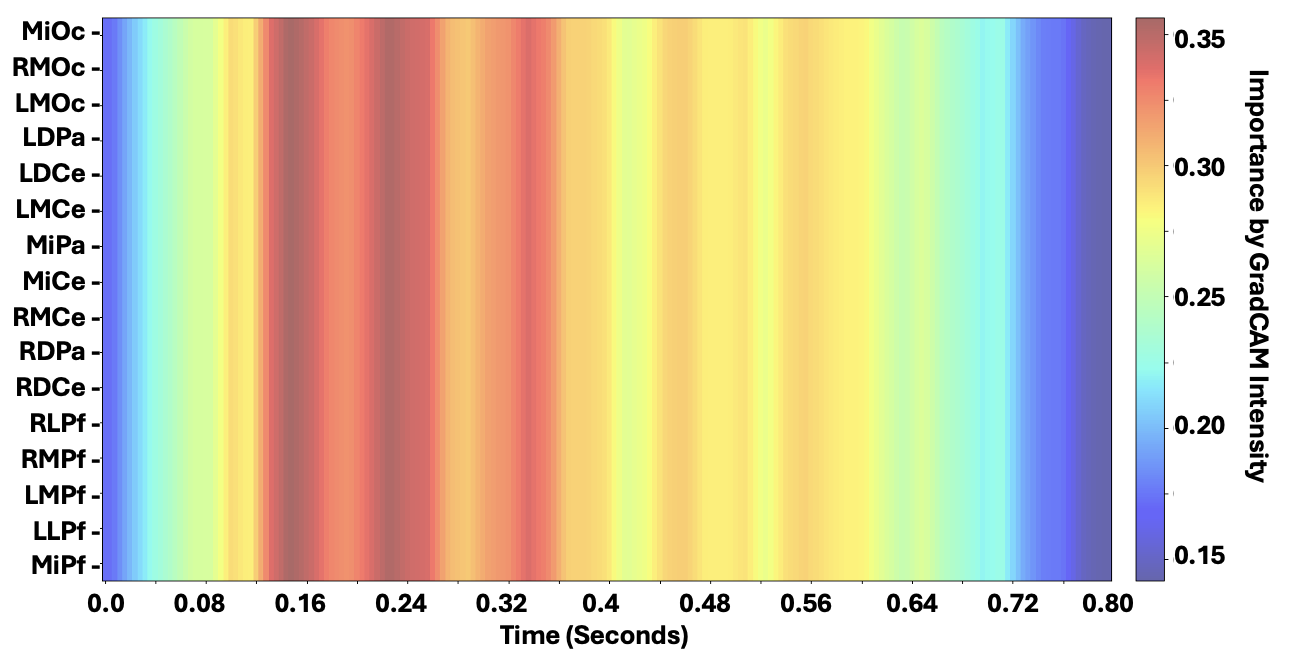} \vspace{-0.4cm}
    \caption{EEGNet calibrated GradCAM}
    \label{fig:snake-order-eegnet-rand-seed-4-calibrated-gradcam}
  \end{subfigure}
    \vspace{-0.4cm} \caption{Supplementary Figure H12. Integrated gradients for calibrated models averaged over discomfort data from balanced accuracy $>75\%$ subjects and GradCAM for EEGNet trained with different channel order. Warmer colors indicate more positive attribution.}
    \label{fig:heatmaps-snake-order}
    \vspace{-0.5cm}
\end{figure*}

\begin{figure*}[h]
    \centering
  \begin{subfigure}{0.5\textwidth}
    \centering
    \includegraphics[width=\linewidth]{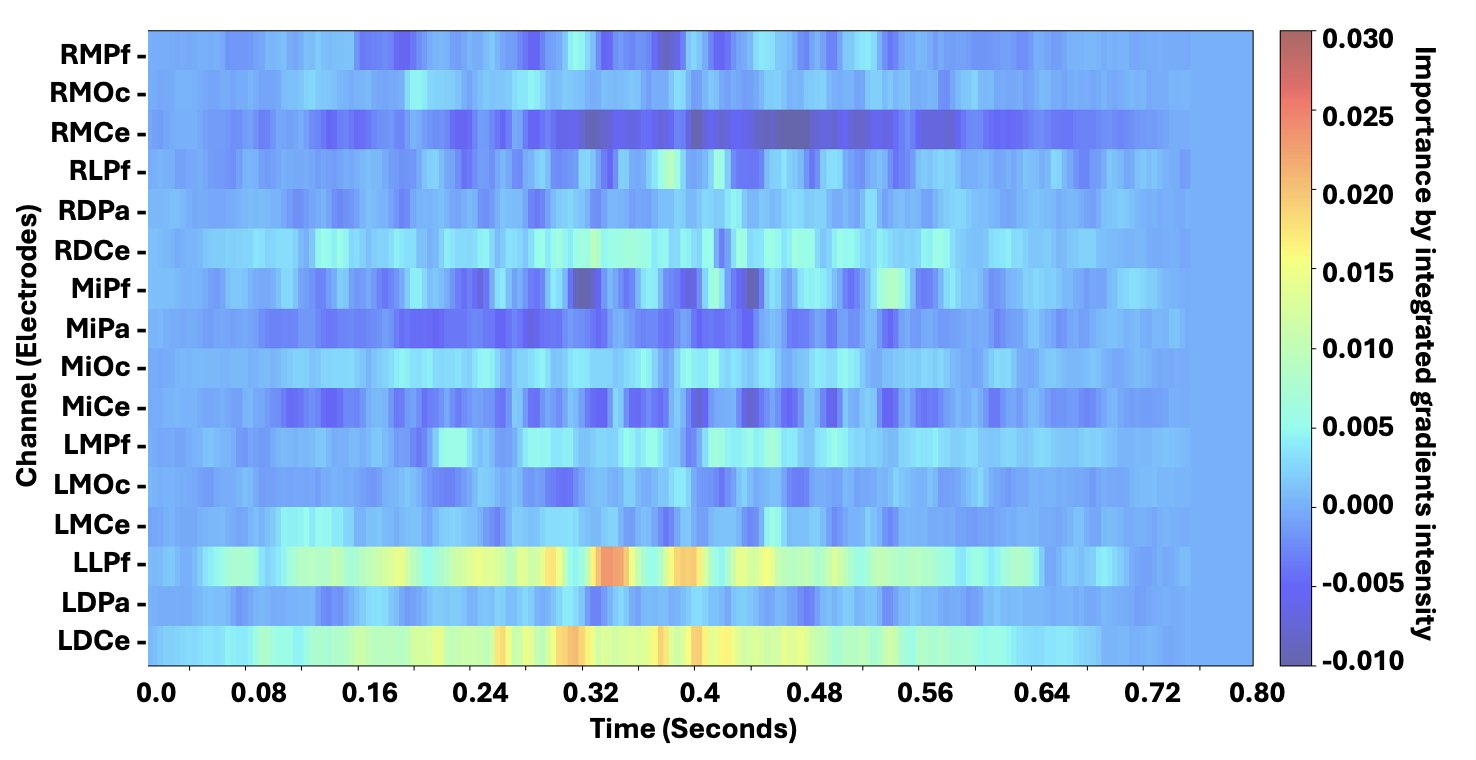}
    \caption{EEG Conformer calibrated integrated gradients}
    \label{fig:conformer-rand-seed-3-calibrated}
  \end{subfigure}\hfill
  \begin{subfigure}{0.5\textwidth}
    \centering
    \includegraphics[width=\linewidth]{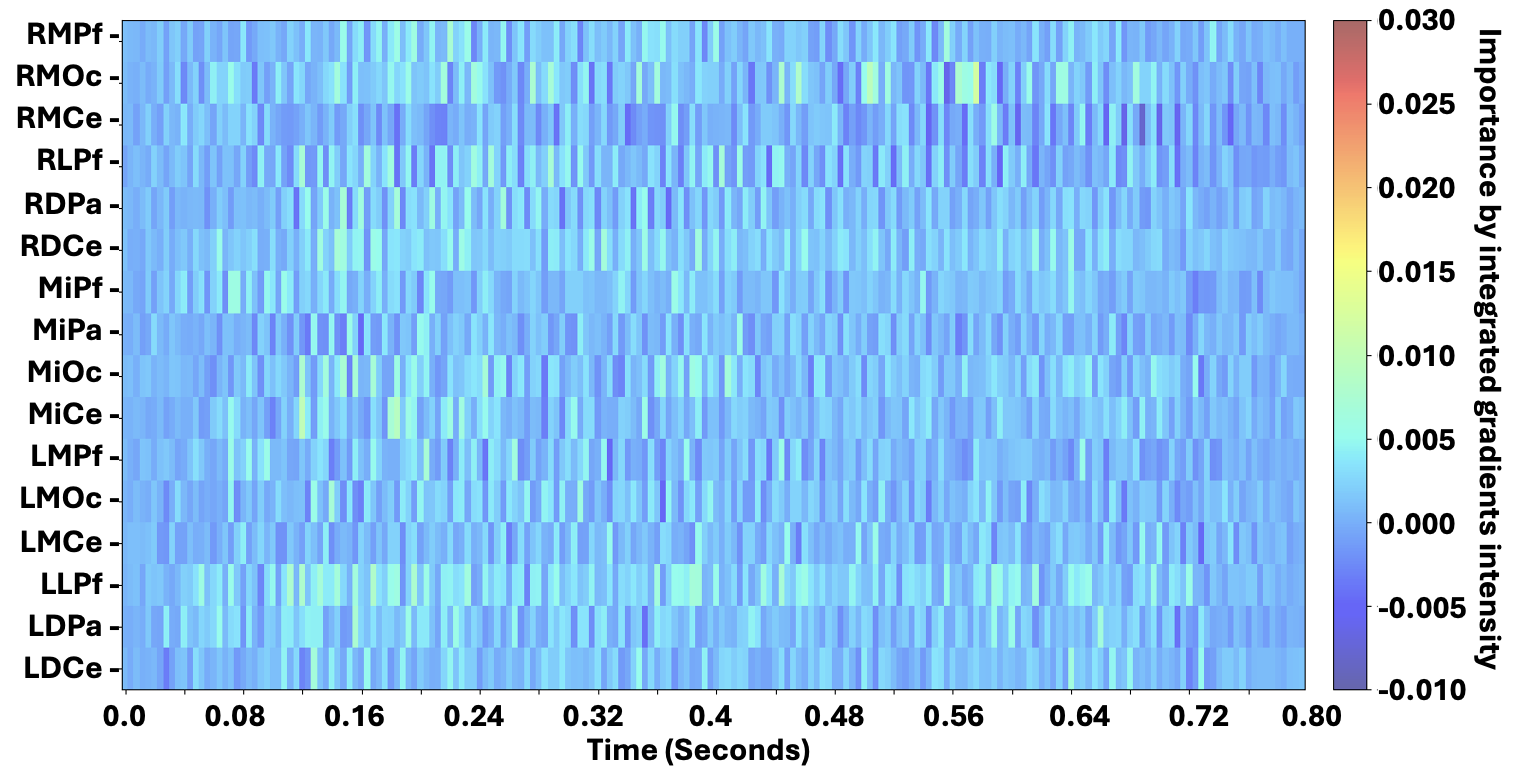}
    \caption{Pre-trained EEG ViT calibrated integrated gradients}
    \label{fig:pretrained-vit-rand-seed-3-calibrated}
  \end{subfigure}

  \vspace{0.75em}

  \begin{subfigure}{0.5\textwidth}
    \centering
    \includegraphics[width=\linewidth]{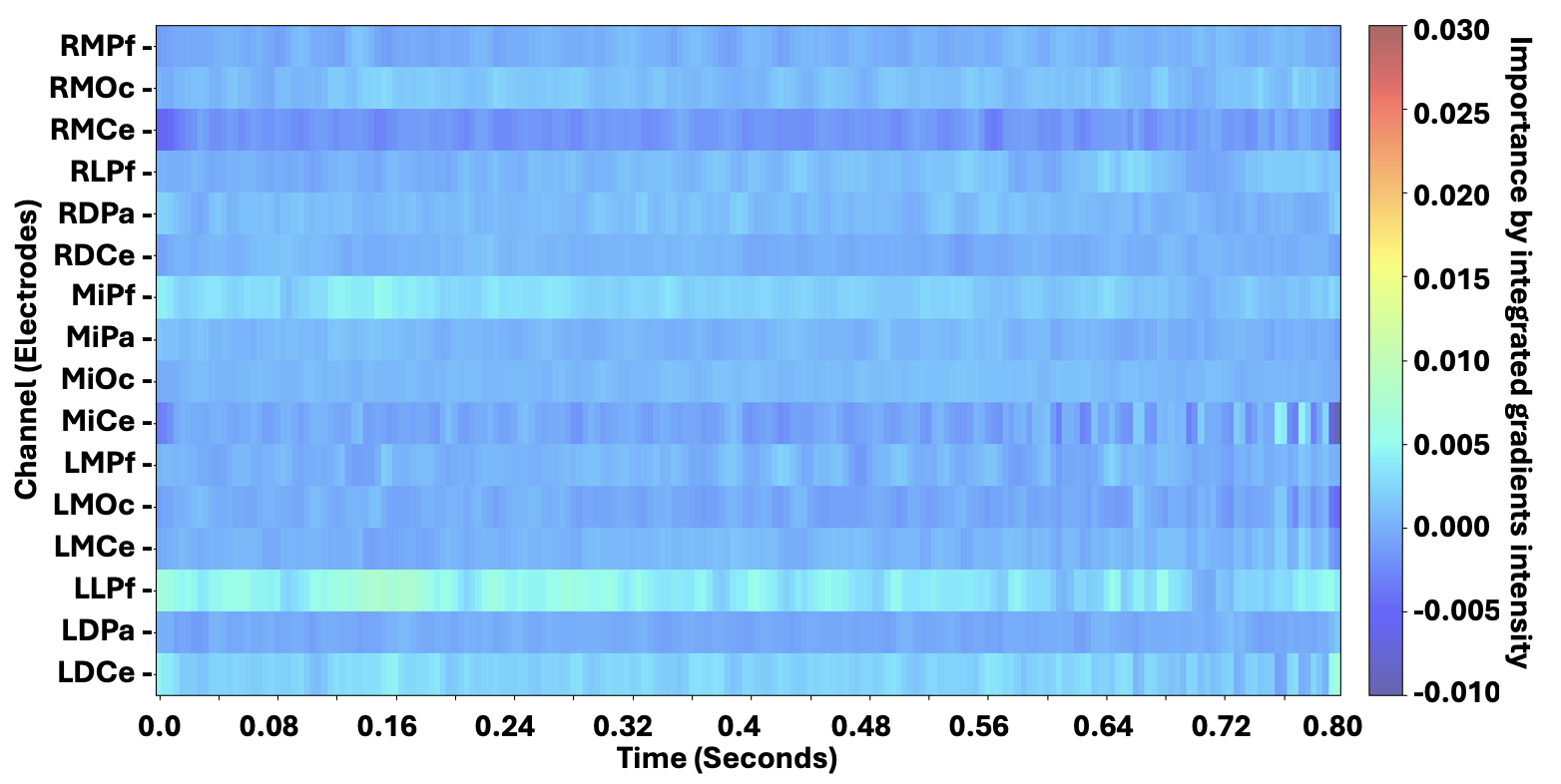}
    \caption{EEGNet calibrated integrated gradients}
    \label{fig:eegnet-rand-seed-3-calibrated-ig}
  \end{subfigure}\hfill
  \begin{subfigure}{0.5\textwidth}
    \centering
    \includegraphics[width=\linewidth]{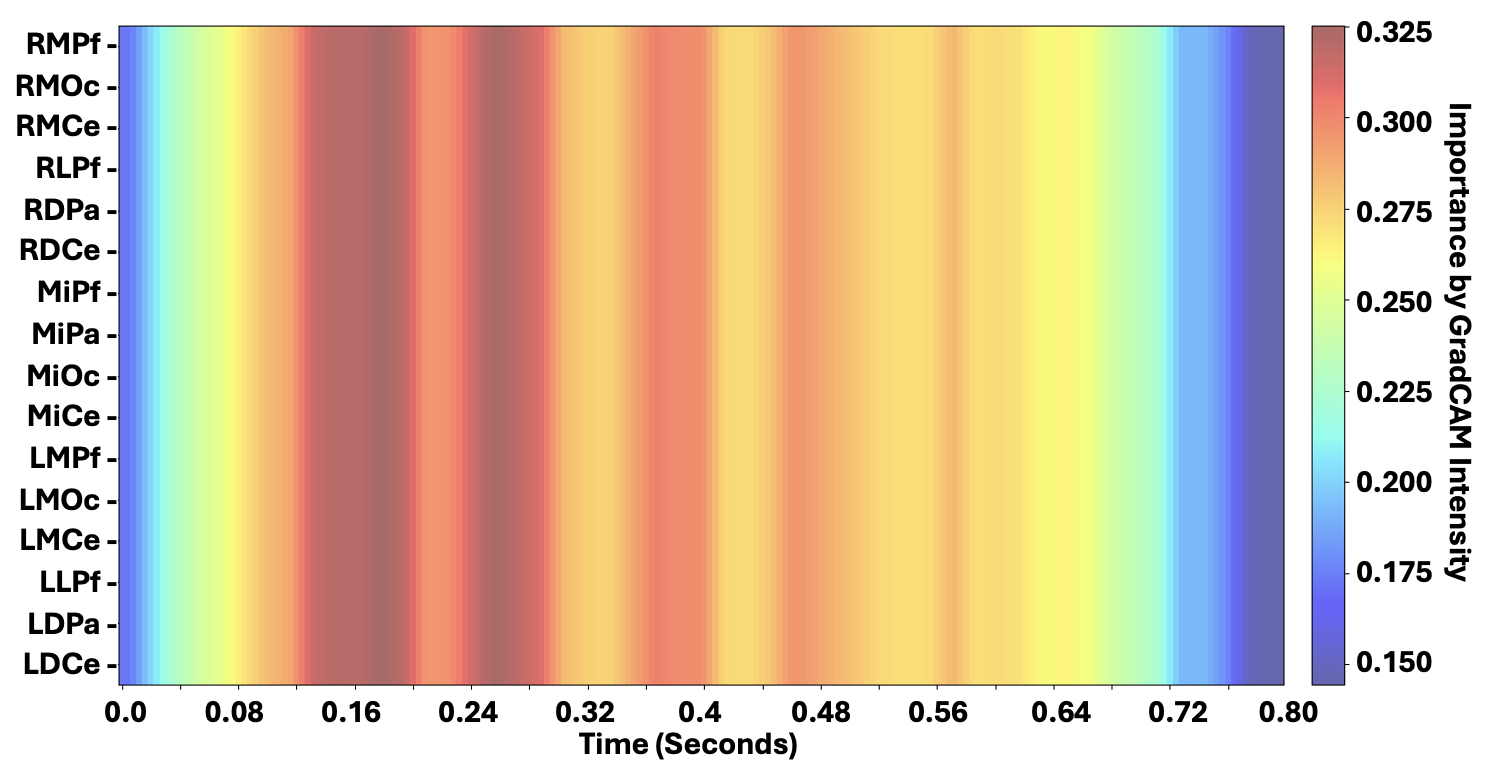}
    \caption{EEGNet calibrated GradCAM}
    \label{fig:eegnet-rand-seed-3-calibrated-gradcam}
  \end{subfigure}
    \caption{Supplementary Figure H13. Integrated gradients for calibrated models averaged over discomfort data from balanced accuracy $>75\%$ subjects and GradCAM for EEGNet with random seed 3. Warmer colors indicate more positive attribution.}
    \label{fig:heatmaps-rand-seed-3}
\end{figure*}

\begin{figure*}
    \centering
  \begin{subfigure}{0.5\textwidth}
    \centering
    \includegraphics[width=\linewidth]{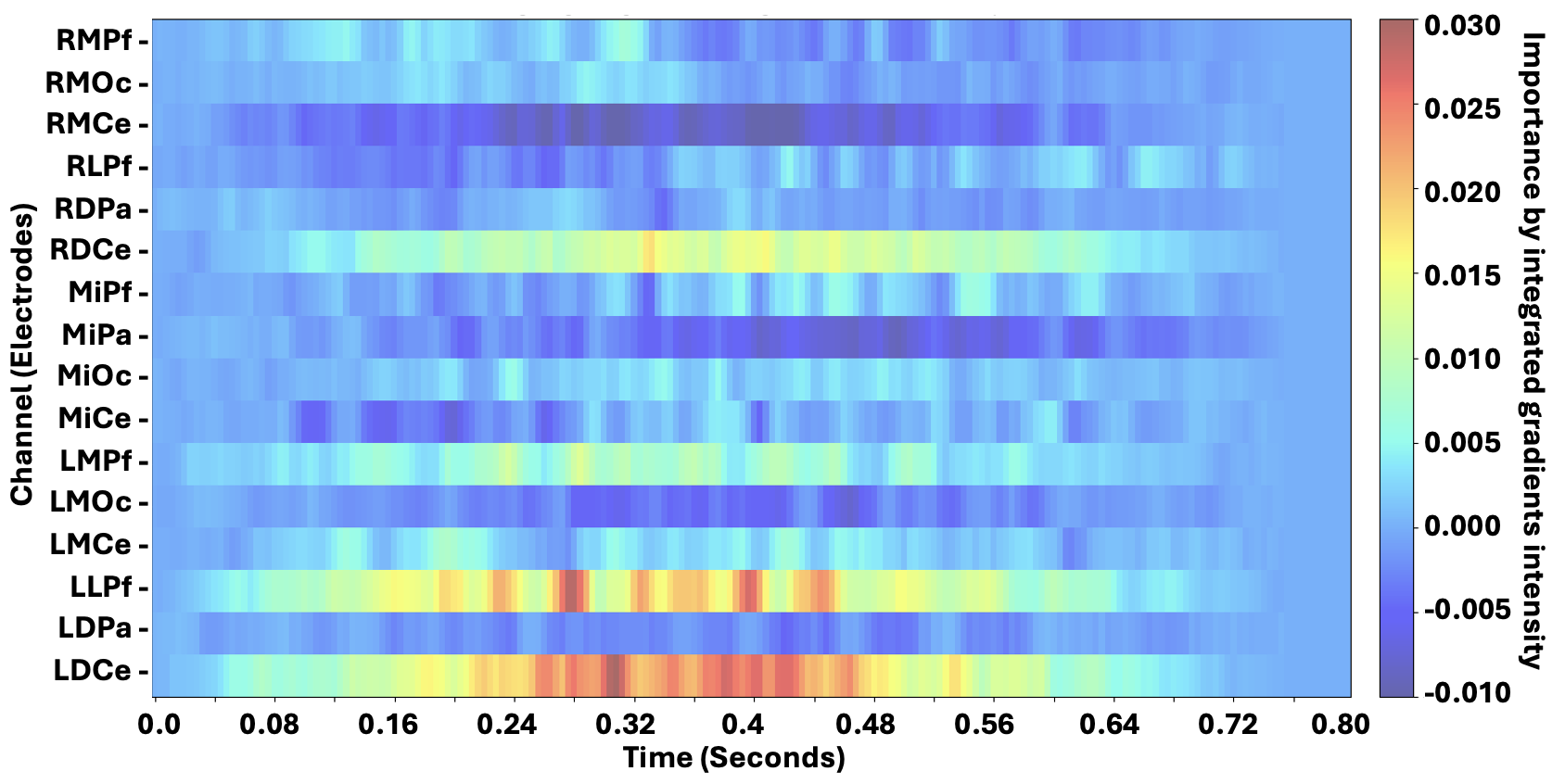}
    \caption{EEG Conformer calibrated integrated gradients}
    \label{fig:conformer-rand-seed-2-calibrated}
  \end{subfigure}\hfill
  \begin{subfigure}{0.5\textwidth}
    \centering
    \includegraphics[width=\linewidth]{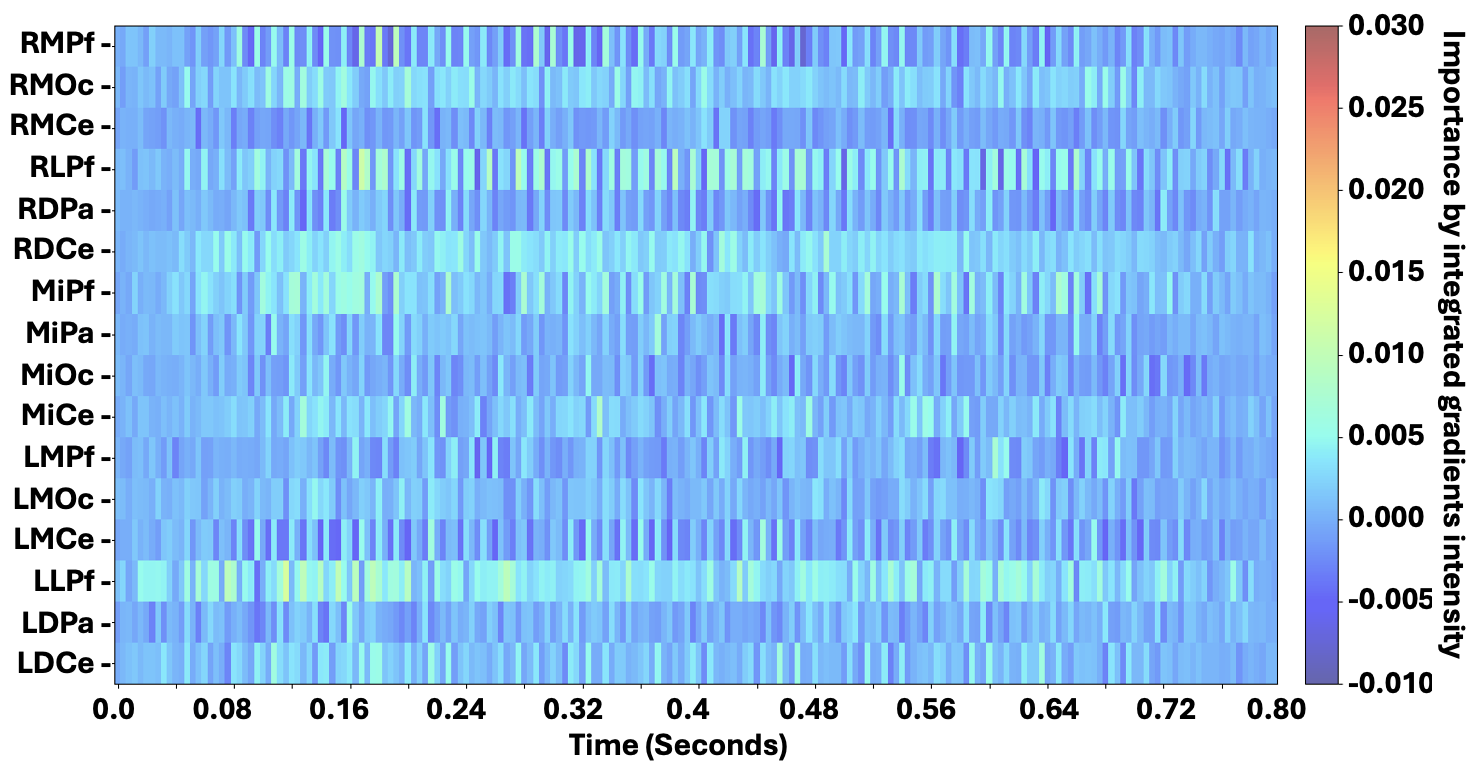}
    \caption{Pre-trained EEG ViT calibrated integrated gradients}
    \label{fig:pretrained-vit-rand-seed-2-calibrated}
  \end{subfigure}

  \vspace{0.75em}

  \begin{subfigure}{0.5\textwidth}
    \centering
    \includegraphics[width=\linewidth]{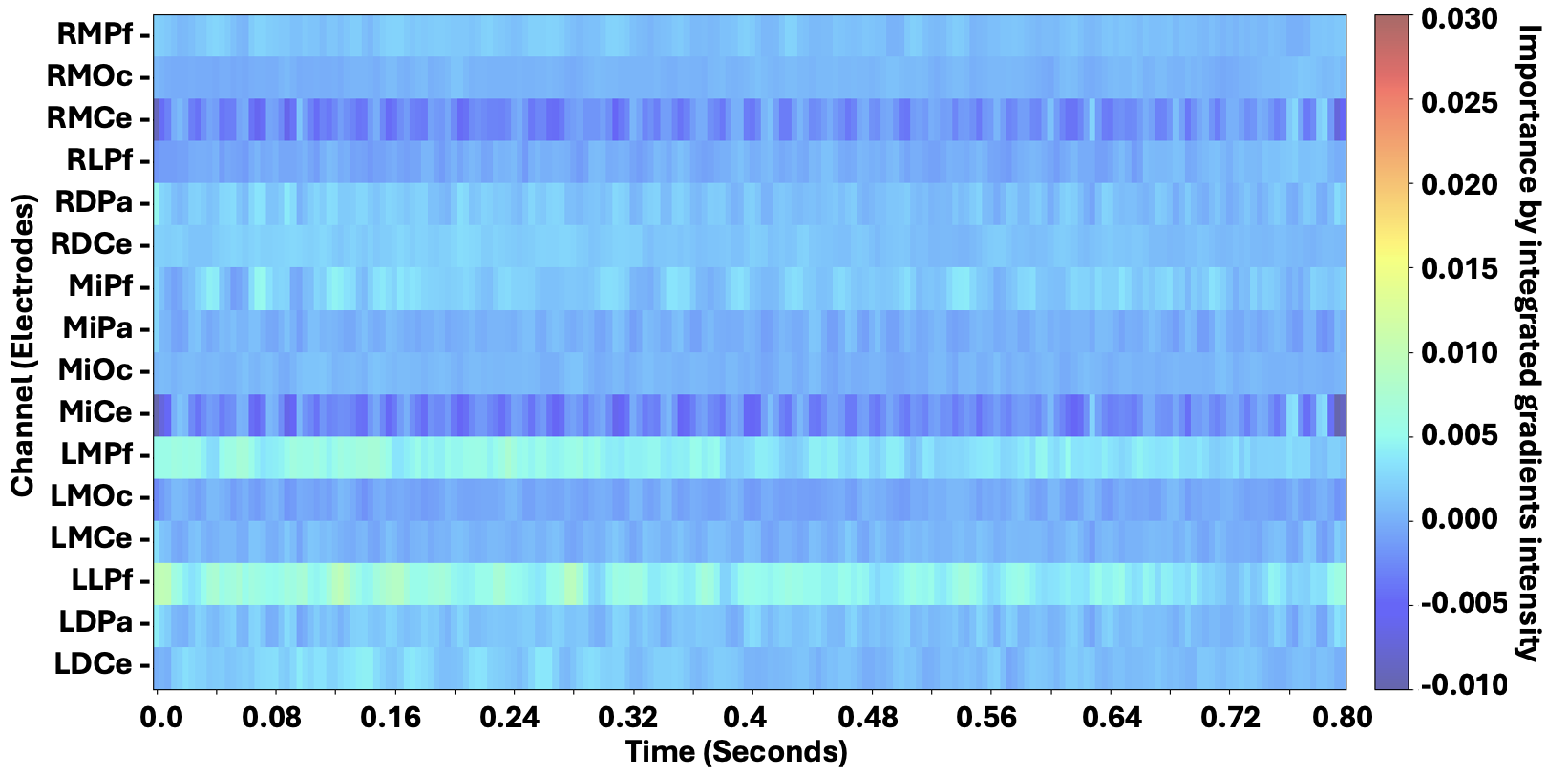}
    \caption{EEGNet calibrated integrated gradients}
    \label{fig:eegnet-rand-seed-2-calibrated-ig}
  \end{subfigure}\hfill
  \begin{subfigure}{0.5\textwidth}
    \centering
    \includegraphics[width=\linewidth]{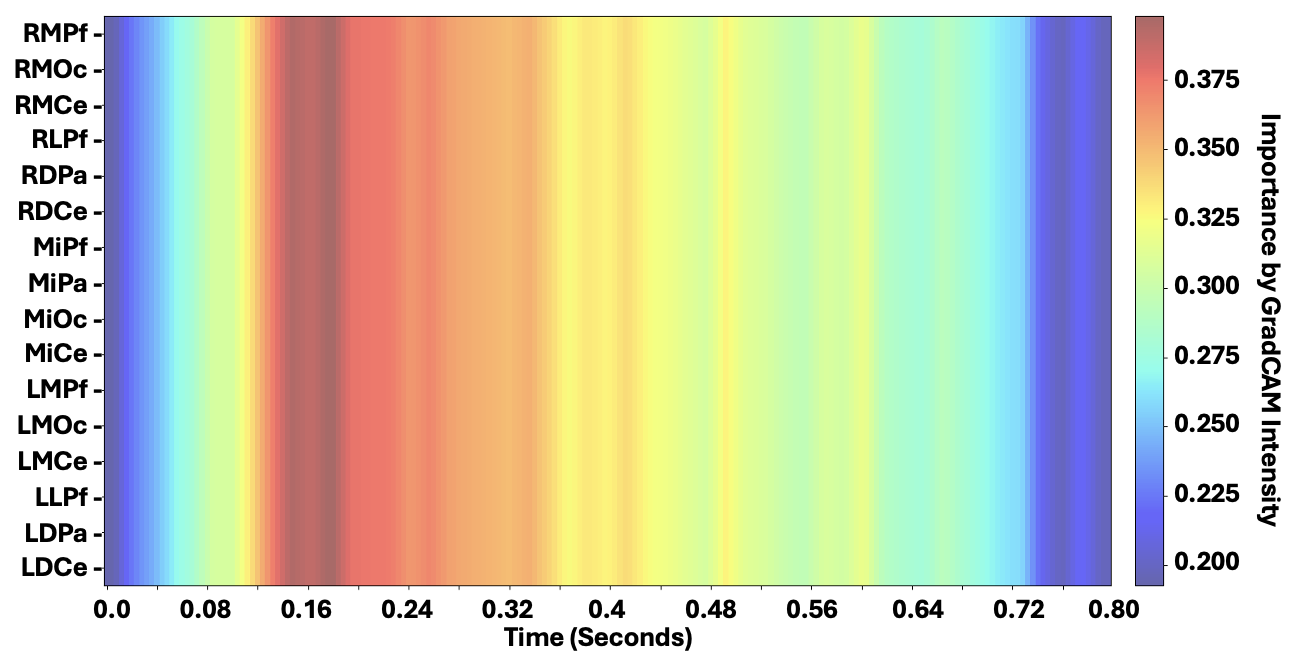}
    \caption{EEGNet calibrated GradCAM}
    \label{fig:eegnet-rand-seed-2-calibrated-gradcam}
  \end{subfigure}
    \caption{Supplementary Figure H14. Integrated gradients for calibrated models averaged over discomfort data from balanced accuracy $>75\%$ subjects and GradCAM for EEGNet with random seed 2. Warmer colors indicate more positive attribution.}
    \label{fig:heatmaps-rand-seed-2}
\end{figure*}

We include additional integrated gradient plots from different random seed experiment trials in Figs. \ref{fig:heatmaps-rand-seed-3} and \ref{fig:heatmaps-rand-seed-2} for ERPVR. The plots for the trial with channel order re-arranged is in Figure \ref{fig:heatmaps-snake-order}. LLPf remains the electrode channel with highest attribution across the models for classifying sickness in ERPVR.

\subsection{Using different SSQ symptom as label}

\begin{figure*}
    \centering
  \begin{subfigure}{0.5\textwidth}
    \centering
    \includegraphics[width=\linewidth]{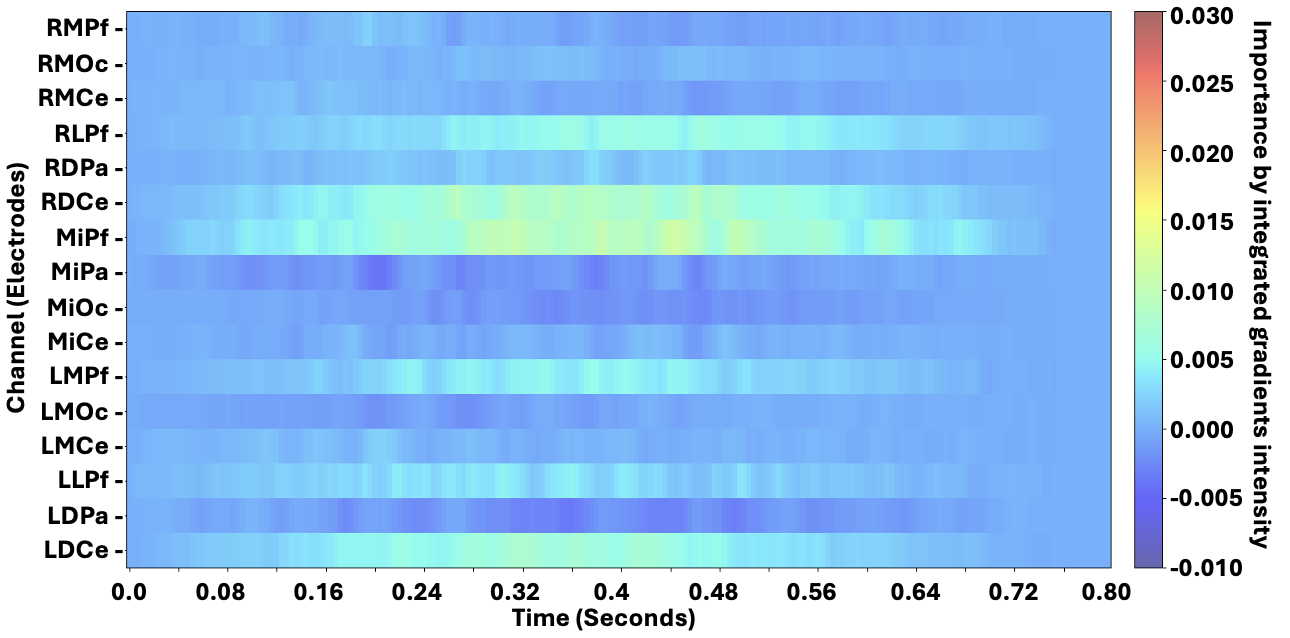}
    \caption{EEG Conformer calibrated integrated gradients}
    \label{fig:conformer-rand-seed-2-calibrated}
  \end{subfigure}\hfill
  \begin{subfigure}{0.5\textwidth}
    \centering
    \includegraphics[width=\linewidth]{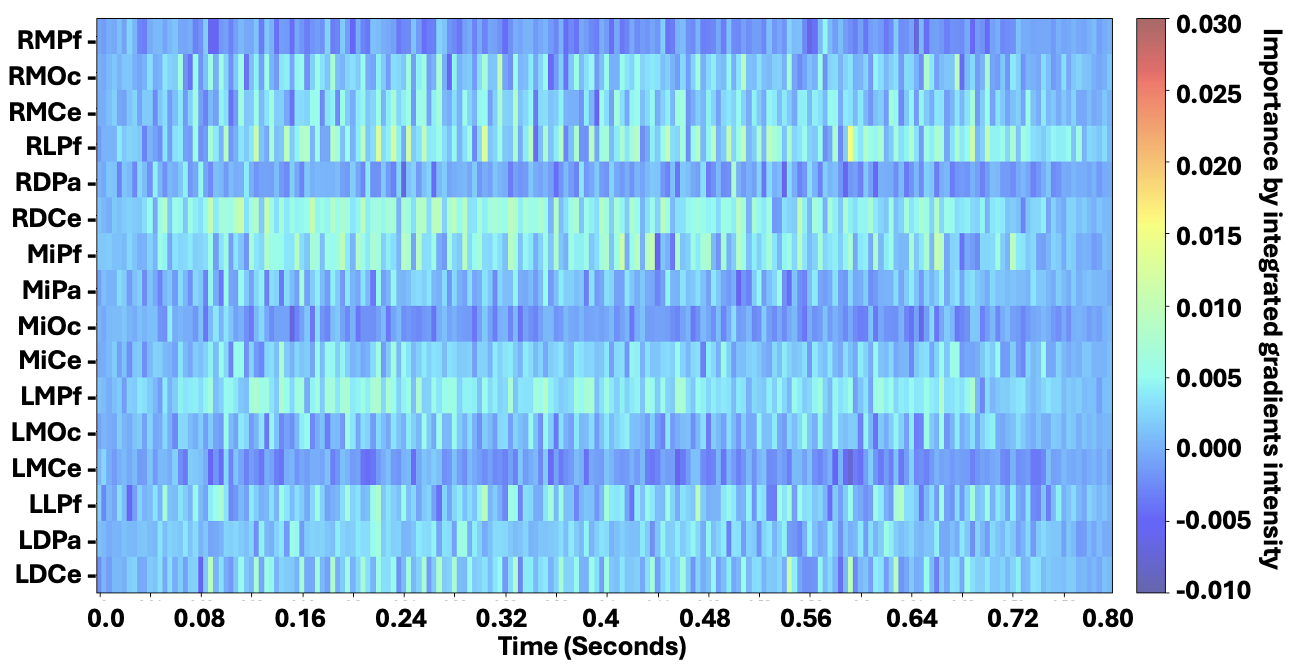}
    \caption{Pre-trained EEG ViT calibrated integrated gradients}
    \label{fig:pretrained-vit-rand-seed-2-calibrated}
  \end{subfigure}

  \vspace{0.75em}

  \begin{subfigure}{0.5\textwidth}
    \centering
    \includegraphics[width=\linewidth]{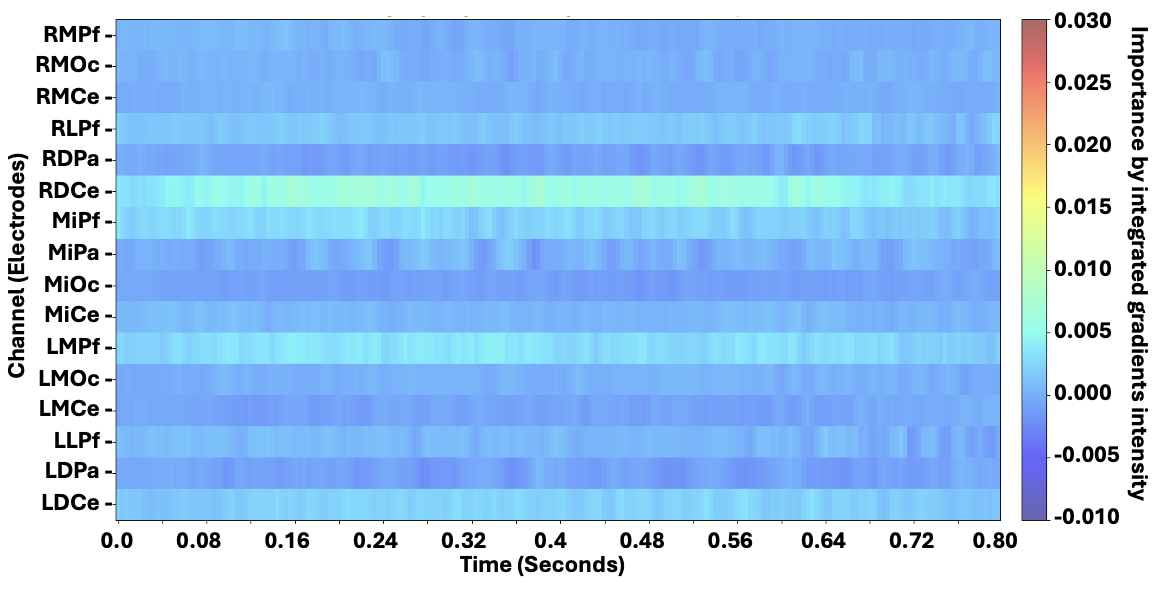}
    \caption{EEGNet calibrated integrated gradients}
    \label{fig:eegnet-rand-seed-2-calibrated-ig}
  \end{subfigure}\hfill
  \begin{subfigure}{0.5\textwidth}
    \centering
    \includegraphics[width=\linewidth]{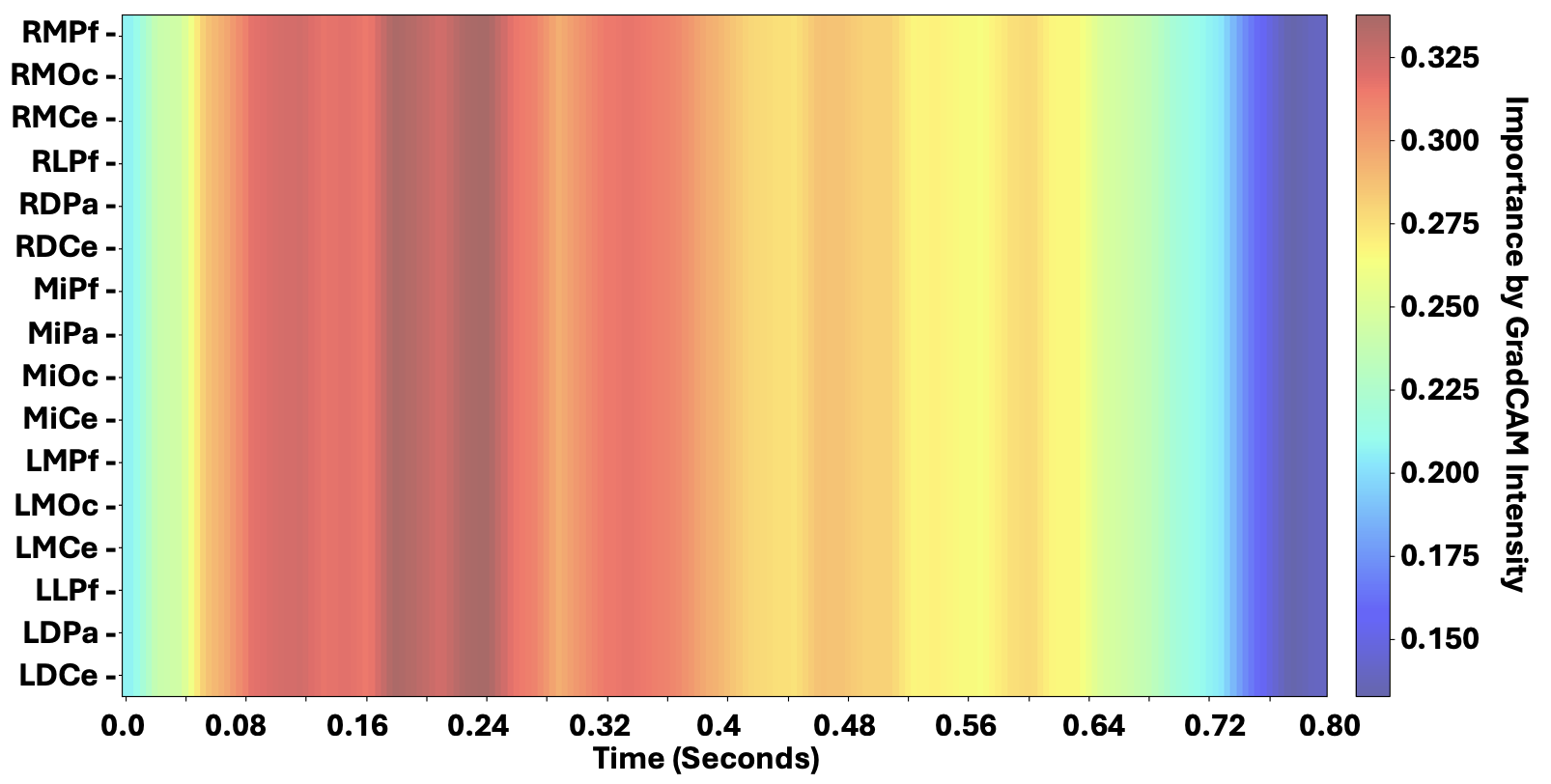}
    \caption{EEGNet calibrated GradCAM}
    \label{fig:eegnet-rand-seed-2-calibrated-gradcam}
  \end{subfigure}
    \caption{Supplementary Figure H15. Integrated gradients for calibrated models averaged over discomfort data from balanced accuracy $>75\%$ subjects and GradCAM for EEGNet for ``Difficulty focusing" labels. Warmer colors indicate more positive attribution.}
    \label{fig:difficulty-focusing}
\end{figure*}

To verify our method works for other kinds of patterns, ``Difficulty focusing," a different SSQ symptom, was used to create binary classification labels for ERPVR in Figure \ref{fig:difficulty-focusing}. The top attribution channels are RDCe and MiPf this time, highlighting different electrodes are useful for detecting ``Difficulty focusing".

\section{ERP Waveform Analysis by Electrode Groups}

\begin{figure*}
    \begin{subfigure}{\textwidth}
        \centering
        \includegraphics[width=\linewidth]{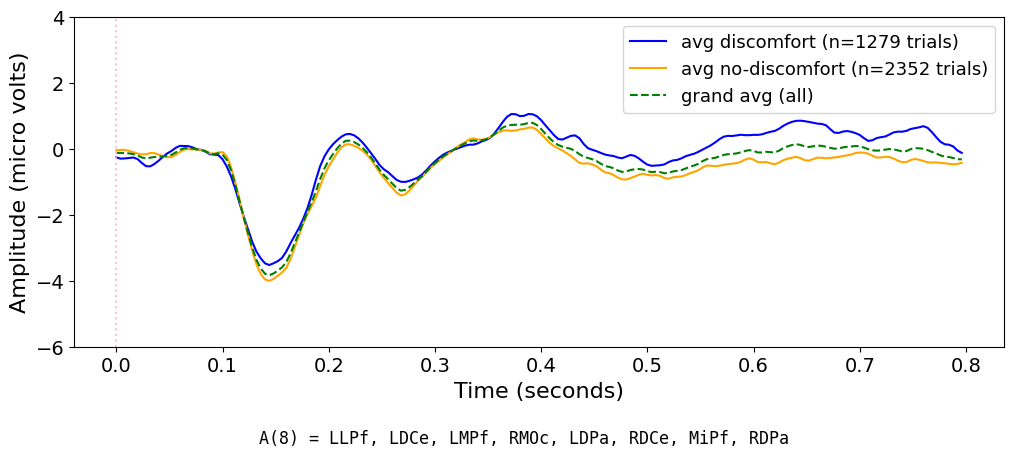}
        \caption{ERPVR ERP Oddballs waveforms averaged by top eight contributing electrodes for discomfort classification, per label.}
    \end{subfigure}
    \begin{subfigure}{\textwidth}
        \centering
        \includegraphics[width=\linewidth]{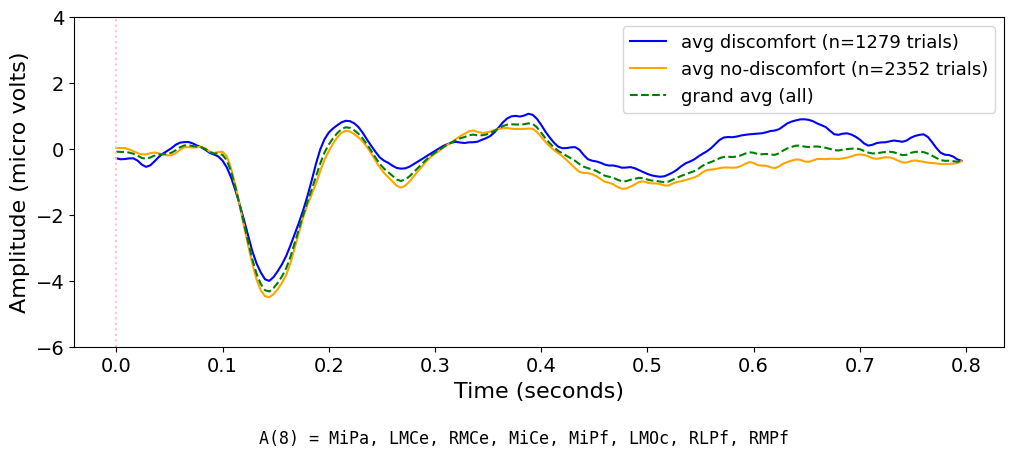}
        \caption{ERPVR ERP Oddballs waveforms averaged by top eight contributing electrodes for no discomfort classification, per label.}
    \end{subfigure}
    \caption{Supplementary Figure I17. ERPVR Oddballs ERP waveforms averaged by identified electrode subset that contributed positively for model classification for each label. Vertical pink dotted line indicates when the audio stimulus was played, around 0.0 seconds.}
    \label{fig:erpvr-top-8-avg-wave-odd}
\end{figure*}

After identifying electrodes of interest from the ERPVR dataset and VectionVR dataset, we plot waveforms by averaging over the identified top 8 consistent electrodes for discomfort vs. no discomfort. For ERPVR no discomfort, the top 8 electrodes that contributed to classification were: MiPa, RMCe, LMCe, MiCe, LMOc, RMPf, MiPf, RDPa. For ERPVR discomfort, the top 8 electrodes that contributed to classification were: LLPf, LDCe, RDCe, RMOc, LDPa, LMPf, MiPf, MiOc. Both are plotted in Supplementary Figure \ref{fig:erpvr-top-8-avg-wave-std} for Standards waveforms. Oddballs are at Supplementary Figure \ref{fig:erpvr-top-8-avg-wave-odd}. For VectionVR for high discomfort, we have: C3, T7, CPz, F8, Fp2, FC1, Cz, Oz and for low discomfort, we have: P2, Fp1, Pz, F7, P4, P8, CP6, F3. The VectionVR averaged waveforms are in Supplementary Figure \ref{fig:vectionvr-top-8-avg-wave-std}.

\begin{figure*}
\begin{subfigure}{\textwidth}
    \centering
    \includegraphics[width=\linewidth]{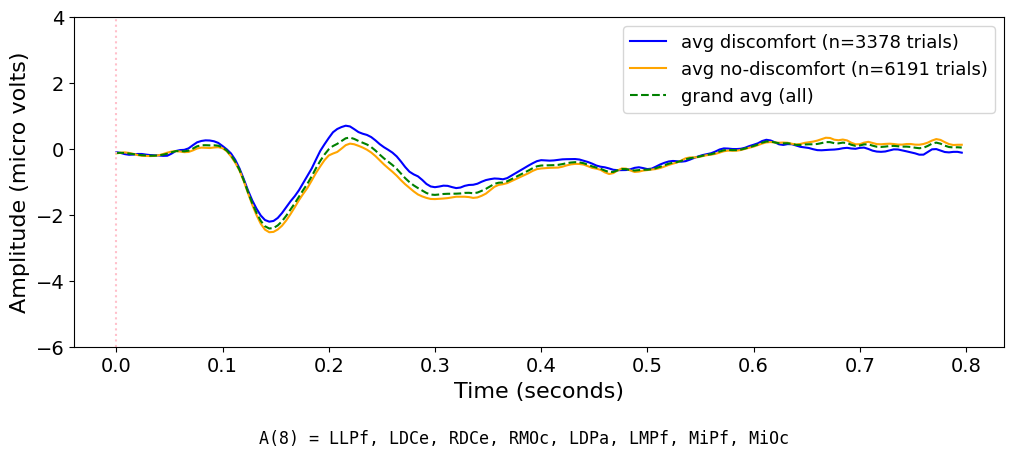}
    \caption{ERPVR ERP Standards waveforms averaged by top eight contributing electrodes for discomfort classification, per label.}
\end{subfigure}
\begin{subfigure}{\textwidth}
    \centering
    \includegraphics[width=\linewidth]{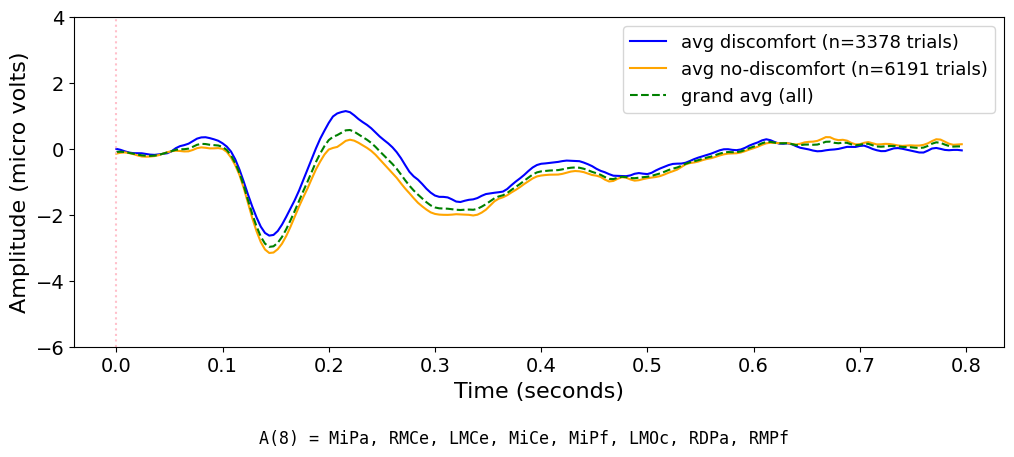}
    \caption{ERPVR ERP Standards waveforms averaged by top eight contributing electrodes for no discomfort classification, per label.}
    \label{fig:erpvr-no-disc-top-8-std}
\end{subfigure}
\caption{Supplementary Figure I16. ERPVR Standards ERP waveforms averaged by identified electrode subset that contributed positively for model classification for each label. Vertical pink dotted line indicates when the audio stimulus was played, around 0.0 seconds.}
    \label{fig:erpvr-top-8-avg-wave-std}
\end{figure*}

\begin{figure*}
    \begin{subfigure}{\textwidth}
        \centering
        \includegraphics[width=\linewidth]{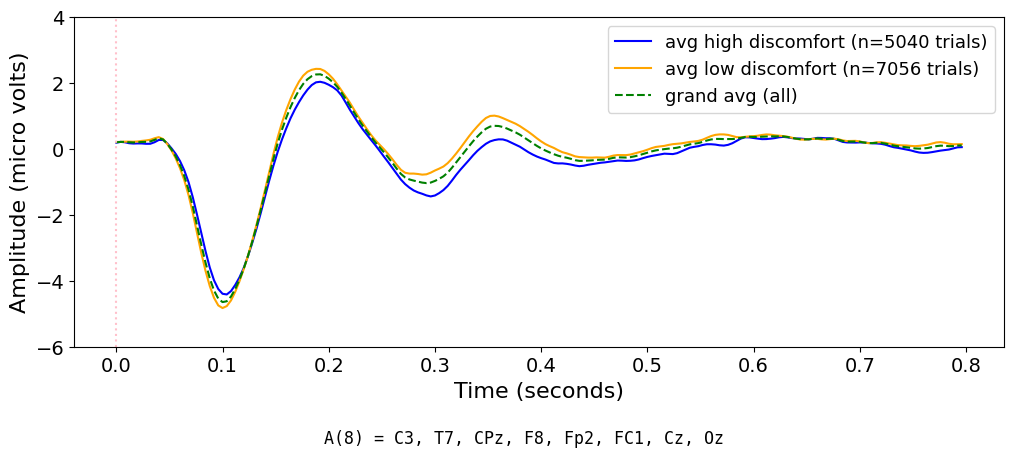}
        \caption{VectionVR ERP Standards waveforms averaged by top eight contributing electrodes for high discomfort classification, per label.}
    \end{subfigure}
    \begin{subfigure}{\textwidth}
        \centering
        \includegraphics[width=\linewidth]{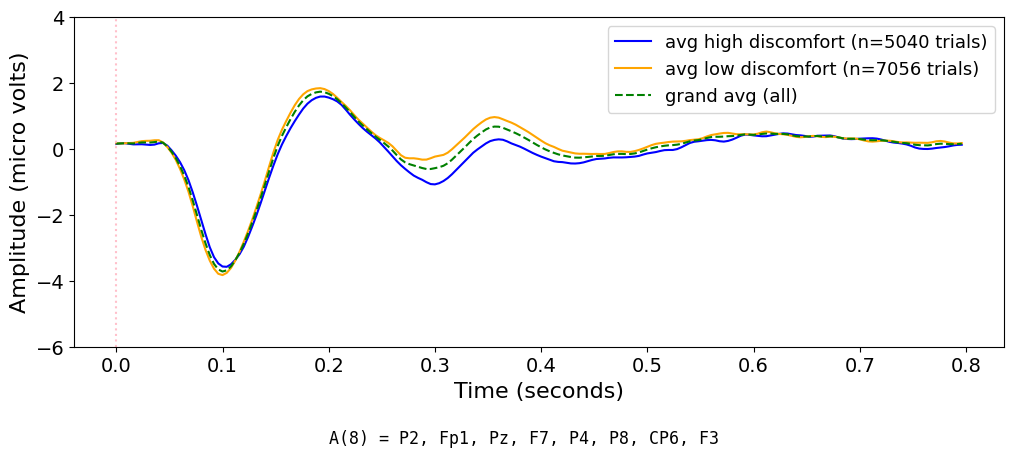}
        \caption{VectionVR ERP Standards waveforms averaged by top eight contributing electrodes for low discomfort classification, per label.}
    \end{subfigure}
    \caption{Supplementary Figure I18. VectionVR Standards ERP waveforms averaged by identified electrode subset that contributed positively for model classification for each label. Vertical pink dotted line indicates when the audio stimulus was played, around 0.0 seconds.}
    \label{fig:vectionvr-top-8-avg-wave-std}
\end{figure*}

\section{ERPVR using SSQ Unweighted Symptom Sum as Label}

To evaluate another labeling method instead of taking a single symptom of SSQ, we tried using the SSQ unweighted symptom sum (USS) as a label, where it is defined as the sum across all 16 SSQ symptoms. Binary labels are created where USS $>0$ is sick and $USS=0$ is nonsick. With this labeling, the subjects split as so: 12 pure-sick, 12 mixed labels, and 5 pure-nonsick. We tested this labeling method with EEGNet and Standards data. Looking at subject 15 who was mixed label under the general discomfort labeling scheme, they also were mixed label under the USS labeling. So as a subject who has mixed labels in both labeling formats, it was interesting to look at how the model performs with the different labeling. However, subject 15's balanced accuracy after calibration dropped greatly (38\% from 44.43\%) when using the USS to set the binary classification labels, compared to the general discomfort labels (81\%). This may be because when using the SSQ unweighted symptom sum, all symptoms are mixed together to compute the sum. With all SSQ symptoms mixed together, it may be harder for the models to extract specific patterns distinguishing between the sick and nonsick labels defined by the USS. Also, the very low balanced accuracy for subject 15 at base LOSO indicates that the training pool shift from the USS labels and the patterns the model learned from this shifted train set is misaligned to subject 15 such that even calibration is not able to recover the model performance on the subject.

\section{Measuring Cybersickness with the P3b}\label{sec:P3b-backgroud-section}

The P3b ERP in EEG (also called the P3 or P300) is one of the most well-studied ERP components~\cite{Polich_2007}. Though there are several theories as to what specific cognitive process the P3b reflects, across decades of research and hundreds of studies, the amplitude of the P3b has consistently been shown to be influenced by task difficulty, cognitive load, and attention~\cite{Ghani_Signal_Niazi_Taylor_2020}. As such, it has been used in myriad studies to investigate attentional allocation~\cite{Vanbilsen_Kotz_Rosso_Leman_Triccas_Feys_Moumdjian_2023,Ghani_Signal_Niazi_Taylor_2020,Polich_2007}.


Attention can be conceptualized as a cognitive resource for prioritizing processing to achieve a goal~\cite{Nobre_Kastner_2014}. 
Over the years, research on attention has shown that attentional changes are more apparent in brain or behavior measures when there is greater competition for processing resources ~\cite{Nobre_Kastner_2014,Polich_2007,Ghani_Signal_Niazi_Taylor_2020}. Thus, a common method to study attention is to use 
a dual-task paradigm~\cite{Kasper_Cecotti_Touryan_Eckstein_Giesbrecht_2014}. In dual-task paradigms, participants are given two tasks that must be performed together, sometimes in different sensory modalities. 
When there is a primary visual task, attentional changes can be measured with the P3b from an auditory secondary task, like the oddball paradigm~\cite{Polich_2007}. Auditory oddball paradigms involve having participants listen to a series of beeps. Infrequently and randomly, a beep will sound different from the other beeps. Participants are asked to pay attention to these series of beeps and respond in some way, for example by keeping a mental count of how many different sounding beeps they heard~\cite{Vanbilsen_Kotz_Rosso_Leman_Triccas_Feys_Moumdjian_2023}. This simple method to produce the P3b and measure attention is reliable and sensitive enough that it has been used in at least 35 studies with more than 1,500 participants to investigate attentional differences in neurological disorders~\cite{Vanbilsen_Kotz_Rosso_Leman_Triccas_Feys_Moumdjian_2023}.

Utilizing this robust and empirically validated attentional measure, we~\cite{Mimnaugh2023} showed that cybersickness impacts attention. Using P3b amplitudes to an auditory secondary task during a primary visual task in virtual reality, we demonstrated that amplitudes of the P3b were directly associated with the amount of cybersickness that participants experienced, with reduced P3b amplitudes during periods of greater sickness. Consequently, the ERPVR EEG data consists of hundreds of periods (trials) of brain activity generated in response to the exact same stimulus (the beep) for each person. Embedded within these trials are very similar patterns of auditory cognitive processing which are subtly modified by the amount of discomfort experienced. Although the brain activity pattern from the auditory stimulus repeats hundreds of times (providing a rich source of data to be parsed algorithmically), the visual stimuli from the VR environment were different during each trial. Thus, the randomness of the visual cognitive processing essentially becomes minor background noise during each largely similar window of auditory stimulus processing. As a result, this ERP data controls for a visual confound that usually exists in EEG cybersickness data.



\end{appendices}

\end{document}